\documentclass[traditabstract]{aa}
\usepackage{longtable,lscape,amssymb}
\usepackage{graphicx}
\usepackage{txfonts}
\usepackage{epsfig}
\usepackage{natbib}

\bibpunct{(}{)}{;}{a}{}{,} 

\newcommand{\Ha}{{H$\alpha$}}
\newcommand{\Hb}{{H$\beta$}}
\newcommand{\Hg}{{H$\gamma$}}
\newcommand{\Hd}{{H$\delta$}}
\newcommand{\Hep}{{H$\epsilon$}}

\newcommand{\pab}{{Pa$\beta$}}
\newcommand{\pag}{{Pa$\gamma$}}
\newcommand{\pad}{{Pa$\delta$}}
\newcommand{\brg}{{Br$\gamma$}}
\newcommand{\brd}{{Br$\delta$}}

\newcommand{\Ll}{{$L_{\rm line}$}}

\newcommand{\Lacc}{{$L_{\rm acc}$}}
\newcommand{\Macc}{{$\dot{M}_{\rm acc}$}}

\newcommand{\Msun}{{$M_{\odot}$}}
\newcommand{\Lsun}{{$L_{\odot}$}}
\newcommand{\Rsun}{{$R_{\odot}$}}

\newcommand{\Mstar}{{$M_{\star}$}}
\newcommand{\Lstar}{{$L_{\star}$}}
\newcommand{\Rstar}{{$R_{\star}$}}
\newcommand{\Teff}{{$T_{\rm eff}$}}

\newcommand{\Mdisk}{{$M_{\rm disc}$}}

\begin{document}

\title{X-Shooter spectroscopy of young stellar objects in Lupus: 
\subtitle{Accretion properties of class~II and transitional objects. \thanks{Based on 
 observations collected at the European Southern Observatory at Paranal, under programs 
 084.C-0269(A), 085.C-0238(A), 086.C-0173(A), 087.C-0244(A), 089.C-0143(A), 095.C-0134(A), 
 097.C-0349(A), and archive data of programmes 085.C-0764(A) and 093.C-0506(A).}
 ~\fnmsep\thanks{Figures \ref{correl1} to \ref{slab5} and Tables \ref{tab:fluxes_EWs_Hae} 
 to \ref{tab:fluxes_EWs_NaI} are available in electronic form at http://www.aanda.org } } 
}

\author{
       J.M.~Alcal\'a\inst{1} 
  \and C.F.~Manara\inst{2} 
  \and A.~Natta\inst{3,4} 
  \and A.~Frasca\inst{5}
  \and L.~Testi\inst{3,6,7}  
  \and B.~Nisini\inst{8}
  \and B.~Stelzer\inst{9}
  \and J.~P.~Williams\inst{10}
  \and S.~Antoniucci\inst{8}
  \and K.~Biazzo\inst{5}
  \and E.~Covino\inst{1}
  \and M.~Esposito\inst{1}
  \and F.~Getman\inst{1} 
  \and E.~Rigliaco\inst{11} 
}

\offprints{J.M. Alcal\'a}
\mail{alcala@oacn.inaf.it}

\institute{ 
      INAF-Osservatorio Astronomico di Capodimonte, via Moiariello 16, 80131 Napoli, Italy
 \and Scientific Support Office, Directorate of Science, European Space Research and 
      Technology Centre (ESA/ESTEC), Keplerlaan 1,2201 AZ Noordwijk, The Netherlands
 \and INAF-Osservatorio Astrofisico di Arcetri, via Moiariello 16, Largo E. Fermi 5, 50125 Firenze, Italy
 \and DIAS/School of Cosmic Physics, Dublin Institute for Advanced Studies, 31 Fitzwilliams Place, Dublin 2, Ireland
 \and INAF-Osservatorio Astrofisico di Catania, via S. Sofia 78, 95123 Catania, Italy
 \and European Southern Observatory, Karl-Schwarzschild-Str. 2, 85748 Garching, Germany
 \and Excellence Cluster Universe, Boltzmannstr. 2, D-85748 Garching, Germany
 \and INAF-Osservatorio Astronomico di Roma, Via di Frascati 33, 00078 Monte Porzio Catone, Italy
 \and INAF-Osservatorio Astronomico di Palermo, Piazza del Parlamento 1, 90134 Palermo, Italy
 \and Institute for Astronomy, University of Hawaii at Manoa, Honolulu, HI, USA
 \and INAF Osservatorio Astornomico di Padova, vicolo dell'Osservatorio 5, 35122 Padova, Italy
 }

\date{Received ; accepted  }

\abstract{ 
The mass accretion rate, \Macc, is a key quantity for the understanding of the
physical processes governing the evolution of accretion discs around young 
low-mass (\Mstar$\lesssim$2.0\Msun) stars and substellar objects (YSOs).
We present here the results of  a study of the stellar and accretion properties of the (almost) 
complete sample of class II and transitional YSOs in the Lupus I, II, III and IV clouds, based on 
spectroscopic data acquired with the VLT/X-Shooter spectrograph.  
Our study combines the dataset from our previous work with new observations of 
55 additional objects. We have investigated 92 YSO candidates in total, 11 of which have been 
definitely identified with giant stars unrelated to Lupus.
The stellar and accretion properties of the 81 bona fide YSOs, which represent more than 90\% 
of the whole class~II and transition disc YSO population in the aforementioned Lupus clouds, have 
been homogeneously and self-consistently derived, allowing for an unbiased study of accretion 
and its relationship with stellar parameters. 

The accretion luminosity, \Lacc, increases with the stellar luminosity, \Lstar, with an overall 
slope of $\sim$1.6, similar but with a smaller scatter than in previous studies.  There is a 
significant lack of strong accretors  below \Lstar$\approx$0.1\Lsun, where \Lacc\ is always lower 
than 0.01\,\Lstar. We argue that the \Lacc-\Lstar\ slope is not due to observational biases, 
but is a true property of the Lupus YSOs. 
The $\log$\Macc--$\log$\Mstar\ correlation shows a statistically significant evidence of a break, 
with a steeper relation for \Mstar$\lesssim$0.2\,\Msun\ and a flatter slope for higher masses. 
The bimodality of the \Macc--\Mstar\ relation is confirmed with four different evolutionary models 
used to derive the stellar mass. 
The bimodal behaviour of the observed relationship supports the importance of modelling self-gravity 
in the early evolution of the more massive discs, but other processes, such as photo-evaporation and 
planet formation during the YSO's lifetime, may also lead to disc dispersal on different timescales 
depending on the stellar mass.

The sample studied here more than doubles the number of YSOs 
with homogeneously and simultaneously determined \Lacc\ and luminosity, \Ll, of many permitted 
emission lines. Hence, we also refined the empirical relationships between \Lacc\ and \Ll\ on a more solid statistical basis.}

\keywords{Stars: pre-main sequence, low-mass -- Accretion, accretion disks -- Open clusters and associations: Lupus}

\titlerunning{Accretion in Lupus YSOs}
\authorrunning{Alcal\'a et al.}
\maketitle

\section{Introduction} 
\label{Sec:intro}

The mass accretion rate, \Macc, is a crucial parameter for the study of the evolution of accretion 
discs around young low-mass (\Mstar$\lesssim$2.0\Msun) stellar and substellar objects (YSOs). It sets 
important constraints for disc evolution models \citep{hartmann98} and disc clearing mechanisms 
\citep[][and references therein]{alexander14}, and is a key quantity for the studies of pre-main 
sequence (PMS) stellar evolution  and planet formation. 
Observationally, \Macc ~can be derived by measuring the flux in excess to the photospheric one due to the 
release of the accretion energy in the form of continuum emission and lines (accretion luminosity \Lacc)
and using the stellar properties 
\citep[see][]{gullbring98, hart98}.
 Continuum excess luminosity has been measured in a number of objects from spectroscopy at 
different resolutions \citep[e.g.][]{gullbring98, HH08, rigliaco12, ingleby13, alcala14, ingleby14, manara14,
manara16a}. 
More often, \Lacc ~has been computed from empirical relations between line luminosity, \Ll , and \Lacc 
~\citep[e.g.][and reference therein]{natta06, fang13, biazzo12, antoniucci14, manara15}.
The results of these  works showed that \Macc ~shows up to three orders of magnitude of unexplained 
spread for stars of similar mass and age and over the mass spectrum. 

During the class~II phase - after the protostar has almost entirely dispersed its envelope
but is still actively accreting from the optically thick accretion disc - the stellar mass undergoes 
negligible changes. 
Therefore, the \Macc ~vs. \Mstar ~relation represents a diagnostic tool for the evolution of 
\Macc ~\citep{clarke06} and for the process driving disc evolution \citep{ercolano14}.  
The distribution of class~II YSOs in the \Macc--\Mstar ~plane has been obtained for a number of 
different star-forming regions (SFRs); in all regions studied so far (e.g. $\rho$-Oph, Taurus, 
$\sigma$-Ori, ONC, Tr37, NGC2264) it has been found that, while there is a  positive correlation 
of \Macc  ~with the stellar mass,  \Macc ~has  a very large scatter, 
sometimes more than 3\,dex for objects with the same \Mstar ~\citep[][ and references therein]{muzerolle05, 
natta06, biazzo12, antoniucci14}.  

Theoretically, both the steep dependence of \Macc ~on \Mstar  ~and the large scatter of \Macc ~values 
are  a somewhat surprising finding, in that it appears to indicate that the accretion processes 
scale not just with \Mstar ~\citep{natta06}. 
Effects such as variability, or the natural decline of \Macc ~with age in viscous disc evolution have 
been ruled out as possible source of the large spread within individual SFRs 
\citep{natta06, costigan14, venuti14}. It appears more likely to be related to a spread in the properties 
of the parental cores, their angular momentum in particular \citep[e.g.][and references therein]{dullemond06}, 
and the disc mass. The scatter of \Macc ~may be also related to a spread of stellar properties, such as 
X-ray and EUV emission \citep{clarke06, ercolano14}, or on the competition between different accretion 
mechanisms, such as viscosity and gravitational instabilities at different stellar masses 
\citep{vorobyov08, vorobyov09, desouza16}. 
These latter authors suggested that two different exponents for the power-law relation 
\Macc~$\propto$~\Mstar$^\alpha$, at different mass regimes, can better describe the 
data than a single power-law. On the other hand, \citet{stamatellos15} claim that very low-mass brown 
dwarfs and planetary-mass objects may follow a different  \Macc-M$_{\rm object}$  ~scaling 
relationship than stars, with their accretion rate being almost independent of the central object mass. 

A third quantity, namely the disc mass, is likely to play an important role, as it is predicted that 
both \Mstar ~and \Macc ~should scale with the disc mass  \Mdisk ~in viscously evolving discs  \citep{hartmann98}.
However, the efforts to observationally confirm such scaling relations have failed in the past 
mainly because of the limited sensitivity of the interferometers used for measuring the bulk of 
dust and gas mass of protoplanetary discs, and because different methodologies to measure the 
stellar and accretion properties produce a large scatter in the relationships. 
A robust \Mdisk--\Mstar ~correlation for Taurus class~II YSOs has been confirmed by \citet{andrews13}, 
and we are now in a position that allows us to study the relationship between these three fundamental 
quantities in a statistically meaningful way for a number of star-forming regions. On one hand, 
VLT/X-Shooter is delivering homogeneous and precise determinations of both accretion and stellar 
properties \citep[e.g.][]{rigliaco12, manara13, alcala14, manara14, manara15, manara16a}.
On the other hand, the Atacama Large Millimeter Array (ALMA) now provides sufficient sensitivity and 
resolution at sub-mm wavelengths to detect and measure the mass of protoplanetary discs around 
YSOs with a mass down to 0.1\,\Msun \citep{ansdell16, pascucci16, barenfeld16}. 

In a previous work \citep[][henceforth A14]{alcala14} we studied the stellar and accretion properties
of 36 accreting YSOs mainly in the Lupus~I and III clouds, spanning a range in mass from $\sim$0.03 to 
$\sim$1.2\,\Msun, but mostly with 0.1\,\Msun~$<$~\Mstar~$<$~0.5\,\Msun. The analysis was based on spectroscopic
data acquired with the VLT/X-Shooter spectrograph. We used the continuum UV-excess emission as a measure 
of the accretion luminosity, \Lacc, hence of \Macc,  and provided improved relationships between \Lacc 
~and the luminosity, \Ll, for a large number of emission lines. 
In A14 we found that the $\log$\Macc--$\log$\Mstar ~correlation has a slope 1.8$\pm$0.2, but a more important
result was that the relationship has a much smaller dispersion ($\sim$0.4\,dex) with respect to previous 
studies. Although the level of accretion was not a criterion for the target selection, the YSOs analysed 
by A14 represent a sub-sample of the total class~II population in Lupus. Therefore, in order to 
confirm or disprove the result avoiding any type of bias, it was necessary to expand our analysis to a sample 
as complete as possible, by including as many class~II sources as possible and using the same methodologies 
as in A14 with X-Shooter to derive the stellar and accretion properties.

In this paper we present a synthesis of the accretion properties of an almost complete sample 
of class~II YSOs in Lupus. Our study combines our previous sample in A14 with
new X-Shooter observations of 55 additional objects classified as class~II and transition disc 
YSOs based mainly on the analysis of their spectral energy distribution (SED) 
\citep[e.g.][and references therein]{merin08, evans09} and/or the presence of strong emission lines in 
spectra with limited resolution and wavelength coverage \citep{comeron08}. 
The stellar and accretion properties of the combined sample have been homogeneously and self-consistently 
derived here, allowing an unbiased study of accretion and its relationship with the stellar parameters. 
The results on the stellar parameters and \Macc ~presented here were combined with those
of the ALMA survey of Lupus protoplanetary discs to study the \Mdisk--\Mstar ~and \Mdisk--\Macc 
~relationships in the papers by \citet{ansdell16} and \citet{manara16b}, respectively.

The new sample, the observations, and data processing are presented in Section~\ref{Sec:spectra}.
In Section~\ref{results} the newly observed sample is characterised in terms of its stellar and 
accretion properties and the results are compared with those in A14. The total sample 
is characterised in Section~\ref{totalsample} in terms of stellar masses and mass accretion rates, 
while the accretion properties  are examined in relation with the stellar 
parameters in Section~\ref{accprop}. The results are then discussed in Section~\ref{discussion}. 
Our main conclusions are summarised in Section~\ref{summary}. The relationships between \Lacc ~and \Ll 
~presented in A14 ~are revisited in Appendix~\ref{correlations} using the total sample. 

\vspace{1cm}

\section{Sample, observations, and data reduction} 
\label{Sec:spectra}

All the data used in this paper were acquired with the X-Shooter spectrograph \citep{vernet11} 
at the VLT. The capabilities of X-Shooter in terms of wide spectral coverage (310-2500 nm), 
resolution and limiting magnitudes allow us to assess simultaneously the mass accretion and outflow, 
and disc diagnostics, from the UV and optical to the near IR.

\subsection{Sample}
\label{sample}

The complete class~II sample in the Lupus~I, II, III and IV clouds, as selected from the 
Spitzer c2d survey \citep{merin08} and from the previous literature \citep[e.g.][]{hughes94, 
comeron08}, contains $\sim$101 objects. Several of these were only candidate YSOs.

The sample studied in this paper consists mainly of two sets of low-mass class~II YSOs in the 
aforementioned Lupus clouds. The first one comprises the 36 objects published in A14, observed 
within the context of the X-Shooter INAF/GTO \citep[][]{alcala11} project; for simplicity we will 
refer to it as the `GTO sample' throughout the paper. One additional source namely Sz105, was 
investigated with X-Shooter during the GTO, but rejected as a legitimate YSO (see below).
The second sample consists of 49 objects observed during ESO periods 95 and 97 
(1~April - 30~September 2015 and 1~April - 30~September 2016, respectively). 
In addition, we include here six objects observed with X-Shooter in other programmes taken from 
the ESO archive. In total, 55  objects were newly analysed here and we will refer to them as 
the `new sample'. The main goal of these new observations was to expand our previous analysis 
in A14 to a more complete sample.
 Among the 101 YSO candidates, there are seven young brown dwarf candidates by \citet{nakajima00} 
which were not observed by us because they are too faint ($J>$17mag) for X-Shooter. We also stress 
that sources with flat SEDs are not considered in our study (however see Appendix~\ref{flat_source}), 
and that we do not include objects of the Lupus~V and VI clouds.

In total, we have investigated 92 Lupus YSO candidates with X-Shooter. The 92 spectroscopically 
studied stars comprise the 36 YSOs and Sz105 investigated in A14 and the 55 objects studied here. 
As will be shown in Section~\ref{nonmembers},  11 of the 92 are confirmed to be giants unrelated to 
the Lupus star forming region. Therefore, the total sample of this paper, reported in Table~\ref{pars},
includes 81 legitimate YSOs.
An additional YSO candidate (SSTc2d\,J155945.3-415457), not included in our X-Shooter  observations, 
was confirmed to be an asymptotic giant branch (AGB) star in a previous work \citep{mortier11}. Thus, assuming 
that the seven \citet{nakajima00} brown dwarf candidates are also legitimate YSOs, the total sample of 
bona-fide class~II YSOs in the Lupus~I, II, III and IV clouds would comprise 89 ($=101-12$) objects. 
Therefore, our 81 YSOs (36 of A14 plus 45 of this paper) represent more than 90\% of the total.  
We note that 12 out of the 81 YSOs 
have been identified with transitional discs based on mid and far IR \citep{merin08, romero12, 
bustamante15} and/or sub-millimeter observations \citep{tsukagoshi14, ansdell16, vandermarel16}.
The study of transitional discs is important for the understanding of
disc evolution in general and of the mechanisms regulating the disc dispersal
in particular \citep[e.g.][]{espaillat14}.
High levels of accretion have been detected in some YSOs with transitional discs in the past
\citep[][and references therein]{espaillat14,alcala14, manara14}. 
It is thus important to also investigate the accretion properties of the Lupus YSOs with 
transitional discs in comparison with those with full discs. Finally, we adopted a distance 
of 150\,pc for objects in the Lupus~I, II and IV clouds, and 200\,pc for those in the Lup~III 
cloud \citep[see][for a discussion on the distance of the Lupus clouds]{comeron08}.

\subsection{Observations}
\label{observations}

As in the GTO, most of the targets in the new sample were observed using
the 1\farcs0/0\farcs9/0\farcs9 slits in the UVB/VIS/NIR arms, 
respectively, yielding resolving powers of 5100/8800/5600. 
Only Sz102 was observed through the 0\farcs5/0\farcs4/0\farcs4 slits in the UVB/VIS/NIR 
arms, respectively, yielding resolving powers  of 9100/17400/10500
In \citet{frasca16} we have measured the resolution using several exposures 
of the ThAr calibration lamp and found that it remained basically unchanged in the 
period between 2011 to 2015. The high resolution mode for the later target was
chosen in order to be able to study the outflows by measuring the gas kinematics 
more accurately (Whelan et al., in prep.) 
Table~\ref{targets} presents the observing log for the new targets. 
In order to achieve the best possible accuracy in flux calibration and account for
slit losses, short exposures (of $\sim$10\% the science exposures) were performed 
using the wide slit of 5\farcs0 right before the science observations. These were 
part of the same observing block for each target, minimising overheads and allowing 
accurate spectrophotometry of the targets.

Most of the targets were observed in one cycle using the A-B nodding mode, 
while five (AKC2006-18, 2MASS\,J16081497-3857145, Lup\,607, 2MASS\,J16085373-3914367, 
and Sz\,108B) were observed in two cycles using the A-B-B-A 
nodding mode. All the 5\farcs0-slit observations were performed in stare mode.
For one target (2MASS\,J16085373-3914367) there was no detection in the UVB arm.

During the observations, the star Sz\,81 showed up in the acquisition image as a visual 
binary with a separation of 1\farcs9 and PA$=20^\circ$. Except for Sz\,102 and the visual 
binary Sz\,81, all targets were observed at parallactic angle in order to minimise the
atmospheric dispersion. Sz102 was observed both with the slit along the known outflow 
\citep[PA$=95^\circ$, ][]{comeron11} and perpendicular to it (PA$=5^\circ$), while 
the components of the visual binary Sz\,81  were observed both by aligning the slit 
at PA$=20^\circ$.

The data gathered from the ESO archive were acquired using the 0\farcs5/0\farcs4/0\farcs4 
slits in the UVB/VIS/NIR arms, respectively, and adopting the  same AB nodding strategy 
as explained above, but with different number of cycles as indicated in Table~\ref{targets}. 
These data were not taken using the wide slit prior to the narrow slit observations. 
Thus, their flux calibration is more uncertain.

Several telluric standard stars were observed with the same instrumental set-up and 
at similar airmass as the targets. Typically, two flux standards per night were 
observed through a 5\,arcsec slit to calibrate the flux.

\subsection{Data reduction}
\label{datared}

The data processing was done using the same methods as for the GTO sample described in A14. 
Here we summarise the procedures. The basic processing of bias- or dark 
subtraction, flat-fielding, optimal extraction, wavelength calibration, and sky subtraction,
and correction for instrumental response was performed using the X-Shooter pipeline v.2.3.0
 \citep{modigliani10}. 
The nodding mode of the pipeline was used for the reduction of the nodding observations, while
the wide-slit observations were reduced using the stare mode.
Post-pipeline processing was done using  IRAF\footnote{IRAF is distributed by the National 
Optical Astronomy Observatory, which is operated by the Association of the Universities for 
Research in Astronomy, inc. (AURA) under cooperative agreement with the National Science Foundation}.
The telluric correction was performed independently in the VIS and NIR spectra, as explained in 
Appendix~A of A14. The X-Shooter scale of $\sim$0.16 arcsec/pix along the slit 
direction allowed us to resolve the components of the binary Sz81, which in turn enabled us 
to extract the spectra of the individual components without any light contamination. 
The flux-calibrated spectra observed with the wide slit were used to correct  
the spectra acquired with the narrow slits for slit losses. 
The correction factors, which depend mainly on seeing variations, are in the ranges 
1--2.9, 1--3.2 and 1--2.7 for the UVB, VIS and NIR arms respectively. Since all the targets 
were observed at low airmass, no wavelength dependence was found in these correction factors.

Finally, photometric data from the literature were used to compare the spectroscopic fluxes 
with the photometric ones. The spectra follow the corresponding SED shape very well, with most 
of them matching the photometric fluxes at the 10\% level.
In a few objects (SSTc2dJ154508.9-341734, MY\,Lup, Sz131, Sz133 and Sz98) we found that the 
flux ratio may be up to a factor 2, meaning 0.3\,dex in log scale,  which is well within 
the expected range of variability for class~II YSOs  \citep[see][and references therein]{venuti14}.
  
In order to estimate the flux losses of the  archive data, observed with the narrow slits, 
we compared the flux of the spectra with NIR photometric fluxes from 2MASS, where the 
variability effects are minimised. The correction factors are consistent with those
based on the spectrophotometry. However, since photometry is not symultaneous with the 
X-Shooter observations, the  spectroscopic flux may be uncertain by a factor of 
about two. In addition, EX\,Lup is the well known prototype of EXors 
\citep[][]{comeron08, sipos09, lorenzetti12, sicilia-aguilar15} hence, large variations 
may be expected. 
However, we find good agreement between the flux of the spectrum after correction for 
slit losses and the photometric flux in the $V$-band, gathered from the AAVSO database 
and quasi-simultaneous to the X-Shooter observation (see Appendix~\ref{EXLup}).

\section{Results}
\label{results}

\subsection{Non-members}
\label{nonmembers}
Nine objects of the new sample (see Table~\ref{targets}), without appropriate spectroscopy 
in the past but previously classified as class~II YSOs, lack the  Li\,{\sc i} $\lambda$670.8\,nm 
absorption line and show narrow photospheric lines, with their spectrum resembling 
more that of a giant than of a PMS star. A detailed analysis of the radial velocity, 
combined with determinations of the surface gravity based on our X-Shooter spectra 
\citep{frasca16}, demonstrated that these objects are indeed background giants. 
Likewise, one additional object (SSTc2dJ161045.4-385455) for which the  Li\,{\sc i} 
line has been detected, was found to be a background Li-rich giant based on discrepant 
surface gravity and radial velocity with respect to the Lupus YSOs. Another object 
namely Sz105, previously classified as class~II YSO candidate based on the Spitzer survey,  
has been rejected as YSO in A14 and confirmed to be a background giant in \citet{frasca16}. 

Thus, including SSTc2d\,J155945.3-415457 classified as an AGB star by \citet{mortier11} 
and not observed by us, there are 12 objects previously classified as class~II YSO candidates, 
which are unrelated to Lupus. It is also worth mentioning that 10 of these were included in 
the 95\% complete ALMA survey of Lupus protoplanetary discs by \citet{ansdell16} and none 
were detected. That survey detected $\sim$70\% of the observed objects in 890\,$\mu$m 
continuum emission. This highlights the importance of combining ALMA discs surveys with 
detailed optical/IR classification of the host star \citep[see also][and Manara et al. 2017]{pascucci16}.

The new sample then consists of the 45 legitimate YSOs listed in Table~\ref{targets}. 
The objects rejected by us as class~II YSOs are listed in the bottom of this table and 
their properties will be discussed in detail in the parallel paper by \citet{frasca16}.
The physical parameters and accretion properties of the 45 confirmed YSOs are derived 
next and compared with those of the GTO sample. The complete list of 81 confirmed 
class~II and transitional YSOs of this work is reported in Table~\ref{pars}.

\onecolumn

{ \small
\setlength{\tabcolsep}{2pt}
\begin{longtable}{l|cc cc ccc |c | c}
\caption[ ]{\label{targets} Observing log for the new sample.}\\
\hline \hline
 Object/other name &  RA(2000)  & DEC(2000)  & Obs. Date &  MJD       & \multicolumn{3}{c}{ \underline {$T_{\rm exp}$ (sec)}}  &   Lupus & Notes\\ 
      &  h \, :m \, :s & $^\circ$ \, ' \, ''   & YY-MM-DD  & (+2400000) &   UVB & VIS & NIR                                   &    cloud & \\            
\hline
Sz65                     & 15:39:27.78  & $-$34:46:17.4 & 2015-06-04   & 57177.017838    &  2x150  & 2x100   &  2x150   & I       & \\  
AKC2006-18               & 15:41:40.82  & $-$33:45:19.0 & 2015-04-20   & 57132.269751    &  4x900  & 4x850   &  4x960   & I       & \\  
SSTc2dJ154508.9-341734   & 15:45:08.88  & $-$34:17:33.7 & 2015-06-15   & 57188.149771    &  2x900  & 2x850   &  2x960   & I       & \\  
Sz68                     & 15:45:12.87  & $-$34:17:30.8 & 2015-05-18   & 57160.210844    &  2x100  & 2x60    &  2x50    & I       & \\  
SSTc2dJ154518.5-342125   & 15:45:18.53  & $-$34:21:24.8 & 2015-06-25   & 57198.978850    &  2x900  & 2x850   &  2x960   & I       & \\  
Sz81A   (SW)             & 15:55:50.21  & $-$38:01:34.0 & 2015-08-19   & 57253.072870    &  2x300  & 2x250   &  2x300   & II      & \\  
Sz81B   (NE)             & 15:55:50.26  & $-$38:01:32.2 & 2015-08-19   & 57253.072870    &  2x300  & 2x250   &  2x300   & II      & \\  
Sz129                    & 15:59:16.48  & $-$41:57:10.3 & 2015-06-26   & 57199.059343    &  2x100  & 2x50    &  2x100   & IV      & \\  
SSTc2dJ155925.2-423507   & 15:59:25.24  & $-$42:35:07.1 & 2015-06-27   & 57200.014418    &  2x900  & 2x850   &  2x960   & IV      & \\  
RY\,Lup                  & 15:59:28.39  & $-$40:21:51.3 & 2015-07-02   & 57205.275027    &  2x100  & 2x50    &  2x100   & IV      & \\  
SSTc2dJ160000.6-422158   & 16:00:00.62  & $-$42:21:57.5 & 2015-04-03   & 57115.345055    &  2x450  & 2x400   &  2x480   & IV      & \\  
SSTc2dJ160002.4-422216   & 16:00:02.37  & $-$42:22:15.5 & 2015-07-01   & 57204.165952    &  2x450  & 2x400   &  2x960   & IV      & \\  
SSTc2dJ160026.1-415356   & 16:00:26.13  & $-$41:53:55.6 & 2015-06-28   & 57201.007970    &  2x900  & 2x850   &  2x960   & IV      & \\  
MY\,Lup                  & 16:00:44.53  & $-$41:55:31.2 & 2015-06-26   & 57199.070898    &  2x150  & 2x100   &  2x150   & IV      & \\  
Sz131                    & 16:00:49.42  & $-$41:30:04.1 & 2015-07-01   & 57204.221691    &  2x450  & 2x400   &  2x960   & IV      & \\  
Sz133                    & 16:03:29.41  & $-$41:40:02.7 & 2015-07-02   & 57205.191567    &  2x900  & 2x850   &  2x960   & IV      & \\  
SSTc2dJ160703.9-391112   & 16:07:03.84  & $-$39:11:11.3 & 2016-06-02   & 57542.198741   &  4x960  & 4x910   &  4x480   & III     & \\ 
Sz90                     & 16:07:10.08  & $-$39:11:03.5 & 2015-07-12   & 57215.095862    &  2x360  & 2x310   &  2x360   & III     & \\  
Sz95                     & 16:07:52.32  & $-$38:58:06.3 & 2015-07-12   & 57215.118419    &  2x360  & 2x310   &  2x360   & III     & \\  
Sz96                     & 16:08:12.62  & $-$39:08:33.5 & 2015-07-03   & 57206.270061    &  2x360  & 2x310   &  2x360   & III     & \\  
2MASSJ16081497-3857145   & 16:08:14.96  & $-$38:57:14.5 & 2015-04-20   & 57132.335763    &  4x900  & 4x850   &  4x960   & III     & \\  
Sz98                     & 16:08:22.50  & $-$39:04:46.0 & 2015-07-02   & 57205.132662    &  2x150  & 2x100   &  2x150   & III     & \\  
Lup607                   & 16:08:28.10  & $-$39:13:10.0 & 2015-05-23   & 57165.214447    &  4x900  & 4x850   &  4x960   & III     & \\  
Sz102                    & 16:08:29.73  & $-$39:03:11.0 & 2015-04-17   & 57129.298225    &  2x1200 & 2x1260  &  2x1200  & III     & \\  
SSTc2dJ160830.7-382827   & 16:08:30.70  & $-$38:28:26.8 & 2015-07-02   & 57205.155128    &  2x150  & 2x100   &  2x150   & III     & \\  
SSTc2dJ160836.2-392302/V1094~Sco  & 16:08:36.18  & $-$39:23:02.5 & 2016-05-13   & 57522.227447   &  4x300  & 4x250   &  4x100   & III &   \\ 
Sz108B                   & 16:08:42.87  & $-$39:06:14.7 & 2015-06-18   & 57191.161119    &  4x900  & 4x850   &  4x960   & III     & \\  
2MASSJ16085324-3914401   & 16:08:53.23  & $-$39:14:40.3 & 2015-07-12   & 57215.136035    &  2x450  & 2x400   &  2x480   & III     & \\  
2MASSJ16085373-3914367   & 16:08:53.73  & $-$39:14:36.7 & 2015-05-23   & 57165.287007    &  4x900  & 4x850   &  4x960   & III     & 1 \\  
2MASSJ16085529-3848481   & 16:08:55.29  & $-$38:48:48.1 & 2015-07-12   & 57215.156936    &  2x900  & 2x850   &  2x960   & III     & \\  
SSTc2dJ160927.0-383628   & 16:09:26.98  & $-$38:36:27.6 & 2015-07-13   & 57216.022778    &  2x900  & 2x850   &  2x960   & III     & \\  
Sz117                    & 16:09:44.34  & $-$39:13:30.3 & 2015-07-13   & 57216.068820    &  2x360  & 2x310   &  2x360   & III     & \\  
Sz118                    & 16:09:48.64  & $-$39:11:16.9 & 2015-07-13   & 57216.097061    &  2x450  & 2x400   &  2x480   & III     & \\  
2MASSJ16100133-3906449   & 16:10:01.32  & $-$39:06:44.9 & 2015-07-13   & 57216.122794    &  2x900  & 2x850   &  2x960   & III     & \\  
SSTc2dJ161018.6-383613   & 16:10:18.56  & $-$38:36:13.0 & 2015-08-18   & 57252.041176    &  2x900  & 2x850   &  2x960   & III     & \\  
SSTc2dJ161019.8-383607   & 16:10:19.84  & $-$38:36:06.8 & 2015-08-08   & 57242.086534    &  2x900  & 2x850   &  2x960   & III     & \\  
SSTc2dJ161029.6-392215   & 16:10:29.57  & $-$39:22:14.7 & 2015-08-13   & 57247.090628    &  2x900  & 2x850   &  2x960   & III     & \\  
SSTc2dJ161243.8-381503   & 16:12:43.75  & $-$38:15:03.3 & 2015-07-10   & 57213.174667    &  2x300  & 2x250   &  2x300   & III     & \\  
SSTc2dJ161344.1-373646   & 16:13:44.11  & $-$37:36:46.4 & 2015-06-26   & 57199.974706    &  2x900  & 2x850   &  2x960   & III     & \\  
                         &              &               &    &    &    &   &      &   & \\ 
\hline
Targets from ESO archive:&              &               &    &    &    &   &      &   & \\ 
 \hline
 \hline
Sz75/GQ\,Lup         & 15:49:12.10  & $-$35:39:05.1 & 2010-05-05   & 55321.270673   &  2x400  & 4x160    &  6x240     & I    & 2 \\ 
Sz76                 & 15:49:30.74  & $-$35:49:51.4 & 2014-04-28   & 56775.245961   &  2x478  & 2x280    &  2x26      & I    & 3 \\ 
Sz77                 & 15:51:46.95  & $-$35:56:44.1 & 2010-05-05   & 55321.376758   &  2x400  & 4x320    &  6x240     & I    & 2 \\ 
RXJ1556.1-3655       & 15:56:02.09  & $-$36:55:28.3 & 2014-04-28   & 56775.268274   &  2x478  & 2x280    &  2x26      & II   & 3 \\
Sz82/IM\,Lup         & 15:56:09.18  & $-$37:56:06.1 & 2010-05-04   & 55320.065259   &  2x300  & 2x120    &  2x200     & II   & 2 \\ 
EX\,Lup              & 16:03:05.49  & $-$40:18:25.4 & 2010-05-04   & 55320.165145   &  3x300  & 4x120    &  6x200     & III  & 2 \\ 
\hline
                         &              &               &    &    &    &   &        &    & \\ 
Objects rejected as      &         &               &    &    &    &   &        & & \\ 
Lupus members:           &         &               &    &    &    &   &        & & \\ 
\hline
\hline
Sz78                     & 15:53:41.18  & $-$39:00:37.1 & 2015-06-26 & 57199.033491 & 2x100  & 2x50  & 2x100  &  &  \\      
Sz79                     & 15:53:42.68  & $-$38:08:10.4 & 2015-06-26 & 57199.046851 & 2x150  & 2x100 & 2x150  &  &  \\      
IRAS15567-4141           & 16:00:07.42  & $-$41:49:48.4 & 2015-07-02 & 57205.171310 & 2x200  & 2x150 & 2x200  &  &  \\      
SSTc2dJ160034.4-422540   & 16:00:34.40  & $-$42:25:38.6 & 2015-07-01 & 57204.204770 & 2x200  & 2x150 & 2x200  &  &  \\      
SSTc2dJ160708.6-394723   & 16:07:08.63  & $-$39:47:21.9 & 2015-07-03 & 57206.237144 & 2x450  & 2x400 & 2x480  &  &  \\      
2MASSJ16080618-3912225   & 16:08:06.18  & $-$39:12:22.5 & 2015-06-04 & 57177.037999 & 2x600  & 2x550 & 2x600  &  &  \\      
Sz105                    & 16:08:36.89  & $-$40:16:20.6 & 2012-04-18 & 56035.154479 & 2x150  & 2x100 & 2x100  &  & 4 \\      
SSTc2dJ161045.4-385455   & 16:10:45.37  & $-$38:54:54.8 & 2016-06-05 & 57545.005854 & 4x960  & 2x910 & 2x480  &  &  \\
SSTc2dJ161148.7-381758   & 16:11:48.66  & $-$38:17:58.1 & 2015-04-04 & 57116.325599 & 2x900  & 2x850 & 2x960  &  &  \\      
SSTc2dJ161211.2-383220   & 16:12:11.20  & $-$38:32:19.7 & 2016-05-11 & 57520.199021 & 2x960  & 2x910 & 2x480  &  &  \\
SSTc2dJ161222.7-371328   & 16:12:22.69  & $-$37:13:27.7 & 2015-06-04 & 57177.038060 & 2x300  & 2x250 & 2x300  &  &  \\      
\hline
\end{longtable}
\tablefoot{ 
~1: no detection in the UVB arm;
~2: from programme 085.C-0764(A) (PI: Guenther);
~3: from programme 093.C-0506(A) (PI: Caceres); 
~4: from GTO sample analysed in A14; 
}

}

\twocolumn

\subsection{Accretion luminosity}
\label{accretiondiagnostics}

The continuum excess emission in YSOs is most easily detected as Balmer continuum 
emission \citep[see][A14 and references therein]{valenti93, gullbring98, 
HH08, rigliaco12, manara16a}. In A14, Balmer continuum emission was evident in  
all the YSOs of the GTO sample.
In the new sample the results are the following: one object (2MASSJ16085373-3914367) 
lacks information in the UVB (see Section~\ref{observations}). Balmer continuum emission 
is seen in 38 objects. All of them, but the K2-type star Sz102, are later than K4. 
Other 3 M-type objects (Lup607, SSTc2dJ154508.9-341734, 2MASSJ16085324-3914401) have 
noisy UVB spectra. In three objects (MY\,Lup, Sz68, and SSTc2dJ160830.7-382827), 
all earlier than K3, the Balmer continuum emission is not evident from the 
spectra. This is because the Balmer continuum emission is more easily seen in 
the spectra of late-type ($>$~K5) YSOs than in the early types due to the higher 
contrast between photospheric emission and continuum emission.

To derive the accretion luminosity, \Lacc, of the new sample we have followed the methods 
described by A14, but using the procedures of \citet{manara13b}. Briefly, the spectrum of 
each class~II YSO was fitted as the sum of the photospheric spectrum of a class~III template 
and the emission of a slab of hydrogen; the accretion luminosity is given by the luminosity 
emitted by the slab. The stellar and accretion parameters are self-consistently derived by 
finding the best fit among a grid of slab models and using the continuum UV-excess emission
and the broad wavelength range covered by the X-Shooter spectra (330\,nm -- 2500\,nm) 
to constrain both the spectral type of the target and the interstellar extinction toward it.
The best fit is found by minimizing a $\chi^2_{\rm like}$ distribution.
The stellar parameters are reported in Table~\ref{pars} and the complete set of plots showing 
the fits for the 44 targets detected in the UVB arm are provided in electronic form in Figures 
from \ref{slab1} to \ref{slab5}. For consistency with the literature and homogeneity with our 
previous work we did not attempt to fit the hydrogen emission lines, but only the continuum 
emission. The adopted class~III templates and \Lacc ~ values corresponding to each YSO are 
reported in Table~\ref{accretion} in Appendix~\ref{stelprop}. For completeness, the \Lacc~ 
values for the GTO sample are also included in this table. From our analysis in A14 we 
estimate that in general the uncertainty on \Lacc ~in log scale is $\sim$0.25\,dex.

For all the  objects we also calculated an average \Lacc ~from the luminosity of several 
emission lines (see Section~\ref{emisslines}) and using the $\log$\Lacc~vs.~$\log$\Ll ~relationships 
reported in A14. In all cases \Lacc ~calculated from the slab modelling is in very good 
agreement with the average \Lacc ~calculated from the lines. This check was particularly 
useful for some cases where the low-S/N in the UVB spectrograph arm made the slab modelling 
difficult.

Based on the slab modelling, UVB excess emission ascribable to accretion is barely
evident in five objects (Lup\,607, MY~Lup, Sz65, Sz68, and SST\,c2dJ160830.7-382827).
The analysis of the emission lines in these objects in Section~\ref{chrom_contrib} 
shows that their excess emission is close to the chromospheric level. Thus, in the 
following we consider these objects as weak accretors, and distinguish them 
in the plots. We do not expect that their \Lacc ~and \Macc ~values are higher than 
what we measured. In principle one could consider them as upper limits, but our results 
are not affected if we assume them as such (see Section~\ref{accprop}). 

Two objects, namely Lup\,607 and SSTc2dJ160703.9-391112, have a rather low accretion 
luminosity (\Lacc$\approx 10^{-5}$\,\Lsun). The former is a weak accretor  
(c.f. Section~\ref{chrom_contrib}), while the other one is sub-luminous, 
hence also sub-luminous in \Lacc ~(see Section~7.4 in A14). 
Finally, in the case of the target 2MASS\,J16085373-3914367, which lacks UVB data 
(see Section~\ref{observations}), \Lacc ~ was calculated from the luminosity of 
7 permitted emission lines detected in the VIS and NIR and using the \Lacc ~-- \Ll 
~ relationships revisited in Appendix~\ref{correlations}. 
The width and intensity of the emission lines in the VIS and NIR, as well as the computed 
\Lacc ~and \Lacc/\Lstar ~values confirm that the object is accreting. 

\subsection{Spectral type, extinction and luminosity of the new sample}
\label{params}

We focus here on the determination of the parameters which are necessary for studying 
the accretion properties, namely spectral type and extinction, effective temperature, 
stellar luminosity, and radius, with the mass determination being deferred 
to Section~\ref{totalsample}. Other properties like radial velocity, surface gravity, 
as well as lithium and other elemental abundances for the whole X-Shooter sample will 
be analysed in parallel papers by \citet{frasca16} and \citet{biazzo16}.

In addition to the self-consistent methods described in Section~\ref{accretiondiagnostics},
we have also used the methods of our previous study in A14 to derive spectral type and 
extinction for the new sample. For the late type (M0 or later) stars spectral types were calculated 
using the spectral indices by \citet{riddick07} and the H2O-K2 index from \citet{rojasayala12} 
for the NIR spectra. For the K-type objects we derived spectral types by direct comparison of 
class~III templates after artificially reddening the templates, until the best match with the 
class~II YSOs is found. Several class~III YSOs, indistinctly quoted here as class~III YSOs 
or class~III templates, were observed throughout the various Italian-GTO star formation runs, 
and their properties were published in separate papers \citep{manara13, stelzer13b}. A few other 
class~III templates, filling-in the gaps of spectral type distribution of the previous templates, 
were used here and will be published in Manara et al., in prep. The use of templates
to derive spectral types for the M-type objects provides basically the same results as
when using spectral indices \citep[see][]{frasca16}. 
Both the A14 methods and those described in Section~\ref{accretiondiagnostics} provide 
consistent results within errors hence, for homogeneity  we adopted the results of the 
self-consistent methods. The spectral types and extinction values 
are reported in Table~\ref{pars}. The uncertainties in spectral type are $\pm$0.5 sub-class for 
the M-type objects and $\pm$1 sub-class for the earlier type stars. The estimated uncertainty 
in extinction is $\le$~0.5\,mag.

\begin{figure}[h]
\resizebox{1.05\hsize}{!}{\includegraphics[]{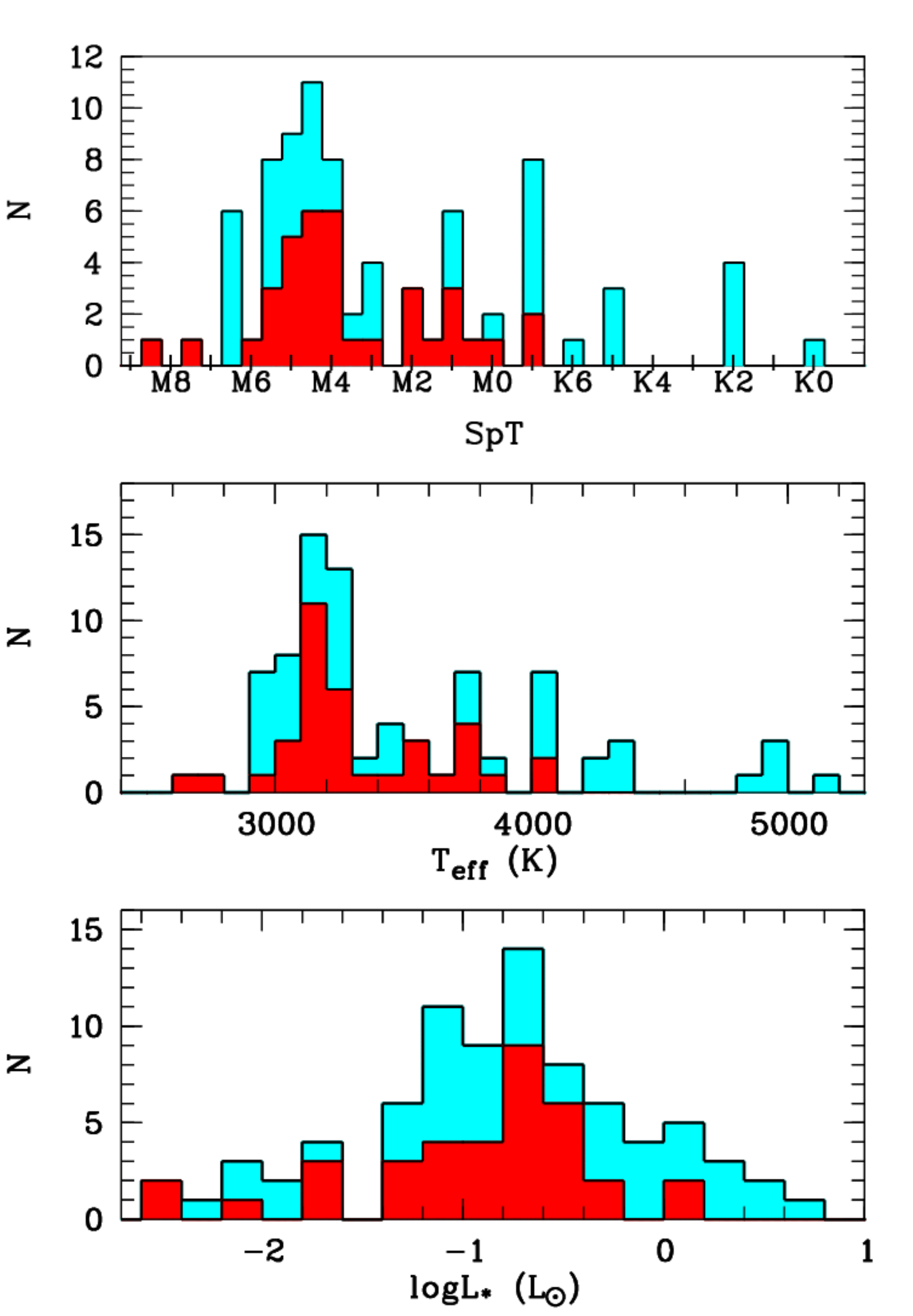}}
\caption{The distributions of spectral type (upper panel), effective temperature (middle panel) 
        and luminosity (lower panel) of the GTO (red histograms) and total (blue histograms) 
        samples.
    \label{histprop}}
\end{figure}

The spectral types of the new sample range from K0 to M7, with an overabundance of M4-M5 objects 
(see Figure~\ref{histprop} upper panel). While the GTO sample did not include objects with spectral 
type earlier than K7, the new sample contains 11 objects earlier than that.
Our spectral types and extinction values are generally consistent within errors with those in 
the literature.


As in A14, the effective temperature, $T_{\rm eff}$, was derived using the temperature 
scales given by \citet{Ken95} for the K-type stars, and by \citet{luhman03} for the 
M-type YSOs. These $T_{\rm eff}$ values, as well as those in the GTO in A14, are in very good 
agreement with those determined using the ROTFIT code \citep[see][]{frasca16}.
The stellar luminosity was derived using our flux calibrated X-Shooter spectra in the same 
way as described in \citet{manara13b}, that is, by direct integration of the spectra and
using the synthetic BT-Settl spectra \citep[][]{allard12}, of the same  $T_{\rm eff}$ as 
the objects, to extrapolate the X-Shooter spectra to wavelengths shorter than 310\,nm  
and longer than 2500\,nm. The error in YSO luminosity was estimated from the 
signal-to-noise (S/N) of the spectra and the error in visual extinction. The details are 
given in Appendix~\ref{stelprop}. 
The error in extinction dominates the overall uncertainty in all cases, although for some 
targets with a low S/N spectrum the contribution of photon noise to the error becomes important.
The stellar radius was calculated from the effective temperature and stellar luminosity.
All the physical stellar parameters are listed in Table~\ref{pars} in Appendix~\ref{stelprop},
which includes the parameters of the GTO sample for completeness.

In comparison with the GTO sample, the new sample extends to \Teff ~values as high as
5100\,K, but about 70\% are cooler than 4000\,K (see Figure~\ref{histprop} middle panel).
Likewise, most of the objects have a luminosity lower than 0.5\,\Lsun, with 11 
exceeding 1\,\Lsun ~(see Figure~\ref{histprop} lower panel).

\subsection{Emission lines}
 \label{emisslines}
A large number of permitted and forbidden emission lines displaying a variety 
of profiles were detected. The analysis of forbidden emission lines 
for the GTO sample has been published by \citet{natta14}, and for the total 
sample in a parallel paper by \citet{nisini16}.

The permitted emission lines studied here are the same as those in Table~4 of A14, 
except for the He\,{\sc i} $\lambda$1082.9nm line. Because its complexity both in 
terms of line profile and interpretation as accretion or wind diagnostic \citep[see][]{edwards06} 
we decided not to include this line here, but defer its analysis to a future paper. Note 
also that the line appears blended with the Si\,{\sc i}~$\lambda$1082.7091nm photospheric line, 
which is very strong in late-type stars.
The number of detections of each  line is given in column five of Table~\ref{linfits} in 
Appendix~\ref{correlations}. For consistency with A14, our analysis is restricted 
to Balmer lines up to H15, Paschen lines up to Pa\,10, and the \brg ~line, as well 
as the helium, calcium, sodium, and oxygen lines.

The flux at the Earth in permitted lines was computed by directly integrating the 
flux-calibrated spectra using the {\em splot} package under  IRAF, and following the 
procedures described in A14, including estimates of upper limits for non-detections. 
The observed fluxes, equivalent widths, and their errors are reported 
in several tables provided in electronic form only (from Table \ref{tab:fluxes_EWs_Hae} 
to \ref{tab:fluxes_EWs_NaI}). The flux errors are those resulting  from the uncertainty 
in continuum placement. 
The estimated $\sim$10\% uncertainty of flux calibration (see Section~\ref{datared}) 
should be added in quadrature. The contribution of the photospheric absorption lines 
of the H$\alpha$, $\ion{Na}{i}$~D lines and the Ca\,{\sc ii}~ IR triplet lines (IRT), 
strongest in the K and early-to-mid M-type objects, were removed in all spectra as 
described by \citet[][]{frasca16}. 
The luminosity of the different emission lines was computed as \Ll ~$ = 4 \pi  d^2 \cdot f_{\rm line}$, 
where $d$ is the YSO distance listed in Table~\ref{pars} and $f_{\rm line}$ is the 
extinction-corrected flux of the lines.
  
Together, the GTO and new sample more than double the number of YSOs in our previous work, 
and have homogeneously and simultaneously determined \Lacc ~and \Ll ~values. Therefore, it 
is worth revisiting the \Lacc -- \Ll ~relationships given in Section~5 of A14. This is reported 
in Appendix~\ref{correlations}. 

\section{The total sample}
\label{totalsample}
The new objects, combined with the GTO targets constitutes our total sample of 81 YSOs for 
the study of accretion in this paper. This sample is complete at more than the 90\% level 
(see Section~\ref{sample}) and is presented in Table~\ref{pars}. In this section we 
characterise it by deriving masses, and mass accretion rates in a homogeneous and 
self-consistent way. 

\subsection{Stellar masses}
\label{stellarmass}

We estimated masses by interpolating PMS evolutionary models (see Appendix~\ref{stelprop}). 
In A14 we have used the \citet{baraffe98}  tracks, which  were suited for deriving 
the mass of all the YSOs because they cover well the range in \Teff ~ and \Lstar ~ of 
the GTO sample. 
As shown in Section~\ref{params} the new sample extends to higher values of  \Teff ~ and \Lstar ~ 
than those of the GTO, i.e. to masses not covered by the  \citet{baraffe98} tracks. 
The \citet{baraffe98} models have been updated by \citet{baraffe15}, but as the previous models, 
they are for masses $\le$ 1.4\,\Msun. The \citet{siess00} tracks include higher masses, but 
their lowest mass is 0.1\,\Msun.

\begin{figure}[h]
\resizebox{1.0\hsize}{!}{\includegraphics[]{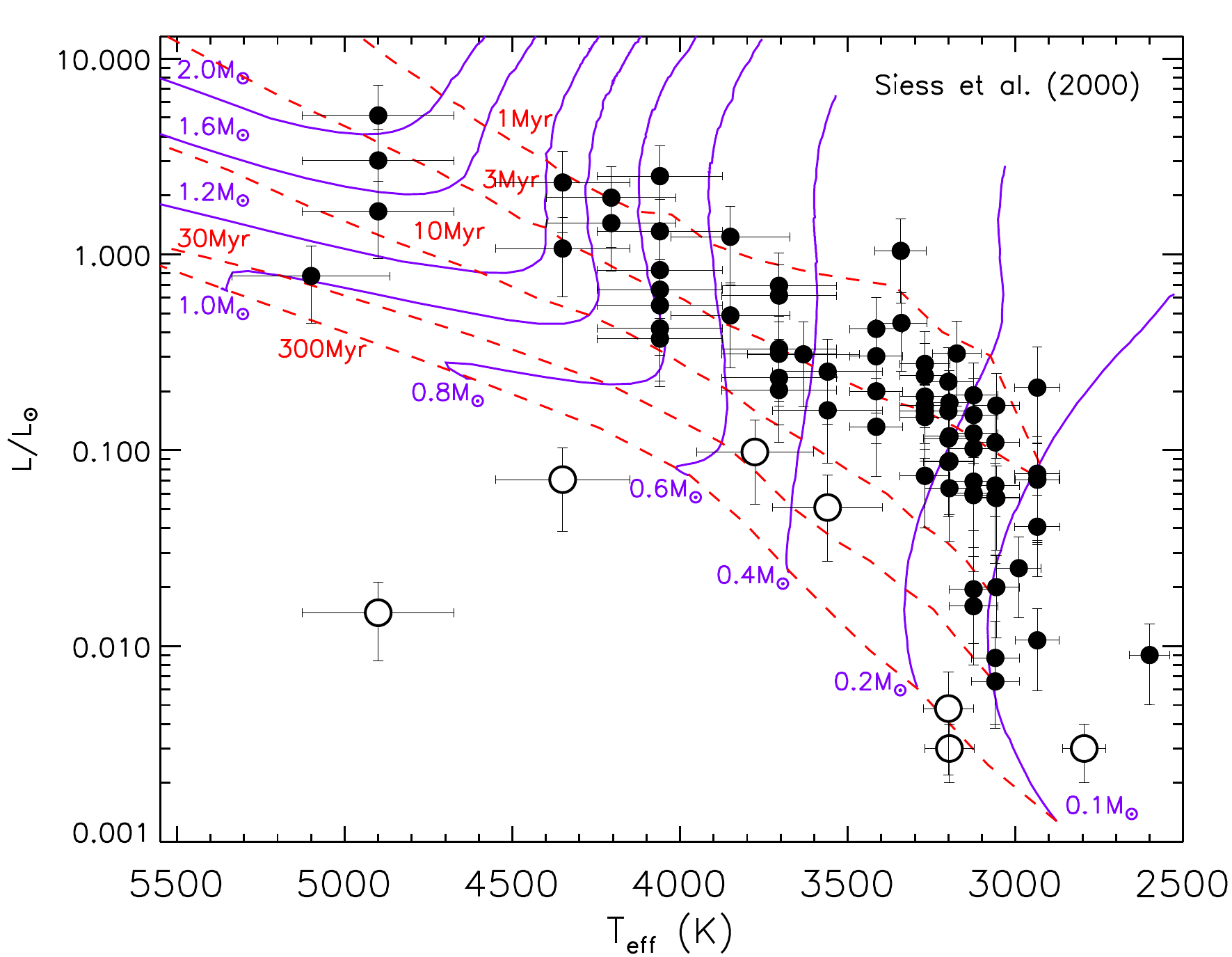}}
\caption{Hertzsprung-Russell diagram for the total sample.  The GTO and
       new samples are represented with the black symbols. The seven 
       sub-luminous objects described in the text are represented with 
       open circles. The dashed lines show the isochrones, reported by 
       \citet{siess00}, while the continuous lines show the low-mass 
       Pre-Main Sequence evolutionary tracks by the same authors as 
       labelled.
    \label{HRD}}
\end{figure}

The Hertzsprung-Russell diagram for the total sample is shown in Figure~\ref{HRD} with the 
PMS evolutionary tracks by \citet{siess00} overplotted. Only three objects of the total sample 
have a mass significantly lower than 0.1\,\Msun ~on these tracks.
In Appendix~\ref{stelprop} we decribe how we compared the resulting masses of the total sample when adopting 
four different models. The \citet{baraffe15} and the \citet{siess00} tracks 
yield very similar results in the overlapping mass range. Therefore, for our 
analysis of accretion in Section~\ref{accprop} we adopted the \citet{siess00} tracks to 
derive masses $\ge$ 0.1\,\Msun ~and those of \citet{baraffe15} for the three objects with 
lower values. The derived masses are given in Table~\ref{pars} in Appendix~\ref{stelprop}.
 More details on the mass determination and its error are provided in that Appendix,
where the \citet{DM97} models are additionally used.


We note that there are seven sub-luminous objects, namely Par-Lup3-4, Lup706, Sz\,123B, Sz\,106, 
Sz\,102, Sz\,133, and SSTc2dJ160703.9-391112. The first four were part of the GTO, while the 
latter three are from the new sample. Both Sz\,102 and Sz\,133 fall below the Zero-Age Main 
Sequence hence we cannot estimate their mass. These two objects were previously known to be 
sub-luminous, with their disc most likely seen edge-on \citep[][]{hughes94}. 
Sz\,102 (also known as Krautter's star) is one of the most 
famous YSOs in Lupus known to host a strong outflow \citep[see][for details]{krautter86, whelan16}. 
The seven sub-luminous objects are represented with open symbols in Figure~\ref{HRD} and are 
flagged in Table~\ref{pars}.
MY\,Lup, the hottest object in the sample, appears rather sub-luminous with respect 
to YSOs of similar spectral type. 
Based on ALMA data, its disc inclination angle has been measured at 73$^{\circ}$ \citep{ansdell16}. 
Thus, we cannot exclude that the star is at least partially obscured by the disc.
As a consequence, the mass of MY\,Lup may be underestimated.

\begin{figure}[h]
\resizebox{1.0\hsize}{!}{\includegraphics[]{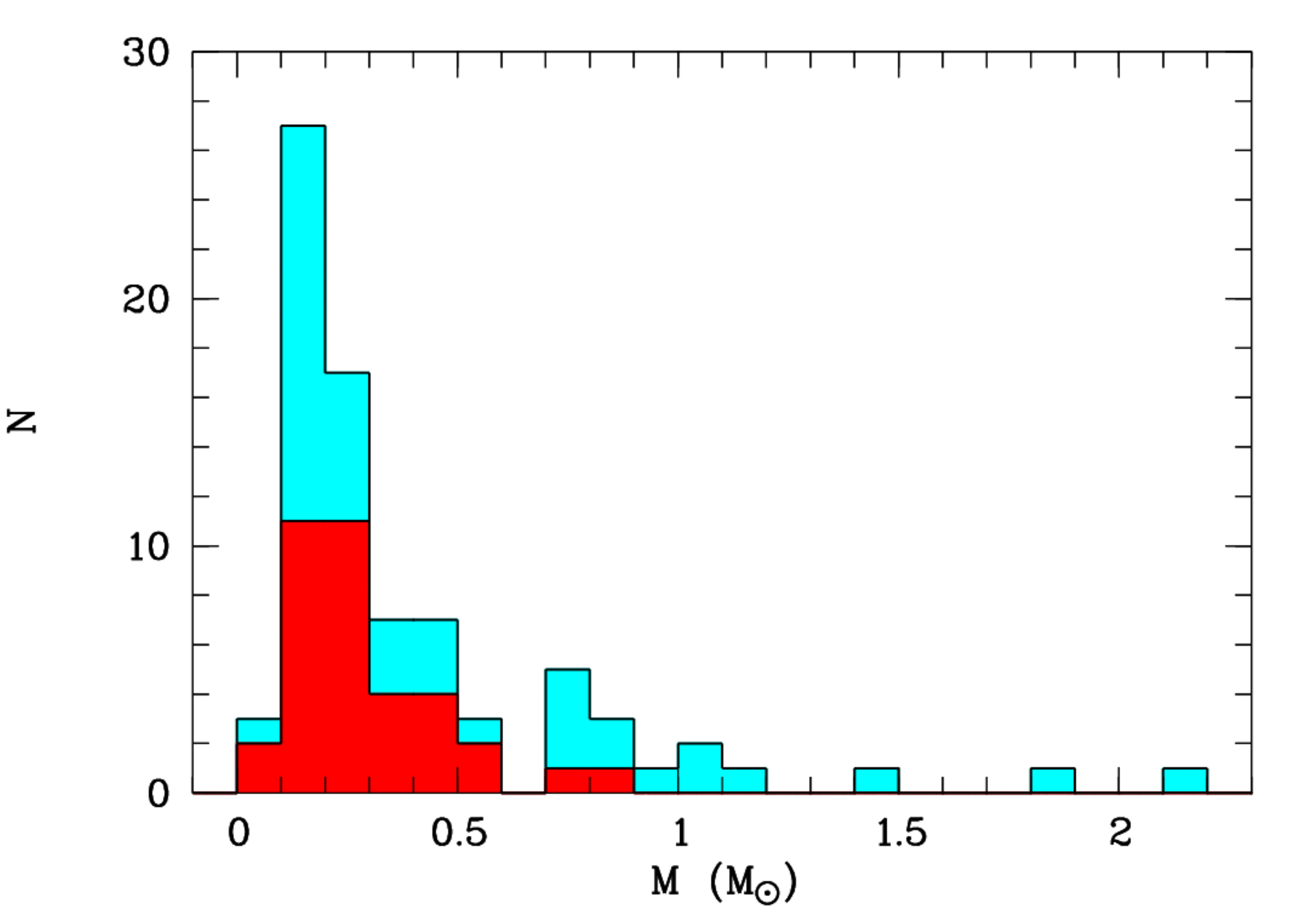}}
\caption{Histograms of \Mstar. The GTO sample is shown with the red histogram, while
        the total sample is shown with the blue one.
    \label{hist_Mstar}}
\end{figure}

The distribution of \Mstar ~ for the total sample, according to the \citet{siess00} tracks, 
is shown in Figure~\ref{hist_Mstar}. All YSOs have masses lower than 2.2\,\Msun, 
with only six having a mass higher than 1\,\Msun, and about 76\% have a mass lower than 0.5\,\Msun.
We note that these numbers do not account for the sub-luminous YSOs Sz\,102 and 
Sz\,133. Apart from the six objects with a mass higher than 1\,\Msun, both the \Mstar 
~distributions of the GTO and new samples are similar, peaking at $\sim$0.2\,\Msun.
With a mass of 0.02\,\Msun, that is, close to the planetary mass regime, 2MASS\,J16085953-3856275 
is the lowest mass object in the total sample.


The distribution of the Lupus stars on the HR diagram suggests an age of $\sim$3\,Myr. We note that 
stars  with \Mstar$\approx$0.1\Msun ~have a very large dispersion in \Lstar, with about 1/2 apparently 
older than 10\,Myr. This effect is seen in many star forming regions, and has been interpreted by \citet{herhill15} 
as an indication that the theoretical isochrones for the lowest mass stars are in reality much steeper 
than what the current models predict.

\subsection{Mass accretion rate of the total sample}
\label{macc_tot_sample}

The accretion luminosities of the total sample given in Table~\ref{accretion} (see Appendix~\ref{stelprop})
were converted into mass accretion rates, \Macc, using the relation

{\setlength{\mathindent}{0pt}
\begin{equation}
\label{Macc}
\dot{M}_{acc} = ( 1 - \frac{R_{\star}}{R_{\rm in}} )^{-1} ~ \frac{L_{acc} R_{\star}}{G M_{\star}}
 \approx 1.25 ~ \frac{L_{acc} R_{\star}}{G M_{\star}} 
,\end{equation}

\noindent
where $R_{\star}$ and $R_{\rm in}$ are the YSO radius and inner-disc radius, respectively
\citep{gullbring98, hart98}.
For consistency  with previous studies \citep[e.g.][and A14]{gullbring98, HH08, rigliaco12, manara16a}, 
here we also assumed $R_{\rm in}$ to be $5\,R_\star$. The $R_{\star}$ and \Mstar ~values 
were taken from Table~\ref{pars}. 
The results on \Macc ~ are listed in Table~\ref{accretion} in Appendix~\ref{stelprop}. As in A14, 
we estimate that the cumulative relative uncertainty in \Macc ~in log scale is about  0.42\,dex. 
The four \Macc ~values reported in Table~\ref{pars} for each YSO correspond to the four evolutionary 
models adopted to derive the mass. The differences on \Macc ~when adopting different models are well 
within the errors.

We derived mass accretion rates in the range from $\sim$5$\times$10$^{-12}$\,\Msun~yr$^{-1}$ 
to $\sim$6$\times$10$^{-8}$\,\Msun~yr$^{-1}$, that is, similar to the \Macc ~range of the GTO sample.
With a \Macc ~$\sim$6$\times$10$^{-8}$\,\Msun~yr$^{-1}$ the strongest accretors in the total sample 
are the $\sim$0.8\Msun ~YSOs  Sz\,83, Sz\,98, and GQ. We note that these numbers do not 
account for the sub-luminous objects. When corrected for disk obscuration Sz\,102 may be among
the strongest accretors. The weakest accretor is the $\sim$0.07\Msun 
~object AKC2006-18 with \Macc ~of 5.8$\times$10$^{-12}$\,\Msun~yr$^{-1}$.
With \Mstar=0.02\Msun, 2MASS\,J16085953-3856275 is close to the planetary mass regime, but its 
mass accretion rate of 2.4$\times$10$^{-11}$\,\Msun~yr$^{-1}$ is similar or higher than that of 
objects with a mass $\approx$0.1\Msun.

\section{Accretion properties of the total sample}
\label{accprop}

In this section we present the results of the accretion properties of the total sample
in relationship with the YSOs stellar parameters. 

\subsection{Accretion luminosity versus YSO luminosity}
\label{Lacc_corr}
The accretion luminosity as a function of stellar luminosity is shown in Figure~\ref{Lacc_vs_Lstar}. 
There is no significant difference between the distribution of points of the new sample in the 
diagram with respect to the GTO sample, although the global dispersion of the \Lacc--\Lstar 
~relationship slightly increased and the trend seems less steep than for the GTO alone. 
Yet, the data points are apparently less scattered than those of previous samples of other 
star forming regions like $\rho$-Oph or $\sigma$-Ori 
\citep[e.g.][ and references therein]{natta06, rigliaco11a, manara15, manara16a}, 
where the scatter may be more than 2\,dex at a given stellar luminosity.


\begin{figure}[h]
\resizebox{1.0\hsize}{!}{\includegraphics[]{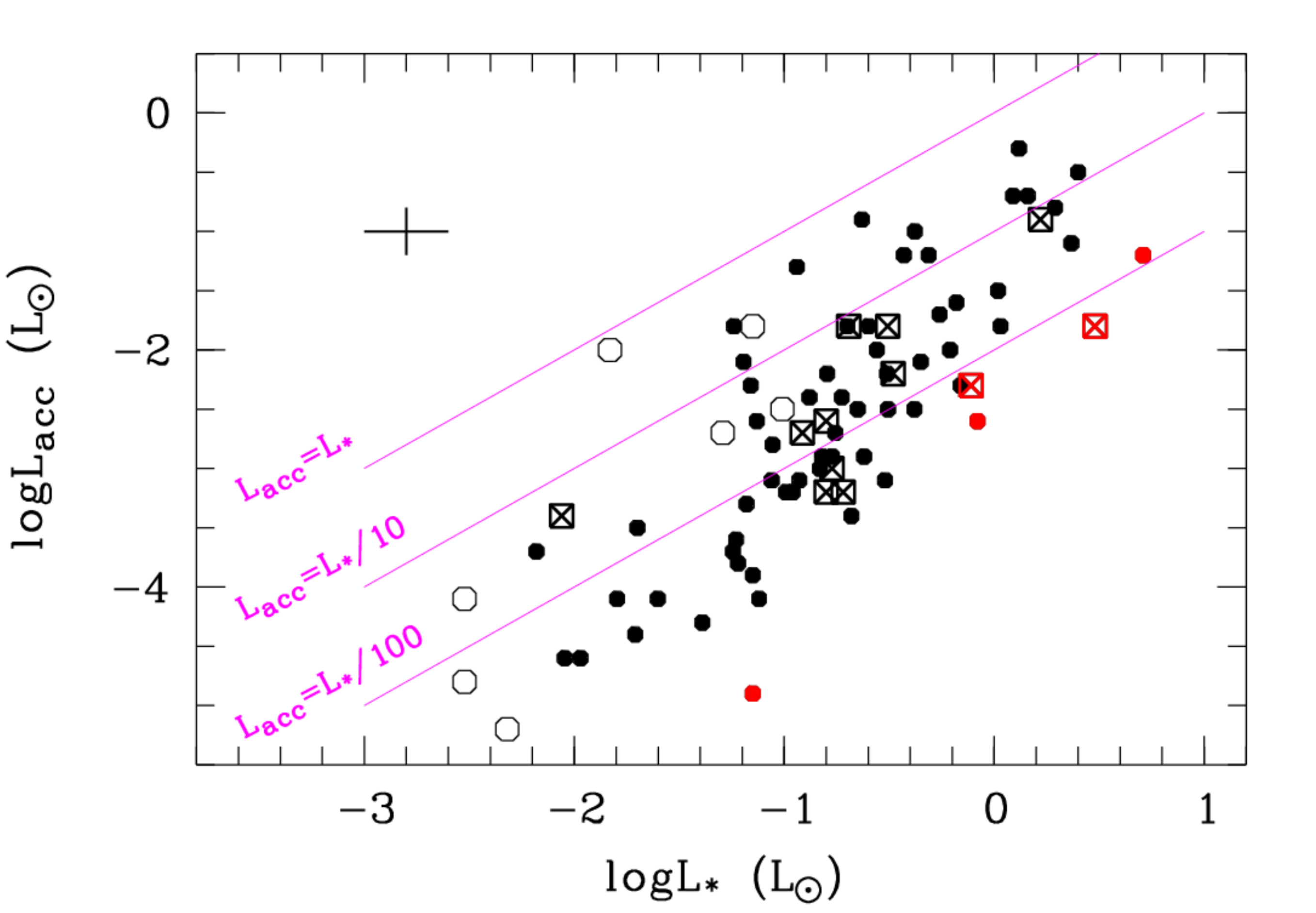}}
\caption{Accretion luminosity as a function of stellar luminosity.
        The transitional discs are shown with crossed squares, 
        while the sub-luminous objects with open circles. 
        The five weak accretors are shown with red symbols.
        The continuous lines represent the three \Lacc ~vs. \Lstar 
        ~relations as labelled. Average error bars are shown in the 
        upper left.  
    \label{Lacc_vs_Lstar}}
\end{figure}

All the Lupus YSOs analysed here fall below the \Lacc=\Lstar  ~boundary, with a small fraction 
of objects ($\sim$12\%) between 0.1 and 1\,\Lsun, and many with \Lacc/\Lstar ~$<$0.01. 
The fact that our sample lacks YSOs with \Lacc$>$~1\Lstar ~is interesting because in other star 
forming regions like Chamaeleon and Taurus there are class~II sources with \Lacc$\ge$\Lstar
~\citep[see][for Chamaeleon~I]{manara16a}. This is rather peculiar for class~II sources
because it is expected that  \Lacc$>$\Lstar ~only in class~I sources in which the level 
of accretion rate is very high. In fact, the luminosity of class~I protostars is mainly 
driven by accretion and not by a photosphere. 
The point closest to the \Lacc=\Lstar ~boundary corresponds to the sub-luminous YSO Sz\,102, 
whereas the (non sub-luminous) object with the lowest $\log{L_{\rm acc}}$ value is Lup\,607, 
but its accretion rate is low and comparable with the chromospheric level. 
The YSOs with transitional discs follow the same trend as the objects with full discs. 
A linear fit to the data in Figure~\ref{Lacc_vs_Lstar} using ASURV \citep{feigelson85}, 
excluding sub-luminous objects and considering the five weak accretors as upper limits 
(See section~\ref{chrom_contrib}), yields:

 \begin{equation}
 \log{L_{\rm acc}}  = (1.26\pm 0.14) \cdot \log{L_{\star}}  - (1.60 \pm 0.13) 
 \label{fit_without_slysos}
 ,\end{equation} 

\noindent
with a standard deviation of 0.7, while considering the five weak accretors as detections 
the fit yields:

 \begin{equation}
 \log{L_{\rm acc}}  = (1.31 \pm 0.13) \cdot \log{L_{\star}}  - (1.54 \pm 0.12) 
 \label{fit_without_slysos}
 ,\end{equation} 

\noindent
with a standard deviation of 0.7. Therefore, considering the five values as upper limits or
real detections has no significant effect on the fits. In the following, we 
consider the values for the weak accretors as detections. The tool of robust regression 
analysis based on the least median of squares (LMS)  \citep[see][]{rousseuw84, rousseuw87} 
implemented in ESO-MIDAS\footnote{European Southern Observatory - Munich Image Data Analysis System} 
yields a slope of 1.55$\pm$0.11. Thus, the $\log$\Lacc--$\log$\Lstar ~relationship 
for the total sample in Lupus is steeper than the \Lacc/\Lstar~$=$constant lines, as 
found in previous works for YSOs in other star forming regions \citep[e.g.][]{natta06, rigliaco11a}. 

\begin{figure}[h]
\resizebox{1.0\hsize}{!}{\includegraphics[]{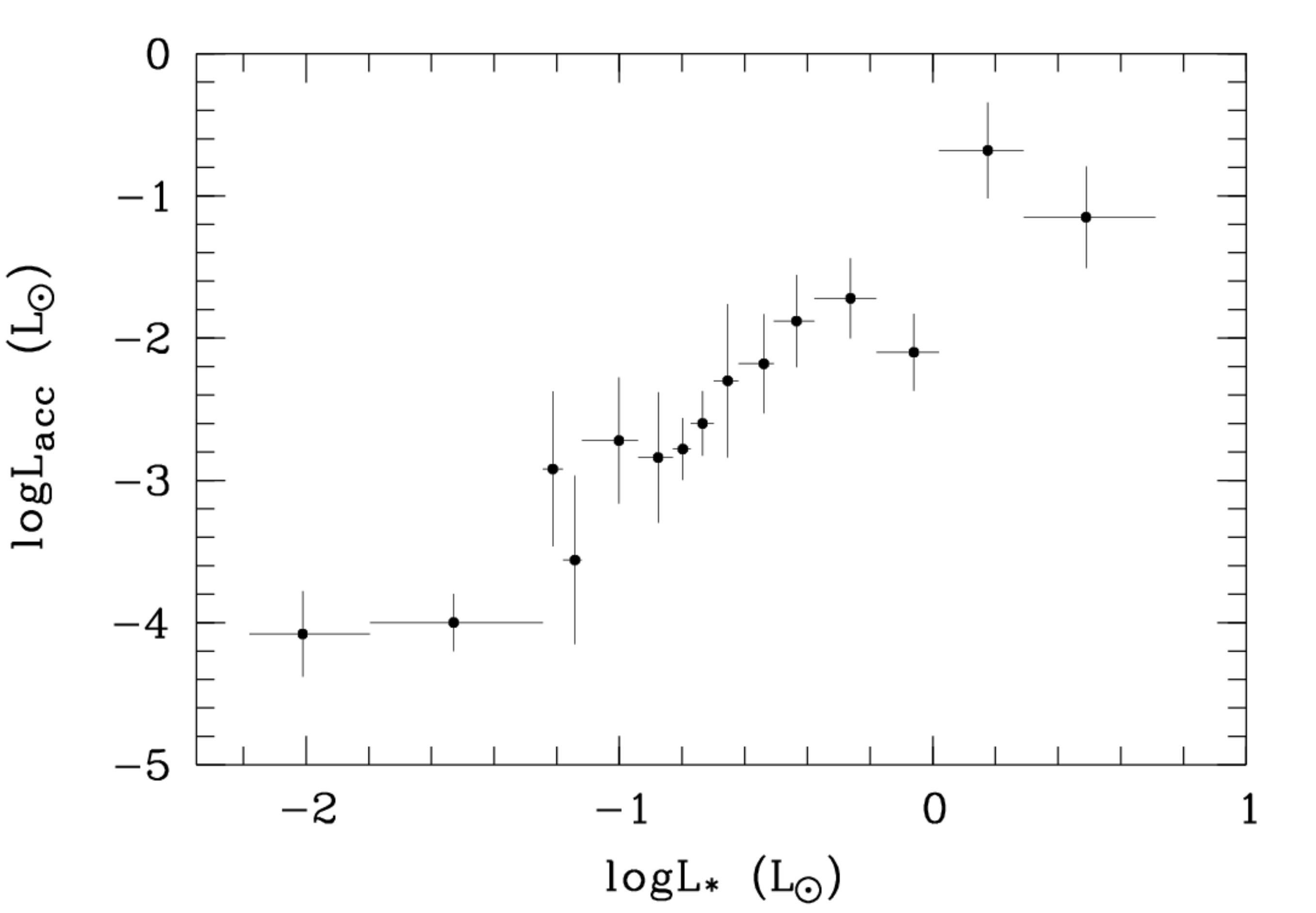}}
\caption{Median values of accretion luminosity as a function of binned stellar luminosities.
        Each point represents the median of 5 \Lacc~values with similar \Lstar.  The horizontal 
        bars show the intervals of $\log$\Lstar. 
    \label{Lacc_Lstar_median}}
\end{figure}

Interestingly, the distribution of points in the \Lacc--\Lstar ~ plane in Figure~\ref{Lacc_vs_Lstar} 
shows some evidence of a break at $\log$\Lstar ~values between $-1.2$ ~and $-1.0$,
which corresponds to a mass between 0.1\,\Msun ~and 0.2\,\Msun ~at the 3\,Myr isochrone 
(see Figure~\ref{HRD}). There are basically no strong accretors at low stellar luminosities; the vast 
majority of the (non sub-luminous) objects with \Lstar ~ lower than a tenth of a solar luminosity 
fall below the \Lacc/\Lstar $=$0.01 line, with only three having \Lacc/\Lstar ~values between 0.1 and 1, 
and six between 0.01 and 0.1. 
To further investigate the behaviour of the $\log$\Lacc--$\log$\Lstar ~relationship we calculated 
median values of \Lacc ~as function  of \Lstar (Figure~\ref{Lacc_Lstar_median}). 
The width of each of the 15 bins has been chosen to have a similar number of stars in each bin (5). 
The binning was done over the total sample, but excluding the sub-luminous objects. The plotting 
errors were estimated as  $\sqrt{\frac{\pi}{2}}$~$\sigma_{mean}$/$\sqrt{n}$ \citep[see][]{kendall77}, 
where $\sigma_{mean}$ is the standard deviation over the mean and $n$ is the sample size
in each bin (five). 
The binned $\log$\Lacc--$\log$\Lstar ~relationship rises rapidly with a slope $1.7\pm0.2$ for $-1.6\lesssim\log$(\Lstar/\Lsun)$\lesssim-0.4$
and flattens at \Lstar$\ga -0.4$ (slope $\approx$ 1.0), while remaining more or less flat for  $\log$\Lstar ~values 
below $-$1.6. However, we stress that the latter behaviour is affected by incompleteness of the sample at very 
low \Lstar ~values, that is, in the sub-stellar regime.

\subsection{Accretion rate versus mass}
\label{Macc_corr}

In Figure~\ref{Macc_Mstar} the mass accretion rate is shown as a function of the stellar mass. 
When including the new sample, the scatter of the \Macc-\Mstar ~relationship increases with 
respect to the scatter of the GTO sample alone (see also Figure~8 in A14). A linear fit 
to all the data taking into account upper limits, but excluding the sub-luminous objects, yields a 
slope 1.8$\pm$0.2, with a dispersion of 0.7 (see Table~\ref{MaccMs_linfits} in Appendix~\ref{stelprop}) . 
Although increased with respect to the value for the GTO sample alone \citep[0.3 using the][tracks]{siess00}, 
the dispersion of the \Macc-\Mstar ~relationship is still less than in previous investigations 
in the literature 
\citep[][and references therein]{muzerolle03, mohanty05, natta06, HH08, rigliaco11a, antoniucci11, biazzo12}. 
Therefore, at first approximation, we can conclude that for the class~II and transitional YSOs in 
Lupus \Macc~$\propto$~\Mstar$^{1.8(\pm0.2)}$, in agreement with the results of a number of previous 
studies of other star forming regions 
\citep{natta06, muzerolle05, HH08, rigliaco11a, antoniucci11, biazzo12, manara16a}.
Using other evolutionary models to derive \Mstar ~and \Macc ~ yields similar results for the 
slope of the relationship, although the \citet{baraffe98} tracks tend to provide a slightly 
less steep (slope 1.6$\pm$0.2) relationship and the scatter varies significantly depending 
on the adopted evolutionary tracks (see Appendix~\ref{stelprop}). 


\begin{figure}[h]
\resizebox{1.03\hsize}{!}{\includegraphics[]{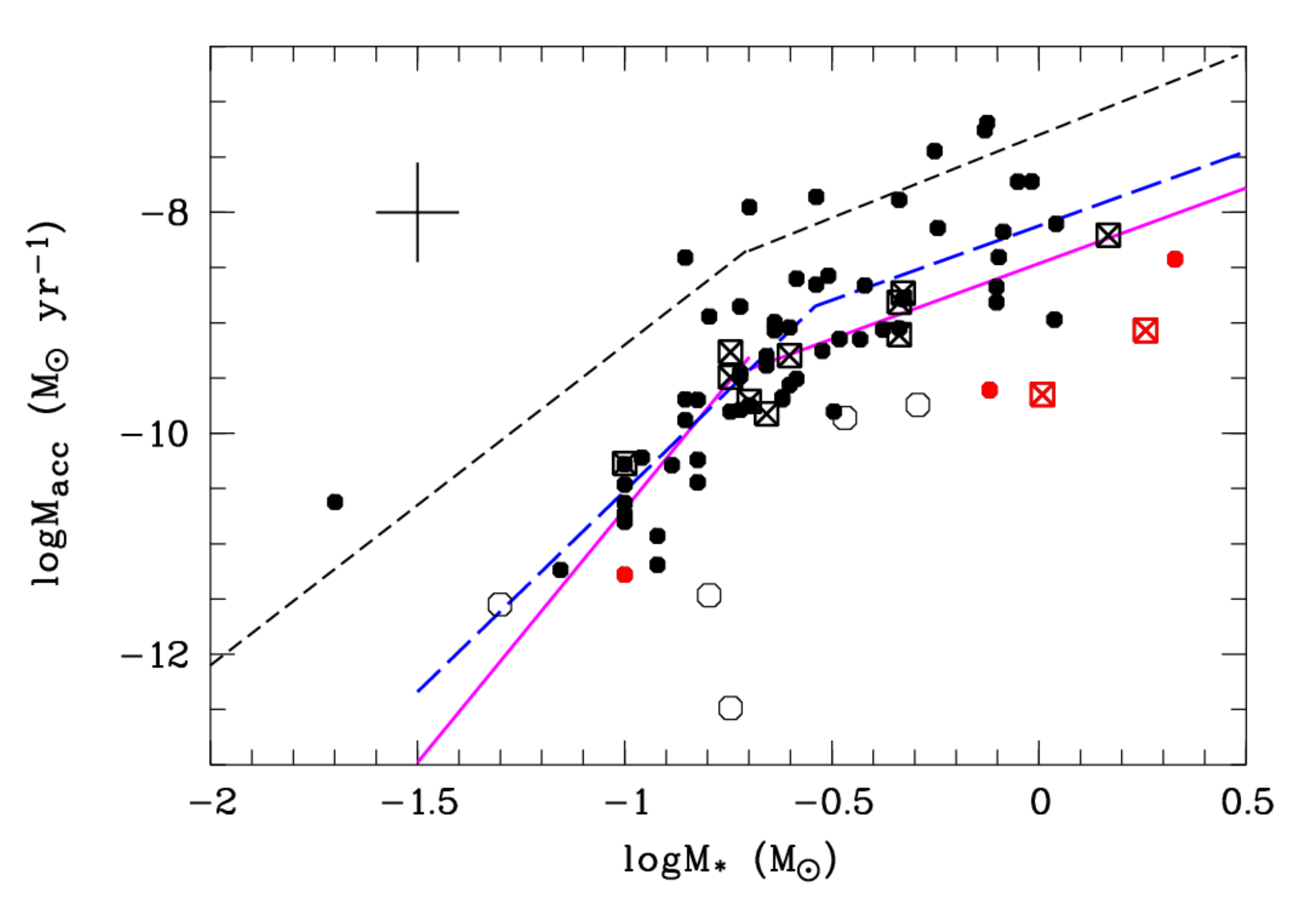}}
\caption{Mass accretion rate \Macc ~as a function of mass for the total sample in $\log$ scale.
        Plotting symbols are the same as in Figure~\ref{Lacc_vs_Lstar}. The average errors 
        in $\log$\Mstar ~and $\log$\Macc ~are shown in the upper left. The black dashed line
        shows the double power-law theoretically predicted by  \citet{vorobyov09}.  The 
        continuous magenta lines represent the fits to the data as in Equations~\ref{fit_with_lmysos}
        and \ref{fit_with_hmysos}. The long-dashed blue line shows the robust double-linear fit
        following the prescription by \citet{manara17} as explained in the text.
     \label{Macc_Mstar}}
\end{figure}

The distribution of points in Figure~\ref{Macc_Mstar} also shows some evidence of a break 
at $\log$\Mstar ~values between $-1$ and $-0.7$ (i.e. 0.1\,\Msun ~and 0.2\,\Msun). 
More interesting, the range of \Macc ~in log scale for the sub-sample with \Mstar$<$ 0.2\Msun 
~covers about 3.5\,dex  ~in less than about 1\,dex in \Mstar, ~whereas in comparison the 
higher-mass sub-sample covers a narrower range of \Macc  ~($\sim$2.7\,dex) in a wider range 
of mass  ($>$1\,dex). 
The larger range in $\log$\Macc ~in the low-mass sub-sample in comparison with the range 
for the high-mass sub-sample is confirmed by the  Kaplan-Meier \citep[K-M ;][]{kaplan58}  
distributions shown in Figure~\ref{KM_logMacc}.  
The difference between the \Macc ~distribution of the sub-samples can be indeed quantified 
by the K-M distributions. The slope (-0.43$\pm$0.01) of the K-M distribution for the 
high-mass sub-sample is slightly steeper than for the low-mass subsample (slope=-0.38$\pm$0.01), 
and significantly steeper than the K-M distribution of the objects with $\log$\Macc$\le -10.0$ 
(slope=-0.28$\pm$0.02).
All these arguments suggest that the distribution of \Macc ~ as a function of mass for the 
high-mass subsample remains flatter than for the low-mass sub-sample, meaning a 
bi-modal behaviour of the $\log$\Macc--$\log$\Mstar ~relationship.

\begin{figure}[h]
\resizebox{1.03\hsize}{!}{\includegraphics[]{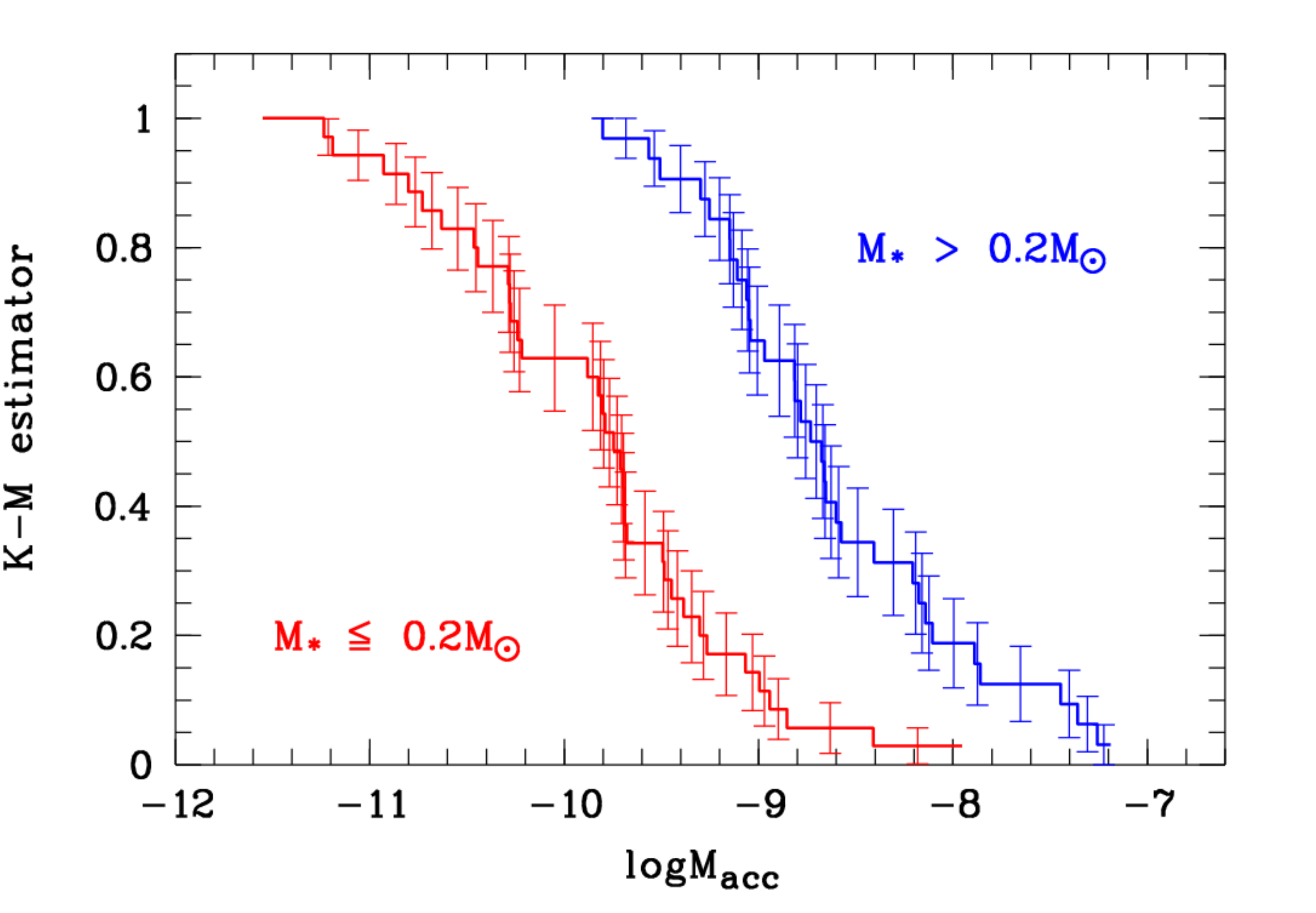}}
\caption{The Kaplan-Meier distribution of $\log$\Macc ~ for the low-mass (\Mstar$\le$0.2\Msun)
        and high-mass (\Mstar$>$0.2\Msun) sub-samples are shown in red and blue,
        respectively. 
     \label{KM_logMacc}}
\end{figure}

The suggestion by \citet{vorobyov08} that the \Macc~$\propto$~\Mstar$^2$ relationship can be 
explained on the basis of self-regulated accretion by gravitational torques in self-gravitating 
discs led these authors to conclude that the relationship is better described as a double power-law, 
with the break occurring at \Mstar$\approx$0.2\,\Msun, \citep{vorobyov09}. The double power-law 
suggested by these authors is shown in Figure~\ref{Macc_Mstar} with the black dashed line. 
Although the theoretically predicted \Macc ~values are generally higher than the measured ones, 
they are rather consistent with the upper envelope of the Lupus relationship. As pointed out by 
\citet{vorobyov09}, the theoretical \Macc ~values may be somewhat overestimated with respect to 
the observed ones. They explained this effect in terms of the adopted values of viscosity in 
the models. The objects in Figure~\ref{Macc_Mstar} falling above the modelled values are the 
strongest accretors at a given mass and are also among the more luminous on the HR diagram. 
Separate linear fits to the data, setting 0.2\,\Msun ~as dividing line, and using the tool of 
robust regression analysis based on the LMS method \citep[see][]{rousseuw84, rousseuw87} yield 
the following results:

  \begin{equation}
  \log{\dot M_{\rm acc}}  = 4.58 (\pm0.68) \cdot \log{M_{\star}}  - 6.11 (\pm0.61)   
  \label{fit_with_lmysos}
  ,\end{equation} 

  \noindent 
  and

  \begin{equation}
  \log{\dot M_{\rm acc}}  = 1.37 (\pm0.24) \cdot \log{M_{\star}}  - 8.46 (\pm0.11)  
  \label{fit_with_hmysos}
  ,\end{equation} 

\noindent 
for the low and high mass regimes, respectively. These fits, shown as magenta lines 
in Figure~\ref{Macc_Mstar}, still resemble the theoretical behaviour, but with a steeper
slope for the low-mass regime and the measured values being about 1\,dex below the 
predicted ones.  
We have performed a further fit setting the breakpoint of the relationship as a free parameter 
and following the prescription outlined in \citet{manara17}. These authors performed a statistical 
test to demonstrate that a double-power law is a slightly better description of the $\log$\Macc-$\log$\Mstar 
~relationship than a single-power law. The break point of the robust double-linear fit is at  
\Mstar$=$0.29\Msun, and the slopes for the low and high mass regimes are 3.64 and 1.35, respectively. 
The double-linear fit is shown with the long-dashed blue line in Figure~\ref{Macc_Mstar}.
As shown in  Appendix~\ref{stelprop} the break of the \Macc--\Mstar ~relationship 
is evident independently of the PMS evolutionary track used to derive \Mstar ~and \Macc.
Thus, we conclude that at high masses the relationship is flatter than at low masses.  

The steeper slope for the low-mass regime might be biased by the small number statistics 
at \Mstar$\lesssim$0.2\,\Msun.
A similar binning approach as for the $\log$\Lacc--$\log$\Lstar ~relationship in the previous 
section, yields the binned $\log$\Macc--$\log$\Mstar ~relationship shown in Figure~\ref{Macc_Mstar_binned}.
In this figure the two-slope relationship by \citet{vorobyov09} is overplotted with a dashed 
line, but shifted by $-1$\,dex in \Macc. Although the predicted \Macc ~values are higher than 
our measurements, the Lupus results are qualitatively consistent with those models.

\begin{figure}[h]
\resizebox{1.03\hsize}{!}{\includegraphics[]{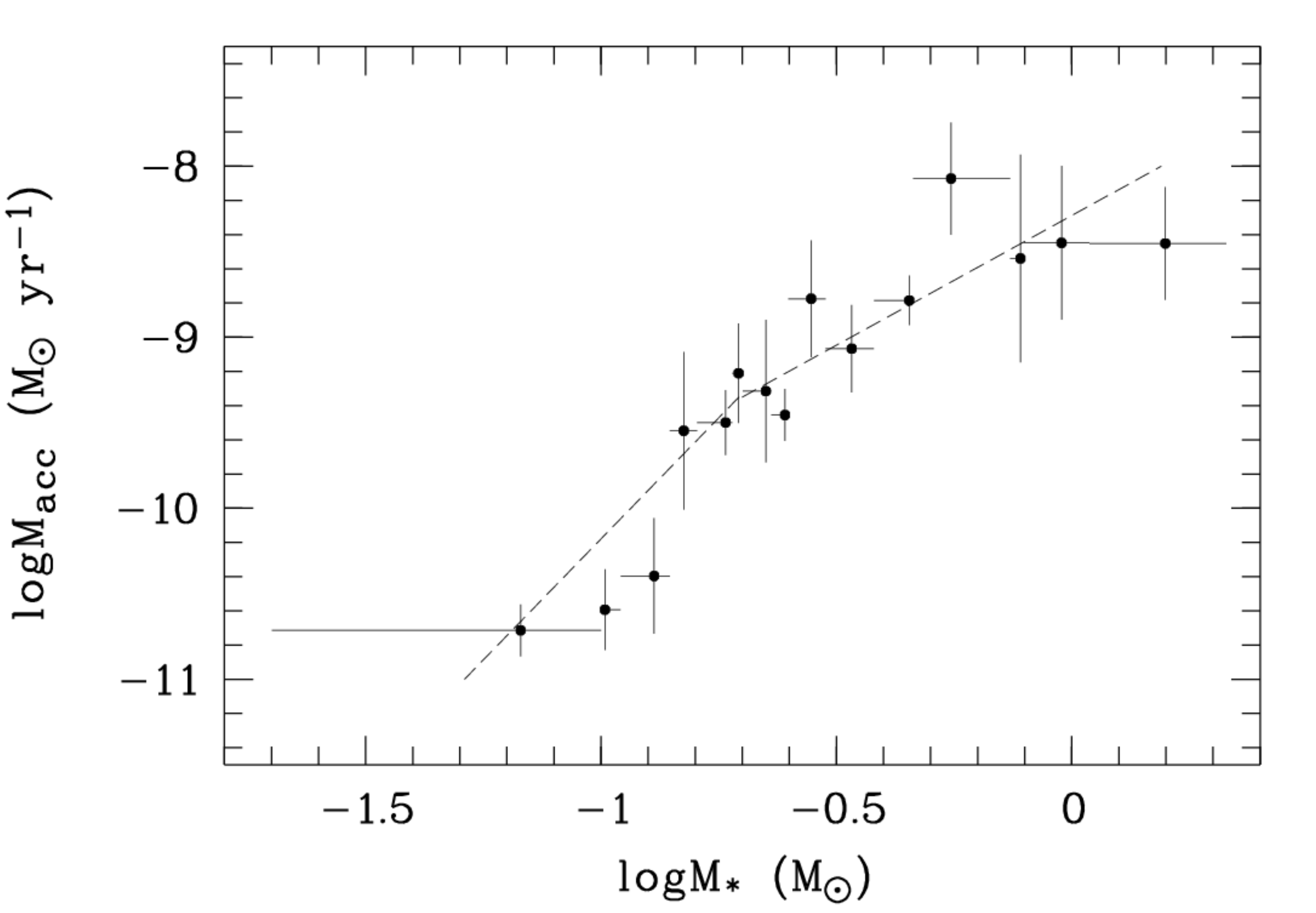}}
\caption{Median values of accretion rate as a function of binned stellar mass.
        Each point is the median of 15 \Macc~values with similar \Mstar. The
        plotting errors were computed in the same way as in Figure~\ref{Lacc_Lstar_median}. 
        The horizontal bars represent the intervals in $\log$\Mstar. The dashed line shows the two 
        power-law relationship by \citet{vorobyov09}, vertically shifted by $-1$\,dex.
     \label{Macc_Mstar_binned}}
\end{figure}

Finally, another interesting result is that most of the YSOs with transitional discs are
well mixed with those of full discs in the $\log$\Macc--$\log$\Mstar ~plot, suggesting that
their accretion properties are in general similar to those of YSOs with full discs, in 
agreement with previous results \citep[][]{manara14, espaillat14}. We note, 
however, that there are no transitional YSOs exhibiting levels of accretion as high as those 
displayed by some YSOs with full discs and at a given mass some of them (e.g. MY\,Lup 
and SSTc2dJ160830.7-382827) are among the weakest accretors in the total sample. 

\section{Discussion}
\label{discussion}

A detailed observational study of accretion and its evolution requires complete and 
homogeneous samples of YSOs. The sample of class~II and transitional YSOs studied here 
is complete at a level of more than 90\% with respect to the total sample of this type
of objects in the Lupus~I, II, III and IV clouds. 
The stellar and accretion properties of the sample have been  self-consistently derived, 
allowing an unbiased study of the accretion and its relationship with the stellar 
parameters. We have shown that the accretion luminosity and stellar luminosity of the 
Lupus class~II  and transitional YSOs are correlated with a lower scatter in comparison 
with previous studies of YSOs in other star forming regions 
\citep[e.g.][ and references therein]{natta06, rigliaco11a}. A similar low scatter has been 
found recently for the \Lacc--\Lstar ~relationship of the Chamaeleon~I YSOs \citep{manara16a}, 
analysed with similar methodologies as here. 

The Lupus correlation between \Lacc ~and \Lstar, when fitted with a single power-law, is  
similar within the errors to that found in previous work \citep[e.g.][]{natta06}. 
\citet{clarke06} pointed out that the distribution of points in the \Lacc -- \Lstar ~plane 
more or less fills a region that is bounded by the \Lacc = \Lstar ~relation at high \Lacc,
and claimed that the relation is the result of a combination of detection biases 
at low values of \Lacc, roughly following a power-law \Lacc~$\propto$~\Lstar$^{1.6}$. 
The Lupus results suggest that the relation is real, as argued, for example, by \citet{ercolano14}.
The relation \Lacc-\Lstar ~in Lupus, however, shows that a single power-law may not the 
best description of the data, which show evidence of a break,  with the relationship being 
steeper at low \Lstar ~values than at high \Lstar ~values. This effect was not seen in other 
regions, but is also observed in the Chamaeleon I X-Shooter survey by \citet{manara17}.

The approximation \Macc$~\propto$~\Mstar$^\alpha$, with $\alpha \approx +2$ for the Lupus YSOs 
is consistent with the previous results for YSOs in other star forming regions. 
The steep relation and sread of the \Macc-\Mstar ~correlation has been interpreted as the imprints
of the initial angular momentum of the parental cores where the star-disc systems 
were formed \citep[e.g.][]{dullemond06}. 
The spread of the relationship has also been ascribed to a spread of stellar properties, such 
as X-ray and EUV emission \citep{muzerolle03, ercolano14}. These latter authors in particular 
conclude that the observed \Macc--\Mstar ~relation in YSOs is consistent with being a simple 
consequence of disc dispersal by X-ray photoevaporation.
Variability is another possible source of spread in the \Macc ~values. \citet{costigan12} have 
shown that the typical variability of \Macc ~in Chamaeleon I targets is generally $\le$0.4 dex. 
Independent studies in other star forming regions confirm similar values \citep[e.g.][]{biazzo12, venuti14}. 
Therefore, this effect can be relevant to explain some of the observed spread of \Macc ~values.
In fact the average size of the error bars of the binned $\log$\Macc--$\log$\Mstar ~relationship 
shown in Figure~\ref{Macc_Mstar_binned} is also on the order of 0.3-0.4\,dex.
The level of variability, however, may be different depending on the evolutionary status of 
the YSO populations. This may contribute explaining a different spread of \Macc ~among YSOs 
of the same mass in star forming regions of different age.

However, the new result from our analysis is that  there is evidence of a break of the scaling
relations at low \Mstar ~and \Lstar ~values. The homogeneous methods used here and the completeness 
of the sample allow us to conclude that the bi-modality of the \Macc--\Mstar ~relation of Lupus 
is real, regardless of the evolutionary models used to derive the stellar mass. It is worth 
noting that a similar behavior has been confirmed for the young stellar population in the 
L1641 region \citep{fang13a} and in Chamaeleon~I \citep{manara17}.

The break of the empirical relationship in Lupus resembles the theoretical prediction by 
\citet{vorobyov09}. In these models the gravitational instability due to the self-gravity of 
the discs in the early phase of disc evolution limits the disc mass in the higher mass 
(\Mstar$\ga$0.2\Msun) objects, effectively setting an upper limit on the mass 
accretion rates in the late evolution, hence flattening the \Macc-\Mstar ~relation in this 
mass regime. The gravitational instability has little effect in the low-mass regime, where 
viscous evolution dominates at basically all times. Therefore, our result that the 
\Macc--\Mstar ~relation flattens at the high mass regime supports the importance of modelling 
self-gravity in the early evolution of the more massive systems, as suggested in \citet{hartmann06}. 
However, as pointed out in \citet{rigliaco11a}, other physical processes, such as photo-evaporation 
and planet formation, may also occur during YSOs lifetime leading to disc dissipation on different 
timescales depending on the stellar mass. 

Interestingly, a break of the \Macc--\Mstar ~relation at the very low-mass regime (\Mstar~$\lesssim$~0.1\Msun) 
has also been predicted by \citet{stamatellos15}. To explain the very high levels of accretion observed 
in substellar and planetary-mass companions to some T~Tauri stars \citep{zhou14}, \citet{stamatellos15}
model the accretion onto very low-mass objects that formed by the fragmentation of the disc around 
the more massive star. During the early evolution the individual discs of substellar companions 
--including those at the planetary-mass regime--  accrete additional material from the gas-rich 
parent disc, hence, their discs are more massive and their accretion rates are higher than if 
they were formed in isolation. Therefore, these very low-mass objects have disc masses and accretion 
rates that are independent of the mass of the central object and are higher than expected from 
the scaling relations of more massive YSOs. These models predict that  \Macc ~is basically 
independent of \Mstar. Our data show a hint for a flattening of both \Lacc--\Lstar ~ 
and \Macc--\Mstar  ~relationships at the very low \Lstar/\Mstar ~end, but the sample lacks a 
statistically significant number of low-mass substellar objects to establish the trend.
It is, however, interesting that our target close to the planetary-mass regime has a relatively 
high \Macc ~ in comparison with the value measured in the lowest mass YSOs in our sample. 

Although the scatter of the Lupus relationship increased with respect to our previous result 
in A14, it is still less than for other samples and its upper envelope follows the same steep 
trend, in contrast to the Taurus YSOs, where the upper envelope of the relationship is flatter 
\citep[see Fig.~1 in][]{hartmann06}. 
The steeper slope of the upper envelope may lead to the idea of a faster disc evolution
of the Lupus low-mass stars than those in Taurus, suggesting that the Lupus YSO population 
might be different from the population in Taurus or other regions. \citet{hughes94} concluded 
that Lupus may be a region of sub-critical star formation where magnetic fields slow the 
collapse of the clouds, leading to low mass accretion rates with the consequence that the 
lowest mass stars in Lupus are less active than similar objects in other regions. On the other 
hand, \citet{galli15} provided evidence that the disc lifetimes may be shorter in Lupus in 
comparison with those in Taurus.

A crucial aspect of the models regarding viscously evolving discs is the  presence of the 
correlations of \Mdisk ~ with \Mstar ~ and \Macc 
~\citep[e.g.][and references therein]{lynden-bell74, hartmann06, dullemond06, tilling08}. 
On the observational side, these scaling relationships have been discussed in the reviews
by \citet[][]{natta07} and \citet[][]{williams11}.
Despite the strong efforts on detecting such correlations, 
previous works \citep[e.g.][]{ricci10, andrews10, olofsson13} failed on finding a scaling between 
the disc mass and the stellar mass, or the mass accretion rate, within the uncertainties of the 
measurements.
Later investigations confirmed a robust \Mdisk~$-$~\Mstar ~correlation for the class~II YSOs 
in Taurus \citet{andrews13} and in the Upper Scorpius OB Association \citet{barenfeld16}, and more 
recently the synergy between the ALMA and X-Shooter projects has also been successful in confirming 
it for YSOs in Chamaeleon~I \citep[][]{pascucci16}.  
The combination of the data presented here with those reported in the ALMA survey of Lupus 
protoplanetary discs have shown significant correlations between \Mdisk ~and \Mstar ~\citep{ansdell16} 
and \Mdisk ~and \Macc~ \citep{manara16b}. The ALMA survey did not include, however, YSOs 
with \Mstar$\le$0.1\Msun, preventing us from investigating whether the scaling relationships are 
different for low-mass substellar objects than for stars.
The characterisation of the physical and accretion properties of candidates to very low-mass 
substellar objects, using future facilities with higher performance than X-Shooter, and 
high-sensitivity observations with ALMA of these objects will provide important clues for 
their formation mechanisms.

\section{Summary and conclusions}
\label{summary}

We have used X-Shooter@VLT to investigate 93 YSOs previously classified as class~II sources in 
the Lupus star forming region. 
The capabilities of X-Shooter in terms of wide spectral coverage, 
resolution and accurate flux allowed us to characterise the sample in terms of stellar and accretion 
properties in a homogeneous and self-consistent way and to accomplish an unbiased study of  
accretion and its relationship with stellar parameters.

Our observations confirm that one of the most important sources of contamination of the samples of 
YSO candidates drawn from photometric surveys are background giants, in agreement with previous works 
\citep[e.g.][]{oliveira09, alcala11, mortier11, comeron13}. We have found that about 10\% of the YSOs 
previously classified as class~II candidates are indeed  unrelated to the Lupus star forming region, 
with an important impact on the disc demography of the star forming region. Without the knowledge 
of this contaminating component the detection rate of the 95\% complete ALMA survey of Lupus protoplanetary 
discs by \citet{ansdell16} would have resulted in $\sim$60\% instead of $\sim$70\%, highlighting the need 
for optical/infrared spectroscopic complementary data to ALMA.

Our study of the 81 confirmed Lupus YSOs allowed us to accomplish a synthesis of the accretion 
properties of the almost ($>$~90\%) complete sample. The accretion luminosity and stellar luminosity 
of the Lupus YSOs are correlated with a lower scatter in comparison with previous studies of YSOs 
in other star forming regions. The  slope of the \Lacc -- \Lstar ~relationship is not driven by 
selection biases and there is a lack of strong accretors at the low YSO luminosity regime, suggesting 
a break of the relationship at \Lstar$\approx$0.1\,\Lsun.  

For the Lupus YSOs we conclude that \Macc$~\propto$~\Mstar$^\alpha$, with $\alpha =  +1.8\pm0.2$, 
but we found evidence of a break of the scaling relations at low \Mstar ~and \Lstar 
~values. The homogeneous methods used here and the completeness of the sample allow us to 
confirm the bi-modality of the \Macc--\Mstar ~relation of Lupus YSOs, independently of the 
evolutionary models used to derive the stellar mass. The bimodal behaviour of the observed 
relationship supports the importance of modelling self-gravity in the early evolution of the 
more massive discs, but other processes such as photo-evaporation and planet formation during 
YSOs lifetime, may also lead to disc dissipation on different timescales depending on the 
stellar mass. Our data show tantalising evidence of relatively constant \Macc ~below 0.1\,\Msun, 
possibly indicating that some of the very low-mass substellar objects may have formed as 
companions of stars by the fragmentation of the circumstellar disc. However, our sample 
lacks a statistically significant number of low-mass substellar objects to confirm the 
result.

The accretion properties of most transitional YSOs are in general similar to those of 
objects with full discs, with a minority of them having accretion rates an order of magnitude 
lower than objects with full discs. However, the highest accretion rates are only seen 
in objects with full discs.

\begin{acknowledgements}
We thank the anonymous referee for her/his comments and 
suggestions. CFM gratefully acknowledges an ESA Research Fellowship.
AN acknowledges funding from Science Foundation Ireland (Grant 13/ERC/I2907). 
We thank M. Tazzari and M. Ansdell for discussions on ALMA data. We also thank the
ESO staff, in particular Markus Wittkowski and Giacomo Beccari for their excellent
support during phase-2 proposal preparation, and the Paranal staff for their
support during the observations. We thank G. Cupani, V. D'Odorico, P. Goldoni and 
A. Modigliani for their help with the X-Shooter pipeline. 
Financial support from INAF under the program PRIN2013 "Disk jets and the dawn of planets"
 is also acknowledged.
This work was partly supported by the Gothenburg Centre for Advanced Studies in 
Science and Technology as part of the GoCAS program Origins of Habitable Planets 
and by the Italian Min- istero dell'Istruzione, Universit\'a e Ricerca through 
the grant Progetti Premiali 2012-iALMA (CUP C52I13000140001).
This research made use of the SIMBAD database, operated at the CDS (Strasbourg, France). 
\end{acknowledgements}





\begin{appendix}

\section{Physical parameters and accretion properties of the total sample adopting different models}
\label{stelprop}

The complete list of the 81 confirmed YSOs and the synthesis of their physical parameters are 
given in Table~\ref{pars}. The quantities have been consistently derived as explained in A14 
for the GTO sample (first 36 raws) and Section~\ref{params} for the new sample, respectively. 
The $\pm$0.5 subclass and $\pm$1 subclass uncertainties for the M-type and earlier type objects 
translate into uncertainties of 0.01\,dex and 0.02\,dex in $\log T_{\rm eff}$, respectively.  
The error in YSO luminosity in $\log$ scale is proportional to the error in flux in log scale. 
The error in $\log$\Lstar ~was then estimated by taking into account the contribution of both 
the signal-to-noise (S/N) of the flux-calibrated spectra and the error in visual extinction, 
which for the purpose of error estimates we assume to be 0.5\,mag for all objects (see also A14). 
The rms of the continuum of the spectra was estimated in four spectral regions adjacent to
the H$\alpha$, H6, H13 and H15 lines; this was taken as the 1$\sigma$ error on the flux at 
the four wavelengths. The uncertainty due to extinction at the corresponding spectral regions 
was computed as 
$ \ln10\cdot 0.4\cdot R_{\rm V} \cdot \Delta A_{\rm V}$, where R$_{\rm V}$ is the extinction curve.
The relative errors due to the flux and extinction were then combined in quadratures, yielding a 
relative error in flux $\Delta$F/F at each spectral region. A weighted average was then computed
providing $\Delta\log$\Lstar $= \frac{1}{\ln10}\cdot\frac{\Delta F}{F}$. The error in luminosity 
was then computed as  $\ln10 \cdot$ \Lstar $\cdot \Delta\log$\Lstar. 

One of our main goals here is the study of the \Macc~ vs. \Mstar ~relation and how it depends
on the adopted PMS evolutionary models. Therefore, we have used the tracks by \citet{DM97}, 
\citet{baraffe98}, \citet{siess00} and \citet{baraffe15} (hereafter DM97, BA98, S00 and B15, 
respectively) to derive masses. 
The errors in the mass were calculated with a Monte Carlo procedure considering the errors in 
\Teff ~and \Lstar ~ on the HR diagram. In each realisation, the value of \Teff ~and \Lstar ~was 
randomly selected in a Gaussian distribution centered on the measured value and with a $\sigma$ 
equivalent to the uncertainty. With these values, the mass was measured using the different 
evolutionary models. A total of 1000 realisations were obtained, and the standard deviation 
of the derived stellar masses is then taken as the error on the mass estimate. The resulting 
masses with their errors are listed in Table~\ref{pars}.

The accretion luminosity for each YSO in the total sample is given in column three of Table~\ref{accretion}.
Using the data in Table~\ref{pars}, and Eq.~\ref{Macc}, these
\Lacc's were converted into the four \Macc ~values listed in the last four columns of Table~\ref{accretion}.
Uncertainties on \Macc ~were derived by error propagation using equation~\ref{Macc}
 in logarithmic form, i.e. $\log$\Macc$=\log(1.25*G) + \log$\Lacc$ + \log$\Rstar $ - \log$\Mstar. 
 Typical errors 0.25\,dex, 0.1\,dex, and 0.1\,dex in \Lacc, \Rstar ~and \Mstar, respectively, 
 yield an uncertainty of $\sim$0.3\,dex in \Macc. The uncertainty on the Lupus YSOs distance is 
 estimated to be $\sim$23\% 
\citep[see][and references therein]{comeron08}, yielding a relative uncertainty of about 0.3\,dex 
in the mass accretion rate \footnote{We note that \Macc~$\propto$~d$^3$, as \Lacc~$\propto$~d$^2$ and 
\Rstar~$\propto$~d.}. Therefore we estimate the cumulative relative uncertainty in $\log$\Macc 
~ to be about 0.42\,dex. Within errors, the mass accretion rates for each object, derived 
using the mass drawn from the different evolutionary models, are practically the same. 

\subsection{The \Macc~ vs. \Mstar ~relation with different PMS evolutionary models}
\label{app_macc_mstar}

Figure~\ref{Macc_Mstar_4models} shows the  $\log$\Macc--$\log$\Mstar ~ plots for our total sample of 
class~II YSOs and transitional discs in Lupus when adopting the four evolutionary models discussed 
in the previous subsection. For a given object the \Macc ~values are practically the 
same hence, the differences in the diagrams are mainly induced by the different \Mstar ~values 
derived from the different models. We found similar results  in a previous work \citep{biazzo14} 
for a sample of YSOs in the L\,1615/L\,1616 cometary cloud in Orion.
The fit corresponding to each model is given in Table~\ref{MaccMs_linfits}. The fits take into account
weak accretors, but exclude the sub-luminous objects. For the three YSOs with \Mstar$<$0.1\,\Msun ~in 
the \citet{siess00} tracks, we used the mass derived from the \citet{baraffe15} models to perform the fit.
The most and less scattered relationships are those drawn form the DM97 and B98 models, respectively, 
but the four fits are similar within errors. 
Importantly, the break of the relationship discussed in Section~\ref{Macc_corr} is evident independently 
of the evolutionary model adopted to derive the mass and mass accretion rate. 

\setlength{\tabcolsep}{6pt}
\begin{table}
\caption[ ]{\label{MaccMs_linfits} Fits to the $\log$\Macc--$\log$\Mstar ~relationship adopting
           different PMS models to derive \Mstar ~and \Macc. The fits are of the form 
           $\log$\Macc$=m\cdot\log$\Mstar$+c$ } 
\begin{tabular}{l|c|c|r}
\hline \hline

Adopted  & $m$ ($\pm$err) & $c$ ($\pm$err) & $\sigma ^{\star}$    \\
PMS model    &              &                &              \\
\hline     
       &                &               &                  \\

 B98   &  1.58 (0.18)   &  -8.57 (0.12) &   0.63           \\
 B15   &  1.85 (0.24)   &  -8.19 (0.16) &   0.72           \\
 S00   &  1.80 (0.23)   &  -8.28 (0.15) &   0.70           \\
 DM97  &  1.92 (0.34)   &  -8.03 (0.24) &   0.87           \\
       &                &               &                  \\

\hline
\end{tabular}
\tablefoot{~\\
$\star$ : standard deviation from linear fit.}
\end{table}


\begin{figure*}[h!]
\resizebox{1.0\hsize}{!}{\includegraphics[]{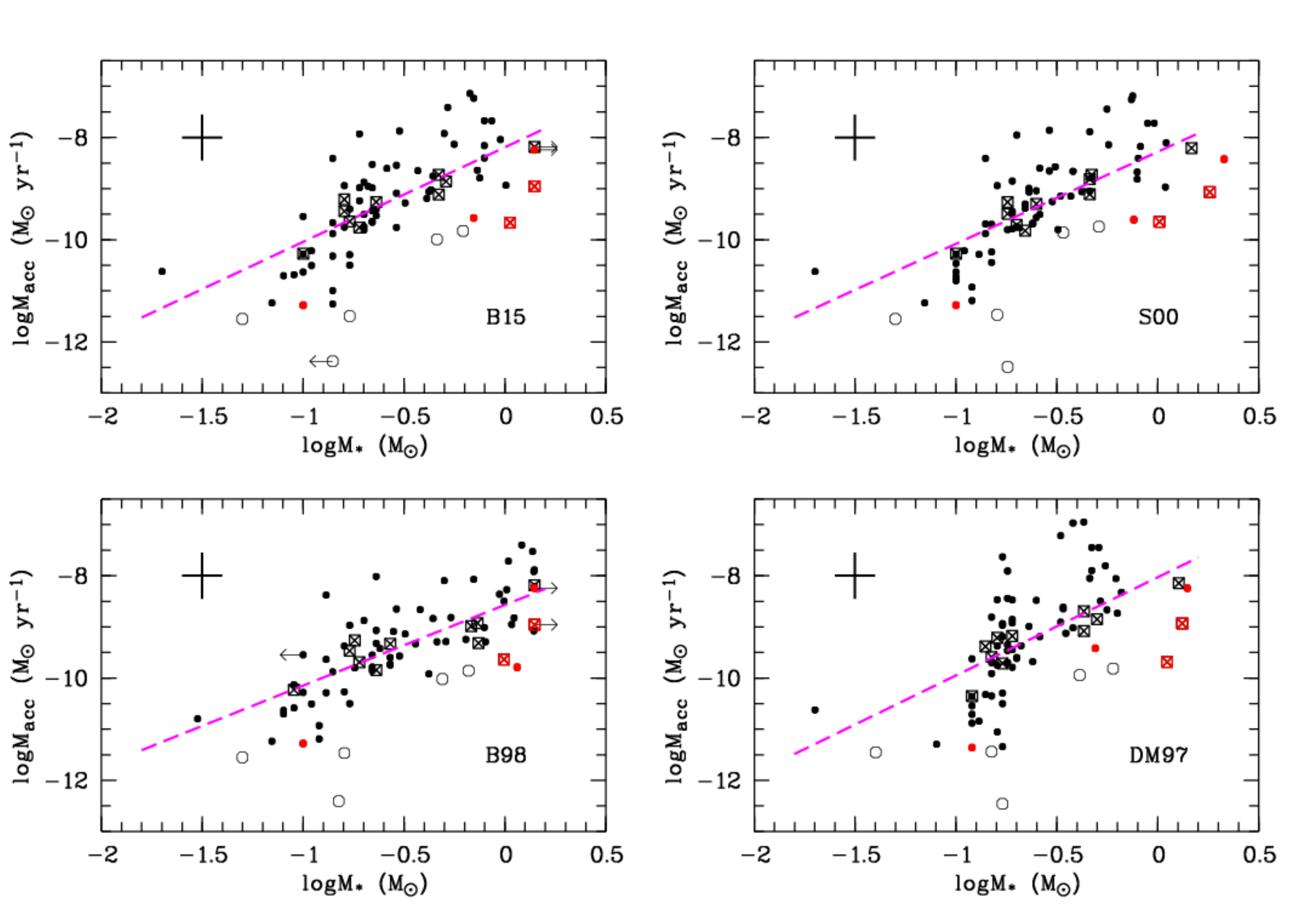}}
\caption{Mass accretion rate \Macc ~as a function of \Mstar ~ in $\log$ scale as determined using the 
        four different PMS evolutionary models described in Section~\ref{stelprop}. The YSOs with 
        transitional discs are distinguished with crossed squares, while the sub-luminous objects 
        are shown with open circles. The low accretors are shown with red symbols (see Section~\ref{chrom_contrib}). 
        The arrows show upper or lower limits on the mass according  to the availability of the 
        tracks in each model. Average errors are shown in the upper left of each panel.
        The purple dashed lines represent the corresponding linear fits as in Table~\ref{MaccMs_linfits}.
     \label{Macc_Mstar_4models} }
\end{figure*}

\onecolumn

\setlength{\tabcolsep}{4pt}

\begin{landscape}
\scriptsize
\begin{longtable}{l|l|c|c|c|c|c|c|c|c|c|l}
\caption[ ]{\label{pars} Confirmed YSOs of the total sample with spectral types, extinction, and physical parameters.}\\
\hline\hline
Object  &   SpT  & $T_{\rm eff}$ (err) & $A_{\rm V}$  & $d$  &  \Lstar ~(err)  &   \Rstar ~(err)  &  \Mstar (B98) ~(err) &  \Mstar (B15) ~(err) & \Mstar (DM97) ~(err) &  \Mstar (S00)~(err)  & Notes \\
        &        & [K]           & [mag.]       & [pc] &  [\Lsun]   &  [\Rsun]   &  [\Msun ]     & [\Msun ]     & [\Msun ]     & [\Msun ]      &   \\
GTO sample: &        &               &              &      &            &               &                &             &              &               &    \\
\hline                                                                           
\hline                                                                           
Sz66                     &  M3   & 3415 ( 79) &  1.00 & 150 & 0.2000 (0.0920) & 1.29  (0.30) &    0.38 (0.07) &    0.29 (0.05) & 0.25 (0.05)   &    0.31 (0.04) &           \\
AKC2006-19               &  M5   & 3125 ( 72) &  0.00 & 150 & 0.0160 (0.0080) & 0.44  (0.10) &    0.12 (0.03) &    0.14 (0.03) & 0.16 (0.03)   &    0.12 (0.03) &           \\
Sz69                     &  M4.5 & 3197 ( 74) &  0.00 & 150 & 0.0880 (0.0410) & 0.97  (0.22) &    0.22 (0.05) &    0.20 (0.04) & 0.18 (0.03)   &    0.19 (0.03) &           \\
Sz71                     &  M1.5 & 3632 (167) &  0.50 & 150 & 0.3090 (0.1420) & 1.43  (0.33) &    0.64 (0.17) &    0.42 (0.12) & 0.38 (0.12)   &    0.42 (0.11) &           \\
Sz72                     &  M2   & 3560 (164) &  0.75 & 150 & 0.2520 (0.1160) & 1.29  (0.30) &    0.54 (0.16) &    0.37 (0.12) & 0.34 (0.12)   &    0.38 (0.09) &           \\
Sz73                     &  K7   & 4060 (187) &  3.50 & 150 & 0.4190 (0.1930) & 1.35  (0.31) &    1.02 (0.14) &    0.79 (0.15) & 0.62 (0.14)   &    0.82 (0.16) &           \\
Sz74                     &  M3.5 & 3342 ( 77) &  1.50 & 150 & 1.0430 (0.4800) & 3.13  (0.72) &    0.50 (0.12) &    0.30 (0.04) & 0.17 (0.03)   &    0.29 (0.03) &           \\
Sz83                     &  K7   & 4060 (187) &  0.00 & 150 & 1.3130 (0.6050) & 2.39  (0.55) &    1.21 (0.16) &    0.67 (0.16) & 0.43 (0.13)   &    0.75 (0.19) &           \\
Sz84                     &  M5   & 3125 ( 72) &  0.00 & 150 & 0.1220 (0.0560) & 1.21  (0.28) &    0.18 (0.04) &    0.16 (0.03) & 0.16 (0.02)   &    0.18 (0.03) &  td       \\ 
Sz130                    &  M2   & 3560 (164) &  0.00 & 150 & 0.1600 (0.0740) & 1.03  (0.24) &    0.51 (0.15) &    0.41 (0.12) & 0.35 (0.13)   &    0.37 (0.09) &           \\
Sz88A                    &  M0   & 3850 (177) &  0.25 & 200 & 0.4880 (0.2250) & 1.61  (0.37) &    0.94 (0.16) &    0.56 (0.14) & 0.46 (0.13)   &    0.57 (0.15) &           \\
Sz88B                    &  M4.5 & 3197 ( 74) &  0.00 & 200 & 0.1180 (0.0540) & 1.12  (0.26) &    0.22 (0.05) &    0.20 (0.03) & 0.18 (0.02)   &    0.20 (0.03) &           \\
Sz91                     &  M1   & 3705 (171) &  1.20 & 200 & 0.3110 (0.1430) & 1.36  (0.31) &    0.73 (0.17) &    0.47 (0.13) & 0.43 (0.13)   &    0.47 (0.12) &  td       \\ 
Lup713                   &  M5.5 & 3057 ( 70) &  0.00 & 200 & 0.0200 (0.0090) & 0.52  (0.12) &    0.09 (0.02) &    0.11 (0.03) & 0.14 (0.03)   &    0.11 (0.02) &           \\
Lup604s                  &  M5.5 & 3057 ( 70) &  0.00 & 200 & 0.0570 (0.0260) & 0.83  (0.19) &    0.13 (0.03) &    0.14 (0.03) & 0.15 (0.02)   &    0.13 (0.02) &           \\
Sz97                     &  M4   & 3270 ( 75) &  0.00 & 200 & 0.1690 (0.0780) & 1.34  (0.28) &    0.27 (0.05) &    0.23 (0.03) & 0.19 (0.03)   &    0.25 (0.03) &           \\
Sz99                     &  M4   & 3270 ( 75) &  0.00 & 200 & 0.0740 (0.0340) & 0.89  (0.20) &    0.24 (0.05) &    0.23 (0.04) & 0.21 (0.03)   &    0.22 (0.03) &           \\
Sz100                    &  M5.5 & 3057 ( 70) &  0.00 & 200 & 0.1690 (0.0780) & 1.43  (0.33) &    0.17 (0.04) &    0.16 (0.02) & 0.14 (0.02)   &    0.18 (0.03) &  td       \\ 
Sz103                    &  M4   & 3270 ( 75) &  0.70 & 200 & 0.1880 (0.0870) & 1.41  (0.30) &    0.28 (0.06) &    0.22 (0.03) & 0.19 (0.03)   &    0.25 (0.03) &           \\
Sz104                    &  M5   & 3125 ( 72) &  0.00 & 200 & 0.1020 (0.0470) & 1.11  (0.26) &    0.18 (0.04) &    0.16 (0.03) & 0.16 (0.02)   &    0.18 (0.03) &           \\
Lup706                   &  M7.5 & 2795 ( 64) &  0.00 & 200 & 0.0030 (0.0010) & 0.22  (0.05) &    0.05 (0.01) &    0.05 (0.01) & 0.04 (0.01)   &    $<$0.10     &  sl       \\
Sz106                    &  M0.5 & 3777 (174) &  1.00 & 200 & 0.0980 (0.0450) & 0.72  (0.17) &    0.66 (0.11) &    0.62 (0.11) & 0.60 (0.12)   &    0.51 (0.11) &  sl       \\
Par-Lup3-3               &  M4   & 3270 ( 75) &  2.20 & 200 & 0.2400 (0.1100) & 1.59  (0.37) &    0.30 (0.06) &    0.22 (0.03) & 0.18 (0.03)   &    0.26 (0.03) &           \\
Par-Lup3-4               &  M4.5 & 3197 ( 74) &  0.00 & 200 & 0.0030 (0.0010) & 0.17  (0.04) &    0.16 (0.02) &    0.17 (0.02) & 0.15 (0.02)   &    0.16 (0.02) &  sl       \\
Sz110                    &  M4   & 3270 ( 75) &  0.00 & 200 & 0.2760 (0.1270) & 1.61  (0.37) &    0.29 (0.07) &    0.22 (0.03) & 0.18 (0.03)   &    0.26 (0.03) &           \\
Sz111                    &  M1   & 3705 (171) &  0.00 & 200 & 0.3300 (0.1520) & 1.40  (0.32) &    0.74 (0.16) &    0.47 (0.13) & 0.43 (0.13)   &    0.46 (0.12) &  td       \\ 
Sz112                    &  M5   & 3125 ( 72) &  0.00 & 200 & 0.1910 (0.0880) & 1.52  (0.35) &    0.19 (0.05) &    0.17 (0.03) & 0.15 (0.02)   &    0.20 (0.03) &  td       \\ 
Sz113                    &  M4.5 & 3197 ( 74) &  1.00 & 200 & 0.0640 (0.0300) & 0.83  (0.19) &    0.20 (0.05) &    0.20 (0.04) & 0.19 (0.03)   &    0.19 (0.03) &           \\
2MASSJ16085953-3856275   &  M8.5 & 2600 ( 60) &  0.00 & 200 & 0.0090 (0.0040) & 0.47  (0.11) &    0.03 (0.01) &    0.02 (0.01) & 0.02 (0.01)   &    $<$0.10     &           \\
SSTc2d160901.4-392512    &  M4   & 3270 ( 75) &  0.50 & 200 & 0.1480 (0.0680) & 1.25  (0.29) &    0.27 (0.05) &    0.22 (0.04) & 0.20 (0.03)   &    0.24 (0.04) &           \\
Sz114                    &  M4.8 & 3175 ( 73) &  0.30 & 200 & 0.3120 (0.1440) & 1.82  (0.42) &    0.32 (0.08) &    0.21 (0.03) & 0.15 (0.02)   &    0.23 (0.03) &           \\
Sz115                    &  M4.5 & 3197 ( 74) &  0.50 & 200 & 0.1750 (0.0800) & 1.36  (0.31) &    0.23 (0.05) &    0.19 (0.03) & 0.17 (0.02)   &    0.22 (0.03) &           \\
Lup818s                  &  M6   & 2990 ( 67) &  0.00 & 200 & 0.0250 (0.0110) & 0.58  (0.13) &    0.08 (0.02) &    0.09 (0.02) & 0.13 (0.02)   &    0.10 (0.01) &           \\
Sz123A                   &  M1   & 3705 (171) &  1.25 & 200 & 0.2030 (0.0930) & 1.10  (0.25) &    0.68 (0.15) &    0.51 (0.14) & 0.50 (0.14)   &    0.46 (0.12) &  td       \\ 
Sz123B                   &  M2   & 3560 (164) &  0.00 & 200 & 0.0510 (0.0240) & 0.58  (0.13) &    0.49 (0.12) &    0.46 (0.11) & 0.41 (0.10)   &    0.34 (0.09) &  sl       \\
SST-Lup3-1               &  M5   & 3125 ( 72) &  0.00 & 200 & 0.0590 (0.0270) & 0.85  (0.19) &    0.16 (0.04) &    0.17 (0.03) & 0.17 (0.02)   &    0.15 (0.03) &           \\

        &        &               &              &      &            &           &                &             &              &               &    \\
\hline
New sample: &        &               &              &      &            &               &                &             &              &               &    \\
\hline                                                                           
\hline                                                                           
Sz65                     &  K7   & 4060 (187) &  0.60 & 150 & 0.8318 (0.3623) & 1.84  (0.40) &    1.15 (0.15) &    0.70 (0.16) & 0.49 (0.13)   &    0.76 (0.18) &           \\
AKC2006-18               &  M6.5 & 2935 ( 66) &  0.00 & 150 & 0.0107 (0.0048) & 0.40  (0.09) &    0.07 (0.01) &    0.07 (0.01) & 0.08 (0.02)   &    $<$0.10     &           \\
SSTc2dJ154508.9-341734   &  M5.5 & 3060 ( 71) &  5.50 & 150 & 0.0575 (0.0283) & 0.85  (0.21) &    0.13 (0.03) &    0.14 (0.03) & 0.16 (0.02)   &    0.14 (0.02) &           \\
Sz68                     &  K2   & 4900 (226) &  1.00 & 150 & 5.1286 (2.1919) & 3.14  (0.67) &    $>$1.40     &    $>$1.40     & 1.40 (0.42)   &    2.13 (0.33) &          \\
SSTc2dJ154518.5-342125   &  M6.5 & 2935 ( 68) &  0.00 & 150 & 0.0407 (0.0181) & 0.78  (0.17) &    0.08 (0.02) &    0.08 (0.02) & 0.12 (0.02)   &    0.10 (0.01) &           \\
Sz81A                    &  M4.5 & 3200 ( 74) &  0.00 & 150 & 0.2239 (0.1103) & 1.54  (0.38) &    0.23 (0.07) &    0.19 (0.03) & 0.17 (0.03)   &    0.23 (0.03) &           \\
Sz81B                    &  M5.5 & 3060 ( 71) &  0.00 & 150 & 0.1096 (0.0638) & 1.18  (0.34) &    0.13 (0.04) &    0.14 (0.03) & 0.15 (0.02)   &    0.15 (0.03) &           \\
Sz129                    &  K7   & 4060 (187) &  0.90 & 150 & 0.3715 (0.1600) & 1.23  (0.27) &    0.99 (0.14) &    0.79 (0.15) & 0.66 (0.14)   &    0.80 (0.16) &           \\
SSTc2dJ155925.2-423507   &  M5   & 3125 ( 72) &  0.00 & 150 & 0.0195 (0.0092) & 0.48  (0.11) &    0.12 (0.03) &    0.14 (0.03) & 0.17 (0.03)   &    0.12 (0.02) &           \\
RY\,Lup                  &  K2   & 4900 (226) &  0.40 & 150 & 1.6596 (0.7077) & 1.79  (0.38) &    1.40 (0.15) &    $>$1.40     & 1.27 (0.26)   &    1.47 (0.22) &  td       \\ 
SSTc2dJ160000.6-422158   &  M4.5 & 3200 ( 74) &  0.00 & 150 & 0.0871 (0.0415) & 0.96  (0.23) &    0.22 (0.05) &    0.20 (0.04) & 0.19 (0.03)   &    0.19 (0.03) &           \\
SSTc2dJ160002.4-422216   &  M4   & 3270 ( 75) &  1.40 & 150 & 0.1479 (0.0666) & 1.20  (0.27) &    0.27 (0.05) &    0.22 (0.04) & 0.20 (0.03)   &    0.24 (0.04) &           \\
SSTc2dJ160026.1-415356   &  M5.5 & 3060 ( 71) &  0.90 & 150 & 0.0661 (0.0397) & 0.91  (0.27) &    0.14 (0.03) &    0.14 (0.03) & 0.15 (0.02)   &    0.14 (0.03) &           \\
MY\,Lup                  &  K0   & 5100 (235) &  1.30 & 150 & 0.7762 (0.3315) & 1.13  (0.24) &    0.99 (0.10) &    1.06 (0.14) & 1.11 (0.14)   &    1.02 (0.13) &  td       \\ 
Sz131                    &  M3   & 3415 ( 79) &  1.30 & 150 & 0.1318 (0.0583) & 1.04  (0.23) &    0.36 (0.07) &    0.32 (0.05) & 0.26 (0.05)   &    0.30 (0.04) &           \\
Sz133                    &  K5   & 4350 (200) &  1.80 & 150 & 0.0708 (0.0323) & 0.47  (0.11) &                &              &               &                &  sl, bz   \\
SSTc2dJ160703.9-391112   &  M4.5 & 3200 ( 74) &  0.60 & 200 & 0.0048 (0.0026) & 0.23  (0.06) &    0.15 (0.03) &          $<$0.10     & 0.18 (0.03)   &    0.17 (0.03)  &  sl , td?  \\
Sz90                     &  K7   & 4060 (187) &  1.80 & 200 & 0.6607 (0.2845) & 1.64  (0.36) &    1.11 (0.15) &    0.73 (0.16) & 0.52 (0.14)   &    0.79 (0.17) &           \\
Sz95                     &  M3   & 3415 ( 79) &  0.80 & 200 & 0.4169 (0.1842) & 1.84  (0.41) &    0.46 (0.09) &    0.29 (0.04) & 0.23 (0.04)   &    0.33 (0.04) &           \\
Sz96                     &  M1   & 3705 (171) &  0.80 & 200 & 0.6918 (0.3234) & 2.02  (0.47) &    0.80 (0.17) &    0.43 (0.10) & 0.33 (0.10)   &    0.46 (0.11) &           \\
2MASSJ16081497-3857145   &  M5.5 & 3060 ( 71) &  1.50 & 200 & 0.0087 (0.0047) & 0.33  (0.09) &    0.09 (0.02) &    0.10 (0.03) & 0.12 (0.03)   &    0.10 (0.02) &           \\
Sz98                     &  K7   & 4060 (187) &  1.00 & 200 & 2.5119 (1.0755) & 3.20  (0.69) &    1.37 (0.13) &    0.70 (0.16) & 0.38 (0.11)   &    0.74 (0.20) &           \\
Lup607                   &  M6.5 & 2935 ( 66) &  0.00 & 200 & 0.0708 (0.0370) & 1.03  (0.27) &    0.10 (0.02) &    0.10 (0.02) & 0.12 (0.02)   &    0.10 (0.01) &           \\
Sz102                    &  K2   & 4900 (226) &  0.70 & 200 & 0.0148 (0.0064) & 0.17  (0.04) &                &              &               &                &  sl, bz   \\ 
SSTc2dJ160830.7-382827   &  K2   & 4900 (226) &  0.20 & 200 & 3.0200 (1.3082) & 2.41  (0.52) &    $>$1.40     &    $>$1.40     & 1.32 (0.35)   &    1.81 (0.28) &  td       \\ 
SSTc2dJ160836.2-392302/V1094\,Sco  &  K6 & 4205 (193) &  1.70 & 200 & 1.9499 (0.8633) & 2.63  (0.63) &    1.39 (0.13) &         0.79 (0.18) & 0.89 (0.24)   &    0.47 (0.14)      &       td?  \\
Sz108B                   &  M5   & 3125 ( 72) &  1.60 & 200 & 0.1514 (0.0813) & 1.33  (0.36) &    0.16 (0.05) &    0.17 (0.03) & 0.16 (0.02)   &    0.19 (0.03) &           \\
2MASSJ16085324-3914401   &  M3   & 3415 ( 79) &  1.90 & 200 & 0.3020 (0.1477) & 1.57  (0.38) &    0.42 (0.08) &    0.29 (0.04) & 0.24 (0.05)   &    0.32 (0.04) &           \\
2MASSJ16085373-3914367   &  M5.5 & 3060 ( 71) &  4.00 & 200 & 0.0066 (0.0028) & 0.29  (0.06) &    0.09 (0.02) &    0.10 (0.03) & 0.12 (0.03)   &    0.10 (0.02) &           \\
2MASSJ16085529-3848481   &  M6.5 & 2935 (68.) &  0.00 & 200 & 0.0759 (0.0414) & 1.07  (0.29) &    0.11 (0.02) &    0.11 (0.02) & 0.12 (0.02)   &    0.10 (0.02) &           \\
SSTc2dJ160927.0-383628   &  M4.5 & 3200 ( 74) &  2.20 & 200 & 0.1148 (0.0501) & 1.10  (0.24) &    0.23 (0.05) &    0.19 (0.03) & 0.18 (0.02)   &    0.20 (0.03) &           \\
Sz117                    &  M3.5 & 3340 ( 77) &  0.50 & 200 & 0.4467 (0.1927) & 2.00  (0.43) &    0.44 (0.09) &    0.26 (0.03) & 0.19 (0.03)   &    0.29 (0.03) &           \\
Sz118                    &  K5   & 4350 (200) &  1.90 & 200 & 1.0715 (0.4663) & 1.82  (0.40) &    1.39 (0.16) &    1.01 (0.19) & 0.63 (0.16)   &    1.09 (0.20) &           \\
2MASSJ16100133-3906449   &  M6.5 & 2935 ( 68) &  1.70 & 200 & 0.2089 (0.1289) & 1.77  (0.55) &    $<$0.10     &    0.10 (0.03) & 0.12 (0.01)   &    0.14 (0.03) &            \\
SSTc2dJ161018.6-383613   &  M5   & 3125 ( 72) &  0.50 & 200 & 0.0603 (0.0315) & 0.84  (0.22) &    0.17 (0.04) &    0.17 (0.03) & 0.17 (0.02)   &    0.15 (0.03) &           \\
SSTc2dJ161019.8-383607   &  M6.5 & 2935 ( 68) &  0.00 & 200 & 0.0708 (0.0378) & 1.03  (0.27) &    0.10 (0.02) &    0.10 (0.02) & 0.12 (0.02)   &    0.10 (0.02) &           \\
SSTc2dJ161029.6-392215   &  M4.5 & 3200 ( 74) &  0.90 & 200 & 0.1585 (0.0698) & 1.29  (0.29) &    0.23 (0.05) &    0.19 (0.03) & 0.17 (0.02)   &    0.22 (0.03) &  td      \\
SSTc2dJ161243.8-381503   &  M1   & 3705 (171) &  0.80 & 200 & 0.6166 (0.2691) & 1.91  (0.42) &    0.79 (0.17) &    0.44 (0.11) & 0.34 (0.11)   &    0.47 (0.11) &           \\
SSTc2dJ161344.1-373646   &  M5   & 3125 ( 72) &  0.60 & 200 & 0.0692 (0.0305) & 0.90  (0.20) &    0.17 (0.04) &    0.16 (0.03) & 0.17 (0.02)   &    0.16 (0.03) &           \\

        &        &               &              &      &            &           &                &             &              &               &    \\
\hline

Targets from ESO archive: & &               &              &      &            &         &                &             &              &               &    \\
 \hline
 \hline
Sz75/GQ\,Lup             & K6   & 4205 (193) &  0.70 & 150 & 1.4454 (0.6260) & 2.26  (0.53) &           1.40 (0.15) &    0.86 (0.19) & 0.96 (0.23)  &    0.51 (0.14) &            \\
Sz76                     & M4   & 3270 ( 75) &  0.20 & 150 & 0.1585 (0.0704) & 1.24  (0.28) &           0.27 (0.05) &    0.23 (0.04) & 0.19 (0.03)  &    0.25 (0.03) &  td        \\
Sz77                     & K7   & 4060 (187) &  0.00 & 150 & 0.5495 (0.2428) & 1.50  (0.36) &           1.08 (0.15) &    0.75 (0.16) & 0.56 (0.14)  &    0.79 (0.17) &           \\
RXJ1556.1-3655           & M1   & 3705 (171) &  1.00 & 150 & 0.2344 (0.1000) & 1.17  (0.27) &           0.70 (0.16) &    0.50 (0.13) & 0.47 (0.14)  &    0.46 (0.12) &            \\
Sz82/IM\,Lup             & K5   & 4350 (200) &  0.90 & 150 & 2.3300 (1.0397) & 2.69  (0.65) &      1.40 (0.00) &    0.95 (0.00) & 0.55 (0.00)  &    1.10 (0.00) &  td       \\ 
EX\,Lup                  & M0   & 3850 (177) &  1.10 & 200 & 1.2303 (0.5302) & 2.49  (0.58) &           1.04 (0.17) &    0.52 (0.13) & 0.33 (0.11)  &    0.56 (0.14) &            \\
 \hline

\end{longtable}
\footnotesize{Notes: \\
 td : YSO with transitional disc \\
 sl : sub-luminous YSO \\
 bz : sub-luminous object falling below the zero-age main-sequence on the HR diagram 
}
\end{landscape}

\setlength{\tabcolsep}{2pt}
\begin{longtable}{llrrrrrl}
\caption[ ]{\label{accretion} Accretion properties of the total sample.} \\
\hline \hline
        &             &       &  &    &  & \\
Object  &   Template  & $\log$\Lacc & $\log$\Macc (B98) & $\log$\Macc (B15)       & $\log$\Macc (DM98)  & $\log$\Macc (S00)   &  Notes  \\
        &             &  [\Lsun]      & [\Msun yr$^{-1}$] & [\Msun yr$^{-1}$] & [\Msun yr$^{-1}$] & [\Msun yr$^{-1}$]       &    \\
GTO sample:           &              &                   &                  &                   &                        &    \\
\hline
\hline
Sz66                    & SO797      &  $-$1.8  &   $-$8.66  &   $-$8.54  &      $-$8.48 &  $-$8.57  &    \\ 
AKC2006-19              & SO641      &  $-$4.1  &  $-$10.93  &  $-$11.00  &    $-$11.05  & $-$10.93  &    \\ 
Sz69                    & SO797      &  $-$2.8  &   $-$9.55  &   $-$9.51  &      $-$9.46 &  $-$9.48  &    \\ 
Sz71                    & TWA15A     &  $-$2.2  &   $-$9.24  &   $-$9.06  &      $-$9.02 &  $-$9.06  &    \\ 
Sz72                    & TWA9B      &  $-$1.8  &   $-$8.81  &   $-$8.65  &      $-$8.61 &  $-$8.66  &    \\ 
Sz73                    & SO879      &  $-$1.0  &   $-$8.27  &   $-$8.16  &      $-$8.06 &  $-$8.18  &    \\ 
Sz74                    & TWA15A     &  $-$1.5  &   $-$8.10  &   $-$7.87  &      $-$7.63 &  $-$7.86  &    \\ 
Sz83                    & SO879      &  $-$0.3  &   $-$7.40  &   $-$7.14  &      $-$6.95 &  $-$7.19  &    \\ 
Sz84                    & SO641      &  $-$2.7  &   $-$9.27  &   $-$9.21  &      $-$9.21 &  $-$9.27  &    \\ 
Sz130                   & TWA2A      &  $-$2.2  &   $-$9.29  &   $-$9.19  &      $-$9.12 &  $-$9.15  &    \\ 
Sz88A                   & TWA25      &  $-$1.2  &   $-$8.36  &   $-$8.13  &      $-$8.05 &  $-$8.14  &    \\ 
Sz88B                   & SO797      &  $-$3.1  &   $-$9.79  &   $-$9.74  &      $-$9.70 &  $-$9.74  &    \\ 
Sz91                    & TWA13A     &  $-$1.8  &   $-$8.92  &   $-$8.73  &      $-$8.69 &  $-$8.73  &    \\ 
Lup713                  & Par-Lup3-2 &  $-$3.5  &  $-$10.13  &  $-$10.22  &    $-$10.32  & $-$10.22  &    \\ 
Lup604s                 & SO925      &  $-$3.7  &  $-$10.29  &  $-$10.32  &    $-$10.35  & $-$10.29  &    \\ 
Sz97                    & Sz94       &  $-$2.9  &   $-$9.60  &   $-$9.53  &      $-$9.44 &  $-$9.56  &    \\ 
Sz99                    & TWA9B      &  $-$2.6  &   $-$9.42  &   $-$9.41  &      $-$9.37 &  $-$9.39  &    \\ 
Sz100                   & SO641      &  $-$3.0  &   $-$9.47  &   $-$9.44  &      $-$9.38 &  $-$9.49  &    \\ 
Sz103                   & Sz94       &  $-$2.4  &   $-$9.09  &   $-$8.99  &      $-$8.92 &  $-$9.04  &    \\ 
Sz104                   & SO641      &  $-$3.2  &   $-$9.80  &   $-$9.75  &      $-$9.75 &  $-$9.80  &    \\ 
Lup706                  & TWA26      &  $-$4.8  &  $-$11.55  &  $-$11.55  &    $-$11.45  & $-$11.55  &    \\ 
Sz106                   & TWA25      &  $-$2.5  &   $-$9.86  &   $-$9.83  &      $-$9.81 &  $-$9.74  &    \\ 
Par-Lup3-3              & TWA15A     &  $-$2.9  &   $-$9.57  &   $-$9.43  &      $-$9.35 &  $-$9.51  &    \\ 
Par-Lup3-4              & SO641      &  $-$4.1  &  $-$11.47  &  $-$11.49  &    $-$11.44  & $-$11.47  &    \\ 
Sz110                   & Sz94       &  $-$2.0  &   $-$8.65  &   $-$8.53  &      $-$8.44 &  $-$8.60  &    \\ 
Sz111                   & TWA13A     &  $-$2.2  &   $-$9.32  &   $-$9.12  &      $-$9.08 &  $-$9.11  &    \\ 
Sz112                   & SO641      &  $-$3.2  &   $-$9.69  &   $-$9.64  &      $-$9.59 &  $-$9.71  &    \\ 
Sz113                   & SO797      &  $-$2.1  &   $-$8.87  &   $-$8.87  &      $-$8.85 &  $-$8.85  &    \\ 
2MASSJ16085953-3856275  & TWA26      &  $-$4.6  &  $-$10.80  &  $-$10.62  &    $-$10.62  & $-$10.62  &    \\ 
SSTc2d160901.4-392512   & Sz94       &  $-$3.0  &   $-$9.73  &   $-$9.64  &      $-$9.60 &  $-$9.68  &    \\ 
Sz114                   & Sz94       &  $-$2.5  &   $-$9.14  &   $-$8.96  &      $-$8.81 &  $-$8.99  &    \\ 
Sz115                   & SO797      &  $-$2.7  &   $-$9.32  &   $-$9.24  &      $-$9.19 &  $-$9.30  &    \\ 
Lup818s                 & SO925      &  $-$4.1  &  $-$10.63  &  $-$10.68  &    $-$10.84  & $-$10.73  &    \\ 
Sz123A                  & TWA2A      &  $-$1.8  &   $-$8.98  &   $-$8.86  &      $-$8.85 &  $-$8.81  &    \\ 
Sz123B                  & TWA15B     &  $-$2.7  &  $-$10.02  &   $-$9.99  &      $-$9.94 &  $-$9.86  &    \\ 
SST-Lup3-1              & SO641      &  $-$3.6  &  $-$10.27  &  $-$10.29  &    $-$10.29  & $-$10.24  &    \\ 
                        &            &        &          &          &           &         &    \\
New sample: &  &  &  & & & & \\
\hline
\hline
Sz65                    & SO879          &$<$$-$2.6 &$<$$-$9.79  &$<$$-$9.57  &$<$$-$9.42   &$<$$-$9.61  &{\bf a} \\ 
AKC2006-18              & Par-Lup3-1     &  $-$4.6  &  $-$11.24  &  $-$11.24  &  $-$11.29   & $-$11.24  &    \\ 
SSTc2dJ154508.9-341734  & Sz107          &  $-$1.8  &   $-$8.38  &   $-$8.41  &   $-$8.47   &  $-$8.41  &    \\ 
Sz68                    & RXJ0438        &$<$$-$1.2 &$<$$-$8.24  &$<$$-$8.24  & $<$$-$8.24   &$<$$-$8.42 &{\bf a} \\ 
SSTc2dJ154518.5-342125  & Par-Lup3-1     &  $-$4.3  &  $-$10.70  &  $-$10.70  &  $-$10.88   & $-$10.80  &    \\ 
Sz81A                   & SO797          &  $-$2.5  &   $-$9.07  &   $-$8.98  &   $-$8.94   &  $-$9.07  &    \\ 
Sz81B                   & SO925          &  $-$3.2  &   $-$9.64  &   $-$9.67  &   $-$9.70   &  $-$9.70  &    \\ 
Sz129                   & TWA6           &  $-$1.2  &   $-$8.50  &   $-$8.40  &   $-$8.32   &  $-$8.41  &    \\ 
SSTc2dJ155925.2-423507  & SO641          &  $-$4.4  &  $-$11.19  &  $-$11.26  &  $-$11.34   & $-$11.19  &    \\ 
RY\,Lup                 & RXJ0438        &  $-$0.9  &   $-$8.19  &   $-$8.19  &   $-$8.14   &  $-$8.21  &    \\ 
SSTc2dJ160000.6-422158  & SO797          &  $-$3.1  &   $-$9.85  &   $-$9.81  &   $-$9.79   &  $-$9.79  &    \\ 
SSTc2dJ160002.4-422216  & Sz94           &  $-$3.0  &   $-$9.75  &   $-$9.66  &   $-$9.61   &  $-$9.69  &    \\ 
SSTc2dJ160026.1-415356  & SO925          &  $-$3.3  &   $-$9.88  &   $-$9.88  &   $-$9.91   &  $-$9.88  &    \\ 
MY\,Lup                 & HBC407         &$<$$-$2.3 & $<$$-$9.64  &$<$$-$9.67  &$<$$-$9.69   &$<$$-$9.65 &{\bf a} \\ 
Sz131                   & CD 36-7429B    &  $-$2.4  &   $-$9.33  &   $-$9.28  &   $-$9.19   &  $-$9.25  &    \\ 
Sz133                   & CD 36-7429A    &  $-$1.8  &            &            &             &          & {\bf c}   \\ 
SSTc2dJ160703.9-391112  & SO797          &  $-$5.2  &  $-$12.41  &  $-$12.38  &  $-$12.46   & $-$12.49  &    \\
Sz90                    & TWA6           &  $-$1.6  &   $-$8.82  &   $-$8.64  &   $-$8.49   &  $-$8.68  &    \\ 
Sz95                    & CD 36-7429B    &  $-$2.5  &   $-$9.29  &   $-$9.09  &   $-$8.99   &  $-$9.15  &    \\ 
Sz96                    & RXJ1121.3-3447 &  $-$2.3  &   $-$9.29  &   $-$9.02  &   $-$8.91   &  $-$9.05  &    \\ 
2MASSJ16081497-3857145  & SO925          &  $-$3.4  &  $-$10.23  &  $-$10.27  &  $-$10.35   & $-$10.27  &    \\ 
Sz98                    & SO879          &  $-$0.5  &   $-$7.52  &   $-$7.23  &   $-$6.97   &  $-$7.26  &    \\ 
Lup607                  & Par-Lup3-1     &$<$$-$4.9 &$<$$-$11.28 &$<$$-$11.28 &$<$$-$11.36   &$<$$-$11.28&{\bf a} \\ 
Sz102                   & CrA75          &  $-$2.0  &            &            &             &          & {\bf c}  \\ 
SSTc2dJ160830.7-382827  & RXJ0438        &$<$$-$1.8 & $<$$-$8.96 & $<$$-$8.96  & $<$$-$8.93  &$<$$-$9.07 &{\bf a} \\ 
SSTc2dJ160836.2-392302/V1094\,Sco & RXJ1543.1-3920 &  $-$0.8  &   $-$7.92  &   $-$7.67  &   $-$7.45   &  $-$7.72  &    \\
Sz108B                  & SO641          &  $-$2.9  &   $-$9.37  &   $-$9.40  &   $-$9.37   &  $-$9.45  &    \\ 
2MASSJ16085324-3914401  & TWA15          &  $-$3.1  &   $-$9.92  &   $-$9.76  &   $-$9.68   &  $-$9.80  &    \\ 
2MASSJ16085373-3914367  &   ---          &  $-$3.7  &  $-$10.58  &  $-$10.63  &  $-$10.71   & $-$10.63  &{\bf b} \\ 
2MASSJ16085529-3848481  & Par-Lup3-1     &  $-$4.1  &  $-$10.51  &  $-$10.51  &  $-$10.54   & $-$10.46  &    \\ 
SSTc2dJ160927.0-383628  & SO797          &  $-$1.3  &   $-$8.01  &   $-$7.93  &   $-$7.91   &  $-$7.95  &    \\ 
Sz117                   & TWA15          &  $-$2.1  &   $-$8.84  &   $-$8.61  &   $-$8.47   &  $-$8.65  &    \\ 
Sz118                   & CD 36-7429A    &  $-$1.8  &   $-$9.08  &   $-$8.94  &   $-$8.73   &  $-$8.97  &    \\ 
2MASSJ16100133-3906449  & Par-Lup3-1     &  $-$3.4  &   $-$9.55  &   $-$9.55  &   $-$9.62   &  $-$9.69  &    \\ 
SSTc2dJ161018.6-383613  & Par-Lup3-2     &  $-$3.8  &  $-$10.50  &  $-$10.50  &  $-$10.50   & $-$10.44  &    \\ 
SSTc2dJ161019.8-383607  & Par-Lup3-1     &  $-$3.9  &  $-$10.28  &  $-$10.28  &  $-$10.36   & $-$10.28  &    \\ 
SSTc2dJ161029.6-392215  & SO797          &  $-$3.2  &   $-$9.84  &   $-$9.76  &   $-$9.71   &  $-$9.82  &    \\ 
SSTc2dJ161243.8-381503  & RXJ1121.3-3447 &  $-$2.0  &   $-$9.01  &   $-$8.76  &   $-$8.64   &  $-$8.78  &    \\ 
SSTc2dJ161344.1-373646  & Par-Lup3-2     &  $-$2.3  &   $-$8.97  &   $-$8.94  &   $-$8.97   &  $-$8.94  &    \\ 
 &  &  &  &  &  &  & \\
Targets from ESO archive:&               &        &          &          &           &         &    \\
 \hline
 \hline
Sz75/GQ\,Lup            & RXJ1540.7-3756 &  $-$0.7  &   $-$7.89  &  $-$7.67  &  $-$7.45  &  $-$7.72  &    \\
Sz76                    & Tyc7760283\_1  &  $-$2.6  &   $-$9.33  &  $-$9.26  &  $-$9.18  &  $-$9.30  &    \\
Sz77                    & Sz94           &  $-$1.7  &   $-$8.95  &  $-$8.79  &  $-$8.67  &  $-$8.81  &    \\
RXJ1556.1-3655          & SO879          &  $-$0.9  &   $-$8.07  &  $-$7.92  &  $-$7.90  &  $-$7.89  &    \\
Sz82/IM\,Lup            & CD\_36\_7429A  &  $-$1.1  &   $-$8.21  &  $-$8.04  &  $-$7.80  &  $-$8.10  &    \\
EX\,Lup                 & SO879          &  $-$0.7  &   $-$7.71  &  $-$7.41  &  $-$7.22  &  $-$7.44  &    \\
\hline
& & & &  &  &  & \\


\end{longtable}
\tablefoot{\\
{\bf a}: considered as weak accretor because \Lacc ~is comparable to the chromospheric level (see Section~\ref{chrom_contrib}).\\
{\bf b}: \Lacc ~calculated from the luminosity of 7 permitted emission lines, using the \Lacc--\Ll ~relationships 
revisited in Appendix~\ref{correlations}.\\
{\bf c}: sub-luminous object falling below the ZAMS.
}

\twocolumn

\section{Accretion luminosity versus line luminosity relationships revisited} 
\label{correlations}

In this paper we more than double the number of YSOs in Lupus with accurately and homogeneously determined 
values of accretion luminosity and line luminosity. The latter was measured for large number of permitted 
emission lines simultaneously observed in a wide spectral range from the UVB to the NIR. It is then worth 
to revisit the relationships between the continuum excess emission and the emission in the individual 
permitted lines that we have derived in A14, on a more statistically significant basis by using the total 
sample. 
 
Figures~\ref{correl1} to \ref{correl6} (on-line material only) show the relationships 
between \Lacc ~~and the luminosity of the permitted emission lines discussed 
in A14 for the GTO sample. Overplotted in these figures are the data corresponding to the
new sample. The line luminosities were calculated as explained in Section~\ref{emisslines}. 
For the reasons discussed in that Section, the He\,{\sc i} $\lambda$1082.9nm line 
is not included in our analysis here. In order to avoid confusion we do not include the 
data from the literature on the plots.

 A comparison of the results shown in Table~4 of A14 and those in Table~\ref{linfits} 
shows that the linear fits of the GTO sample and those presented here for the total sample
are in very good agreement for all the lines.
The linear fits of the  $\log{}$\Lacc ~vs. $\log{}$\Ll ~ relationships were
then recalculated using the package ASURV \citep[][]{feigelson85}, which includes 
censoring of upper or lower limits in the fits.
The results of the new fits (cf. Table~\ref{linfits}) including and excluding 
upper limits are consistent within the errors, but given the good number statistics
the fits were done with detections only. The total number of points and the standard 
deviation of the fits are given in the fifth and sixth columns of Table~\ref{linfits}, 
respectively. For the reasons discussed in A14 no fits were calculated for 
the Br\,8 (\brd) relation. 

The new relationships are very similar to those in A14 with the only difference that
the errors on the parameters of the linear fits are reduced by about 30\%, although the
standard deviation from the fits has generally increased by about 15\% on the average. 
The latter is a natural consequence of the larger number of points included here with respect
to A14. The recommended relationships to calculate \Lacc ~ from \Ll ~are indicated in 
the notes of Table~\ref{linfits}. All the conclusions regarding the physical 
interpretations on these relationships given in A14 are  confirmed here with 
the total sample.

\citet{mendigutia15} suggest that all the \Lacc-\Ll ~relationships are a direct consequence 
of the \Lacc--\Lstar ~correlation and not necessarily related with the physical origin of 
the lines. Whatever the case, these relationships are a useful tool to derive estimates
of the accretion luminosity hence, accretion rate.  
The relations computed here have in general a lower dispersion than those found in the 
literature by applying similar methodologies of fitting the UV excess emission, and in 
general continuum excess emission ~\citep[e.g.][]{HH08, rigliaco12, ingleby13}. However, 
one must keep in mind that each point in the relationships represents an instantaneous 
snapshot of \Lacc ~and \Ll. Note that the results of temporal monitoring of several YSOs 
indicate variability in optically thick line fluxes, without significant changes in the 
corresponding continuum accretion rate \citep[e.g.][]{gahm08, herczeg09}, so that some 
dispersion may still arise from variability even when the observations are simultaneous. 

\setlength{\tabcolsep}{6pt}
\begin{table*}
\begin{center}
\caption[ ]{\label{linfits} Revisited \Lacc -- \Ll ~linear fits$^\dag$. } 
\begin{tabular}{l|r|c|c|l|r|l}
\hline \hline

Diagnostic   &   $\lambda$ & $a$ ($\pm$err) & $b$ ($\pm$err) & N$_{\rm points}^\ddag$ & $\sigma ^{\star}$   & Comments \\
             &     [nm]   &               &               &  {\small GTO + New } &                    &   \\
\hline     
        &            &               &               &                  &                   &  \\

H3 (\Ha)  & 656.2800 &  1.13 (0.05)  &  1.74 (0.19)  & 36 + 6 + 45 &  0.41   &  \\
H4 (\Hb)  & 486.1325 &  1.14 (0.04)  &  2.59 (0.16)  & 36 + 6 + 42 &  0.30   & ~~$\bullet$ \\
H5 (\Hg)  & 434.0464 &  1.11 (0.03)  &  2.69 (0.17)  & 36 + 6 + 41 &  0.29   & ~~$\bullet$ \\
H6 (\Hd)  & 410.1734 &  1.07 (0.04)  &  2.64 (0.18)  & 36 ~~~~~ + 41 &  0.32   & ~~$\bullet$ \\
H7 (\Hep) & 397.0072 &  1.06 (0.04)  &  2.69 (0.18)  & 36 ~~~~~ + 41     &  0.32   &  {\bf 1 } \\
H8        & 388.9049 &  1.06 (0.04)  &  2.73 (0.18)  & 36 ~~~~~ + 39     &  0.30   & ~~$\bullet$ \\
H9        & 383.5384 &  1.04 (0.04)  &  2.78 (0.19)  & 36 ~~~~~ + 38     &  0.31   & ~~$\bullet$ \\
H10       & 379.7898 &  1.04 (0.04)  &  2.83 (0.19)  & 35 ~~~~~ + 36     &  0.30   & ~~$\bullet$ \\
H11       & 377.0630 &  1.06 (0.03)  &  3.02 (0.18)  & 35 + 6 + 34 &  0.28   & ~~$\bullet$ \\
H12       & 375.0151 &  1.04 (0.03)  &  3.07 (0.18)  & 35 ~~~~~ + 34     &  0.28   & ~~$\bullet$ \\
H13       & 373.4368 &  1.03 (0.04)  &  3.13 (0.20)  & 34 ~~~~~ + 32     &  0.28   & ~~$\bullet$ \\
H14       & 372.1938 &  1.03 (0.04)  &  3.25 (0.21)  & 31 ~~~~~ + 29     &  0.28   & ~~$\bullet$ \\
H15       & 371.1977 &  1.05 (0.04)  &  3.43 (0.23)  & 31 ~~~~~ + 28     &  0.29   & ~~$\bullet$ \\

\hline     
         &          &               &              &                &        &   \\
Pa5  (\pab) & 1281.8070  &   1.06 (0.07)  &   2.76 (0.34)  &  29 + 6 + 26 &  0.45  &  ~~$\bullet$ \\ 
Pa6  (\pag) & 1093.8086  &   1.24 (0.06)  &   3.58 (0.27)  &  33 + 6 + 29 &  0.36  &  ~~$\bullet$ \\
Pa7  (\pad) & 1004.9368  &   1.22 (0.09)  &   3.74 (0.43)  &  25 ~~~~~ + 19 &  0.40  &  ~~$\bullet$ \\
Pa8         &  954.5969  &   1.09 (0.12)  &   3.19 (0.59)  &  17 ~~~~~ +  6 &  0.42  &   \\
Pa9         &  922.9014  &   1.18 (0.08)  &   3.71 (0.43)  &  27 ~~~~~ + 25 &  0.44  &    \\
Pa10        &  901.4909  &   1.15 (0.10)  &   3.60 (0.52)  &  26 ~~~~~ + 28 &  0.53  &    \\

\hline
        &            &               &               &                  &          &  \\

Br7  (\brg) & 2166.1210  &  1.19 (0.10) &  4.02 (0.51) &  19 ~~~~~ +  17    &    0.45   &  ~~$\bullet$ \\

\hline
        &            &               &               &                  &          &  \\

He\,{\sc i}            &  402.6191  &   1.05 (0.04)  &  3.66 (0.22)  &  31 ~~~~~ + 28 &  0.26  & ~~$\bullet$ \\
He\,{\sc i}            &  447.1480  &   1.06 (0.04)  &  3.52 (0.22)  &  33 ~~~~~ + 33 &  0.29  & ~~$\bullet$ \\
He\,{\sc i}            &  471.3146  &   0.84 (0.08)  &  2.89 (0.46)  &  16 ~~~~~ + 17 &  0.38  &         \\
He\,{\sc i}Fe\,{\sc i} &  492.1931  &   0.97 (0.04)  &  3.08 (0.24)  &  32 ~~~~~ + 26 &  0.30  & {\bf 2}   \\
He\,{\sc i}            &  501.5678  &   0.99 (0.04)  &  3.49 (0.24)  &  30 ~~~~~ + 22 &  0.27  & ~~$\bullet$ \\
He\,{\sc i}            &  587.5621  &   1.15 (0.04)  &  3.67 (0.21)  &  36 + 6 + 40 &  0.31   & ~~$\bullet$ \\
He\,{\sc i}            &  667.8151  &   1.25 (0.06)  &  4.70 (0.33)  &  36 ~~~~~ + 28 &  0.36   & ~~$\bullet$ \\
He\,{\sc i}            &  706.5190  &   1.18 (0.05)  &  4.47 (0.29)  &  36 ~~~~~ + 26 &  0.34   & ~~$\bullet$ \\
He\,{\sc ii}           &  468.5804  &   1.04 (0.05)  &  3.85 (0.33)  &  28 ~~~~~ + 27 &  0.35   &        \\

\hline
        &            &               &               &                  &        &  \\

Ca\,{\sc ii} (K)     &  393.3660  &  1.03 (0.04)  &  2.50 (0.18)  &  36 ~~~~~ + 45 &  0.33  & ~~$\bullet$ \\
Ca\,{\sc ii} (H)     &  396.8470  &  1.06 (0.03)  &  2.65 (0.16)  &  36 ~~~~~ + 45 &  0.28  & {\bf 3} \\
Ca\,{\sc ii}         &  849.8020  &  0.99 (0.05)  &  2.60 (0.29)  &  34 ~~~~~ + 41 &  0.47  & \\
Ca\,{\sc ii}         &  854.2090  &  0.97 (0.06)  &  2.43 (0.29)  &  32 ~~~~~ + 43 &  0.48  & \\
Ca\,{\sc ii}         &  866.2140  &  0.93 (0.06)  &  2.30 (0.30)  &  29 ~~~~~ + 42 &  0.49  & \\

\hline
        &            &               &               &                       &   &    \\
Na\,{\sc i} &  588.995   & 1.01 (0.06)  & 3.14 (0.36)  & 36  ~~~~~ + 18  &  0.44       &    \\
Na\,{\sc i} &  589.592   & 1.01 (0.06)  & 3.33 (0.38)  & 36  ~~~~~ + 19  &  0.49       &    \\
\hline
        &            &               &               &                  &        &    \\
O\,{\sc i}  &  777.3055  & 1.27 (0.09)  & 4.66 (0.49)  & 14 ~~~~~ + 15  &  0.45  & {\bf 4}    \\
O\,{\sc i}  &  844.6360  & 1.08 (0.12)  & 3.46 (0.62)  & 18 ~~~~~ + 16  &  0.60  &              \\

\hline
\end{tabular}
\tablefoot{
\\
~~~ $\dag$: as in A14 the relations  are of the form ~~~
 $\log{(L_{\rm acc}/L_{\odot})} = a\cdot\log{(L_{\rm line}/L_{\odot})} ~+~ b $.\\
\vspace{0.1cm}
{\large $\ddag$}: number of points for the fit over the total sample. The fits in which the six YSOs in $\sigma$-Ori 
  \citep{rigliaco12} were included in A14 are indicated with "+6".  
\vspace{0.0cm}
$\star$ : standard deviation from linear fit\\
Comments in last column: 
$\bullet$: Suggested relations for deriving \Lacc ~from the line luminosity.
{\bf (1)}~partially blended with  Ca\,{\sc ii}~H ;
{\bf (2)}~He\,{\sc i} + Fe\,{\sc i} blend ;
{\bf (3)}~partially blended with  \Hep ;
{\bf (4)}~O\,{\sc i} $\lambda \lambda$ 777.194, 777.417nm doublet. ~~
}
\end{center}
\end{table*}


\begin{figure}[h]
\resizebox{1.0\hsize}{!}{\includegraphics[]{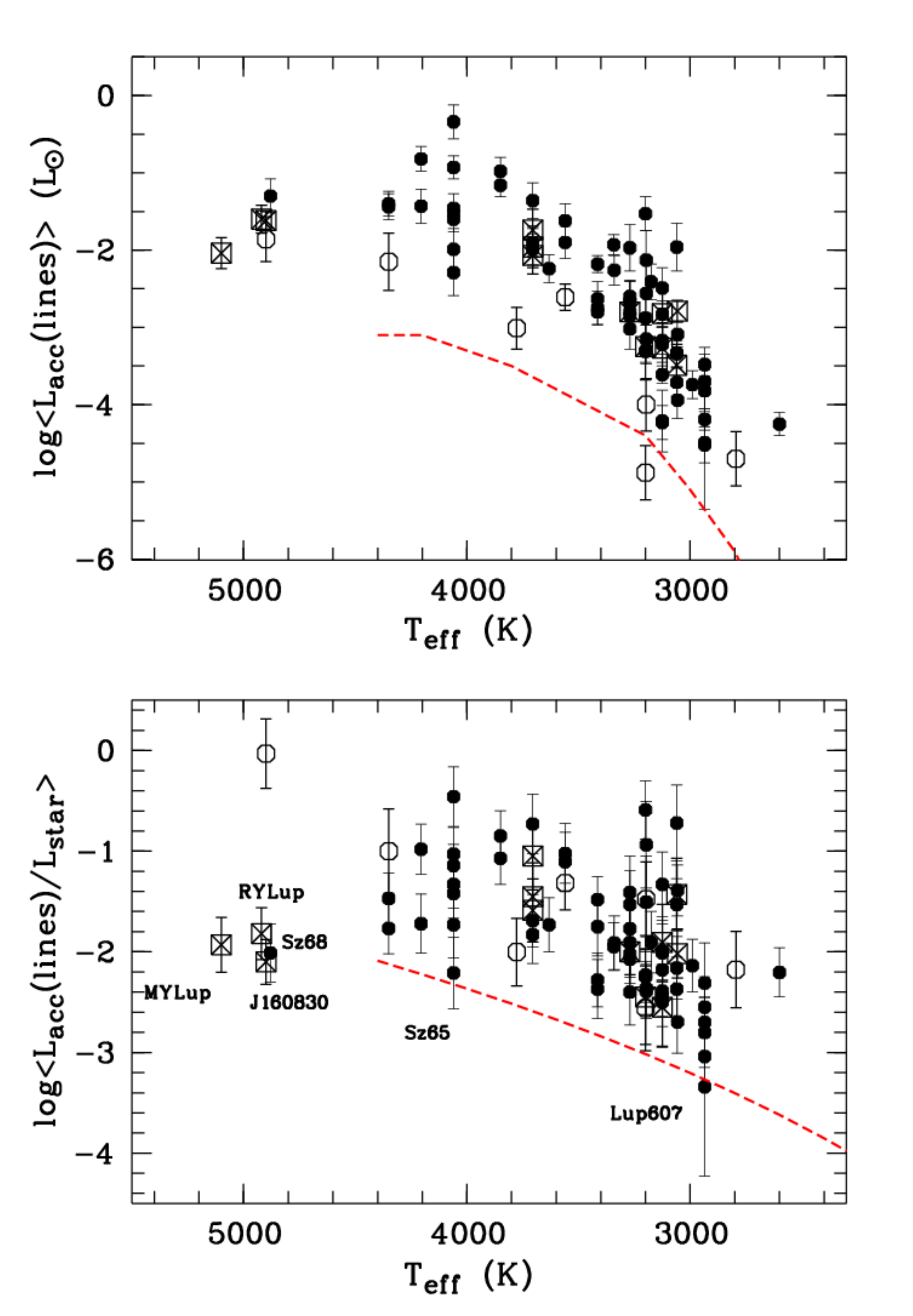}}
\caption{Average accretion luminosity  $\langle$\Lacc(lines)$\rangle$ derived 
       from emission lines  as described in the text (upper panel) 
       and the $\langle$\Lacc(lines)/\Lstar $\rangle$ ratio (lower panel) 
       in logarithmic scale as a function of effective temperature for
       the total sample. 
       The objects with transitional discs are distinguished with crossed 
       squares, while the sub-luminous objects are shown with open circles. 
       The dashed lines in both panels mark the locus below which chromospheric 
       emission is important in comparison with \Lacc. The vertical error bars 
       represent the standard deviation over the average. The weak accretors
       and RY\,Lup are labelled.
    \label{Lacc_Teff}}
\end{figure}

\subsection{Accretion versus chromospheric emission}
\label{chrom_contrib}

An important aspect to be considered when determining the accretion and line luminosity
is the contribution of chromospheric emission. The relative importance of (hydrogen) line 
emission with respect to \Lacc ~is higher for low \Lacc ~values, and chromospheric emission 
may be the dominant process in the lines \citep[][]{ingleby11,rigliaco12,manara13, frasca15}. 
Based on the luminosity of several chromospheric emission lines in the class~III templates, 
\citet{manara13} determined a threshold below which chromospheric emission dominates line 
luminosities.  The threshold depends on YSO effective temperature and age. 

To investigate the possible effects of chromospheric line emission in the new sample, 
we have compared the threshold derived by \citet{manara13} with the accretion luminosity, 
\Lacc(lines), derived by using emission line diagnostics and the revisited
\Lacc--\Ll ~relations.
In Figure~\ref{Lacc_Teff} the $\langle$\Lacc(lines)$\rangle$ values and the 
$\langle$\Lacc(lines)/\Lstar $\rangle$ ratio \citep[as suggested in][]{manara13, mendigutia15} 
are plotted in logarithmic scale as a function of \Teff. 
The dashed lines in the figure show the level of chromospheric noise as determined 
by \citet{manara13}. The lines represent the locus below which the contribution of 
chromospheric emission starts to be important in comparison with energy losses due 
to accretion. 

Except for a sub-luminous object, the accretion level of all the YSOs shown in Figure~\ref{Lacc_Teff} 
is above the chromospheric noise in the \Lacc ~vs. \Teff ~ diagram, but some of the new sample are 
scattered towards lower \Lacc ~values than those of the GTO. When normalising to the stellar luminosity, 
six objects namely RY~Lup, MY~Lup, Sz65, Sz68, SST\,c2dJ160830.7-382827, and the M6.5 type 
star Lup607, fall on the locus of chromospheric noise in the \Lacc/\Lstar ~ vs. \Teff ~diagram. 
Note that these are the objects for which Balmer continuum emission is not evident after the slab 
modelling analysis of Section~\ref{accretiondiagnostics}. Of these, RY~Lup, MY~Lup, and 
SST\,c2dJ160830.7-382827 have transitional discs, whereas the classification of Lup\,607 as 
a class~II YSO is dubious because based on an uncertain SED \citep{merin08}. 
On the other hand, the results of the ALMA survey of Lupus  by \citet{ansdell16} show that their 
resolved transition discs have much higher disc gas masses than the disc gas mass in Sz65 and Sz68. 
Therefore, except for RY~Lup, where the \pab ~and \brg ~lines are clearly detected in emission, 
the other five objects are considered as weak (or dubious) accretors. These objects are flagged in 
Table~\ref{accretion} and are distinguished in the plots.

\section{The flat source SSTc2d\,J160708.6-391408}
\label{flat_source}

This source is interesting because it is probably one of the brightest and least extincted 
\citep[A$_{\rm V}=3.0\pm0.5$\,mag as determined by][]{muzic14} YSOs with a flat SED \citep{merin08, evans09}. 
This makes possible the acquisition of a X-Shooter spectrum with sufficient S/N 
in the three spectrograph arms allowing us to perform the same analysis as for 
the class~II sources.
We have thus observed SSTc2d\,J160708.6-391408 in 2016-06-05 following the 
same observational strategy as for the class~II sources. The data reduction procedures, 
as well as the analysis to determine the stellar and accretion properties was the same 
as for the class~II sources. 

The X-Shooter spectrum of SSTc2d\,J160708.6-391408 is very rich in permitted and forbidden 
emission lines, and shows a strong continuum UV-excess emission (see Figure~\ref{spec_flat}).
The extinction corrected flux and equivalent width of permitted lines are reported in 
the Tables~\ref{tab:fluxes_EWs_Hae} to \ref{tab:fluxes_EWs_NaI}.
We classify the star as M5, but noticed that the results of the slab modelling are also
consistent with a M3 type. The reason for this uncertainty is related to the fact that this 
object is strongly accreting, and thus strongly veiled, while having a relatively high extinction 
due to a still partially optically thick envelope. This represents an extreme case which can 
hardly be reproduced by a model including only the photospheric and the accretion emission.
Assuming a M5 type, a distance of 200\,pc and the \citet{siess00} tracks, we derived the 
following stellar and accretion properties:
T$_{\rm eff}=3125\pm72$\,K ; A$_{\rm V}=3.60$\,mag ;  \Lstar$=0.0107\pm0.0061$\Lsun ~; 
\Mstar$=0.13\pm0.03$\,\Msun ~, \Lacc$=6.31 \times 10^{-3}$\,\Lsun ~and \Macc$=5.89 \times 10^{-10}$\,\Msun ~yr$^{-1}$.
Adopting M3, yields basically the same results on the accretion properties, but increases 
\Lstar ~and \Mstar ~by a factor of about 2. Our results place SSTc2d\,J160708.6-391408 in 
a rather anomalous position on the $\log$\Lacc-$\log$\Lstar ~plot with respect to other YSOs, 
with a quite high \Lacc/\Lstar ~ratio of 0.6, but \Lstar ~may be underestimated (see below).


\begin{figure}[h]
\resizebox{1.03\hsize}{!}{\includegraphics[]{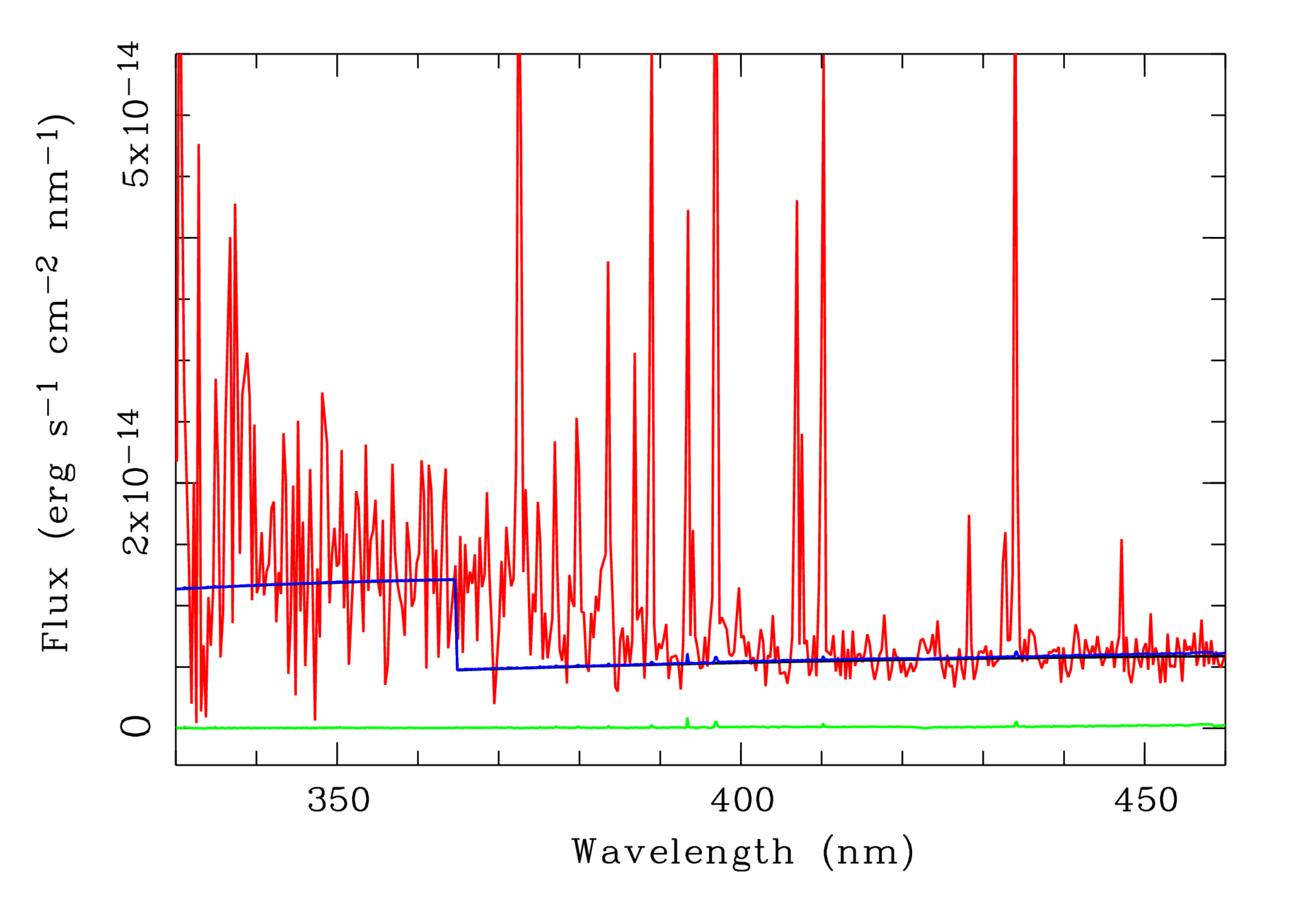}}
\caption{Extinction-corrected X-Shooter spectrum of the flat source SSTc2d\,J160708.6-391408 (red).
        The continuum is fitted with a combination of a photospheric template (green) and the synthetic 
        continuum spectrum from a hydrogen slab. The total fit is represented with the blue line.
     \label{spec_flat}}
\end{figure}

Based on their VIMOS@VLT data, \citet{muzic14} classify the star as  M1.75, in agreement with 
the result by \citet{frasca16} after applying the ROTFIT code, which classifies the object as M2.
Although these results would be more in line with our M3 estimate, we worn that it is not
straightforward to calculate and include veiling in these analyses.
Determinations of \Lstar ~by these and other authors \citep[e.g.][]{comeron09} are similar to 
our result. All these estimates make the object rather under-luminous with respect to other 
YSOs of similar spectral type. 

Flat sources may be interpreted as YSOs with infalling envelopes \citep{calvet94} hence, 
SSTc2d\,J160708.6-391408 may still be on a stage of accretion from an envelope of gas and dust  
in which part of the stellar radiation is reprocessed. 
The above calculations of \Lstar ~do not account for these effects hence, may underestimate 
the luminosity of the YSO. The bolometric luminosity of 0.18\Lsun, as derived by \citet{evans09} 
for this source, would imply a \Lacc/$L_{bol}$ ~ratio of 0.04, i.e. quite consistent with the 
value for YSOs of similar mass. 

Further results on the analysis of the X-Shooter spectrum of this object will be presented in 
the papers by \citet{frasca16} for the stellar parameters and by \citet{nisini16} for the analysis 
of forbidden emission lines.

\section{EX\,Lup}
\label{EXLup}

The X-Shooter spectrum of EX\,Lup is very rich in emission lines and displays strong UVB 
continuum emission (See Figure~\ref{slab5}). The spectrum shows narrower emission lines in 
comparison with other spectra of the same object acquired during burst  \citep[e.g.][]{sicilia-aguilar15}. 
Thus, it is most likely that the object was not in burst during the X-Shooter observation. 

Our \Macc ~determination of 3.6$\times$10$^{-8}$\,\Msun/yr is much higher than the one estimated 
by \citet[][]{sicilia-aguilar15}: these authors used the same X-Shooter data as us, but they
analysed the reduced 1-D spectra gathered from the ESO Phase-3 data release and to our knowledge 
did not correct for slit losses, despite the narrow slits used during the observations. 
Two main reasons may explain the large discrepancy. First, \citet[][]{sicilia-aguilar15} adopt
A$_{\rm V}=0$\,mag, while our best fit to the spectrum yields A$_{\rm V}=1.1$\,mag. Second,
we have applied a factor of 2.5 to correct for slit losses, and since the object is quite variable, 
we may overestimate the absolute flux of the spectrum. In order to investigate this, we have used 
the AAVSO database to check for photometric observations closest in time to the date of the 
X-Shooter acquisition. We found that the $V$ magnitude of EX\,Lup was 13.7\,mag and 
13.4\,mag, in JD\,2455307.90278  and JD\,2455331.98958, respectively, i.e. 13.5\,mag when 
interpolating to the observing date May 4, 2010 (or JD\,2455320.164155). This can be converted
into a fluxe of 1.6$\times$10$^{-13}$\,erg~s$^{-1}$~cm$^{-2}$~nm$^{-1}$, which is in agreement whithin 
less than a factor 1.5 with the flux of the X-Shooter spectrum, after our correction for slit losses. 
Likewise, calculating the "synthetic" V magnitude from the spectrum with the Johnson $V$ passband
we derived a $V=13.03$\,mag.
 
It is worth noting that the $\log$\Macc$\approx -$9.4 ~estimate by \citet[][]{sicilia-aguilar15} 
would imply a  $\log$(\Lacc/\Lstar)$=-2.3$, and given the \Teff$=3850$\,K, would place EX\,Lup  
very close to the chromospheric noise level (see Section~\ref{chrom_contrib}) and among the 
lowest accretors in Lupus, with a position on the $\log$\Macc-$\log$\Mstar ~diagram comparable 
to the one of our weakest (or dubious) accretors.
All this is at odds with the strong Balmer continuum emission detected in the X-Shooter spectrum 
(see Figure~\ref{slab5}) an with the substantial veling of 0.43 and 0.36 that we measured at 580\,nm 
and 670\,nm, respectively. In addition, the $\log$\Lacc(lines)$=-0.98\pm$0.18 we calculated 
from the luminosity of 34 emission lines and the \Lacc-\Ll ~relationships in A14, is in good agreement
with the $\log$\Lacc(slab)=$-0.70\pm$0.25 derived from our slab modelling. 
Assuming that our extinction estimate is wrong and fixing A$_{\rm V}=0$\,mag would drop our 
$\log$\Macc ~estimate only by about 0.5\,dex. 

Therefore, we think that the flux of the emission lines, measured by \citet[][]{sicilia-aguilar15}
in the ESO Phase-3 1D spectrum to calculate \Lacc, was underestimated.

It has been shown that EX\,Lup may reach \Macc ~values as high as 10$^{-7}$\,\Msun/yr during busrts
\citep[][and references there in]{sicilia-aguilar15}. Our \Macc ~estimate is thus consistent with 
the fact that the object was not in a burst stage during the X-Shooter acquisition. 


\section{On-line material} 

\onecolumn
\begin{figure} 
\resizebox{0.9\hsize}{!}{\includegraphics[]{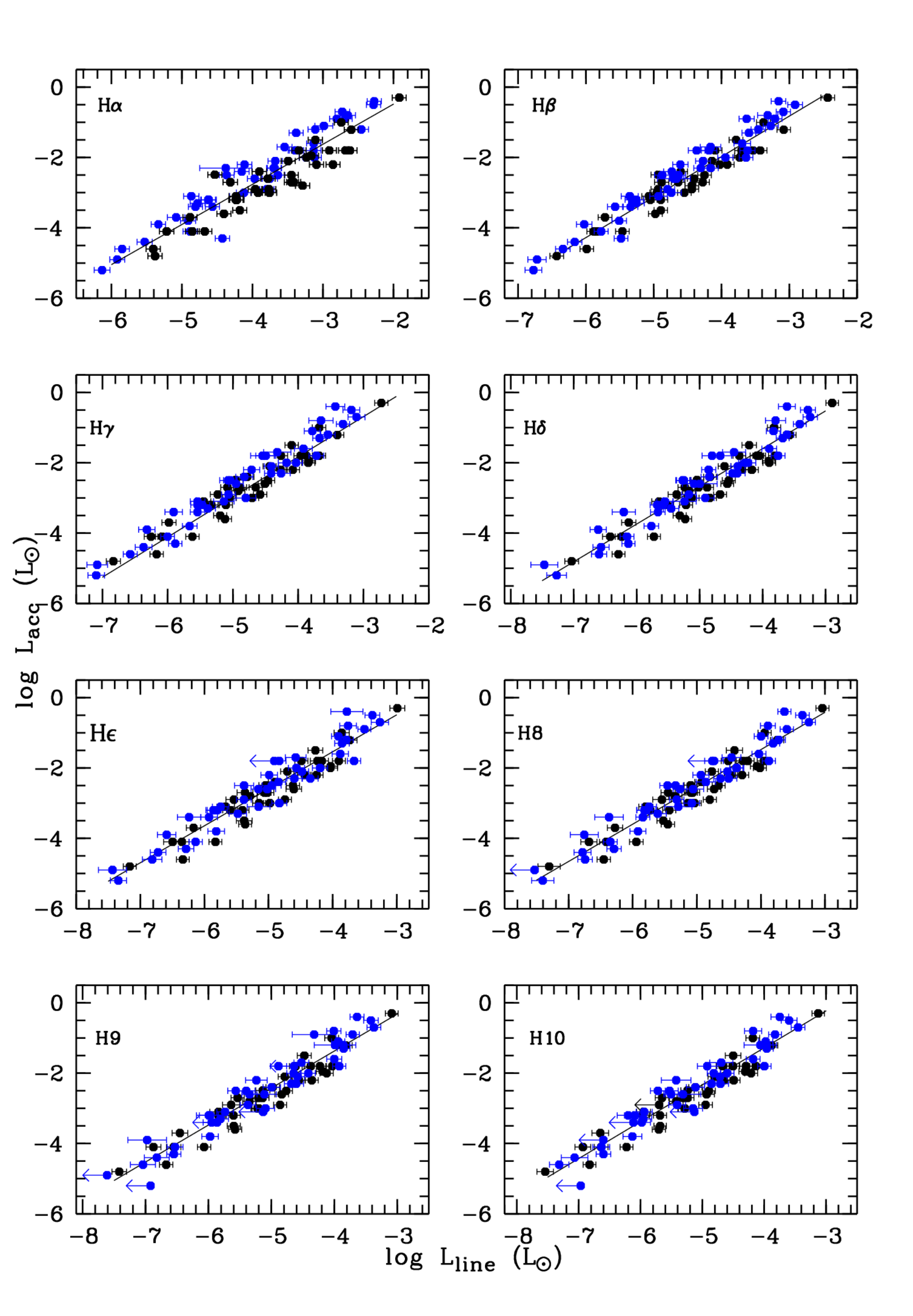}}
\caption{Relationships between accretion luminosity and line luminosity for 
     the several diagnostics as labelled in each panel. The YSOs of the GTO
     and new samples are represented as black and blue dots, respectively. 
    \label{correl1}}
\end{figure}

\begin{figure} 
\resizebox{0.9\hsize}{!}{\includegraphics[]{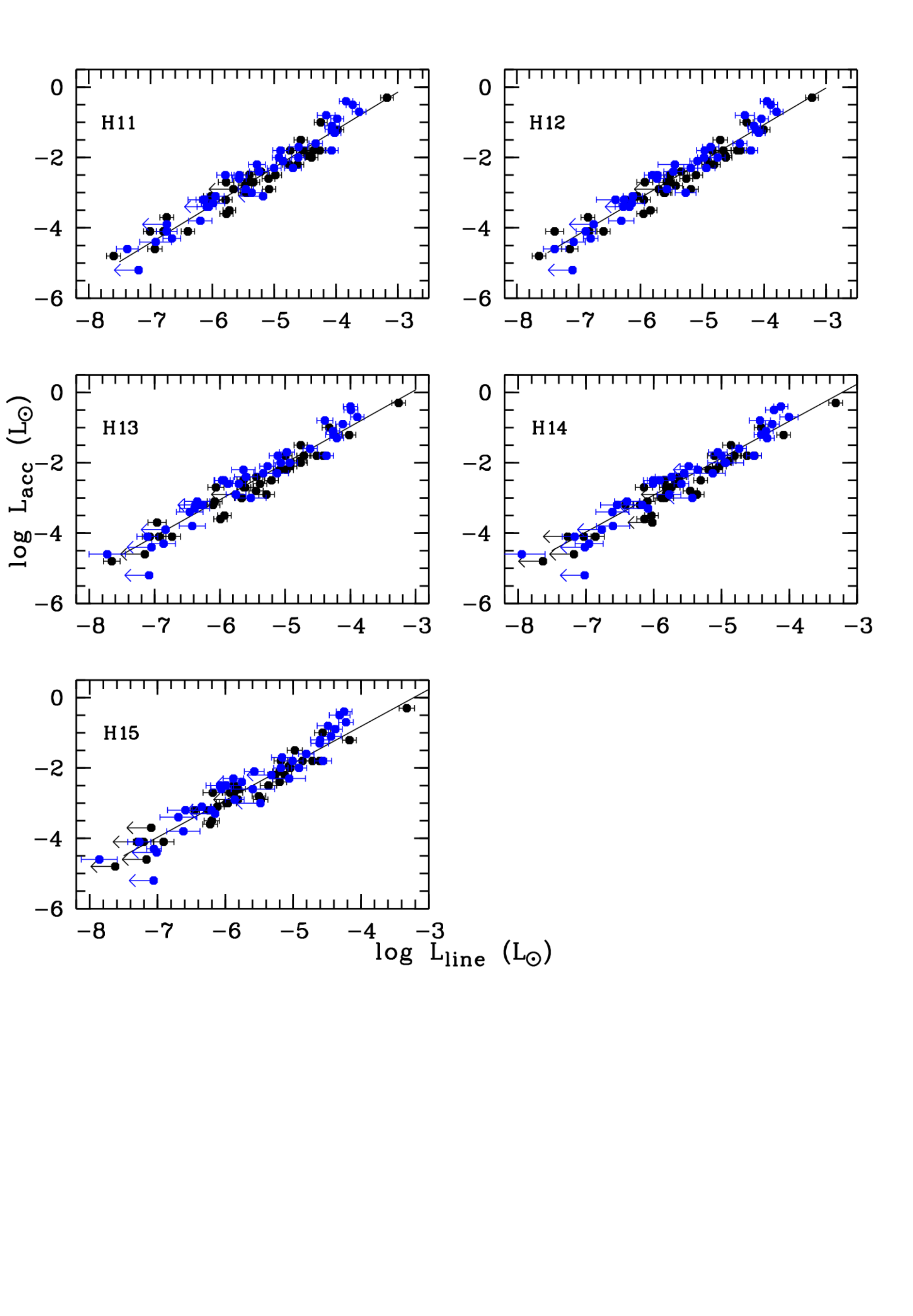}}
\caption{Relationships between accretion luminosity and line luminosity for 
     the several diagnostics as labelled in each panel. Plotting symbols
     are as in Figure~\ref{correl1}.
    \label{correl2}}
\end{figure}

\begin{figure} 
\resizebox{0.9\hsize}{!}{\includegraphics[]{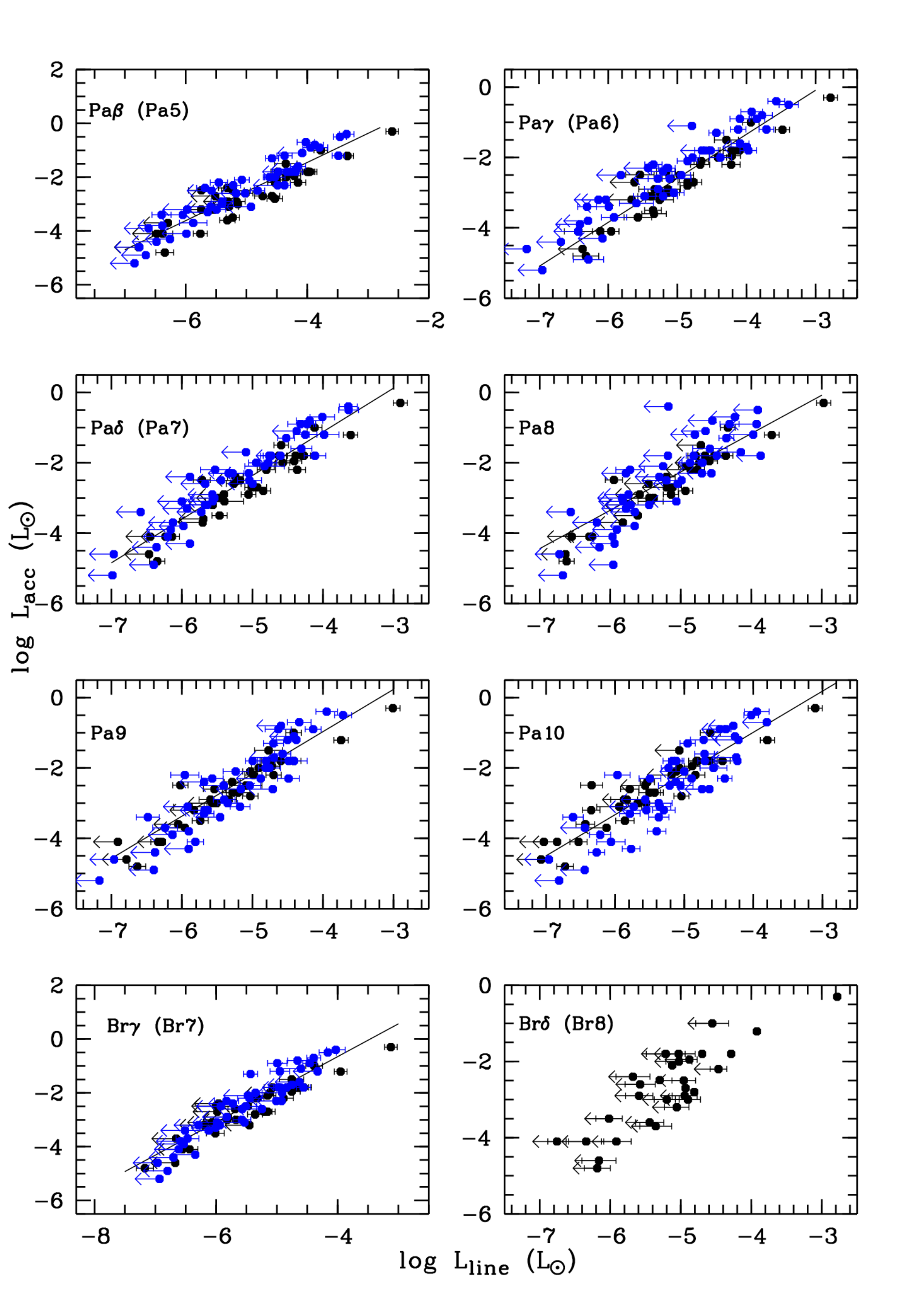}}
\caption{Relationships between accretion luminosity and line luminosity for 
      the several diagnostics as labelled in each panel. Plotting symbols
      are as in Figure~\ref{correl1}.
    \label{correl3}}
\end{figure}

\begin{figure} 
\resizebox{0.9\hsize}{!}{\includegraphics[]{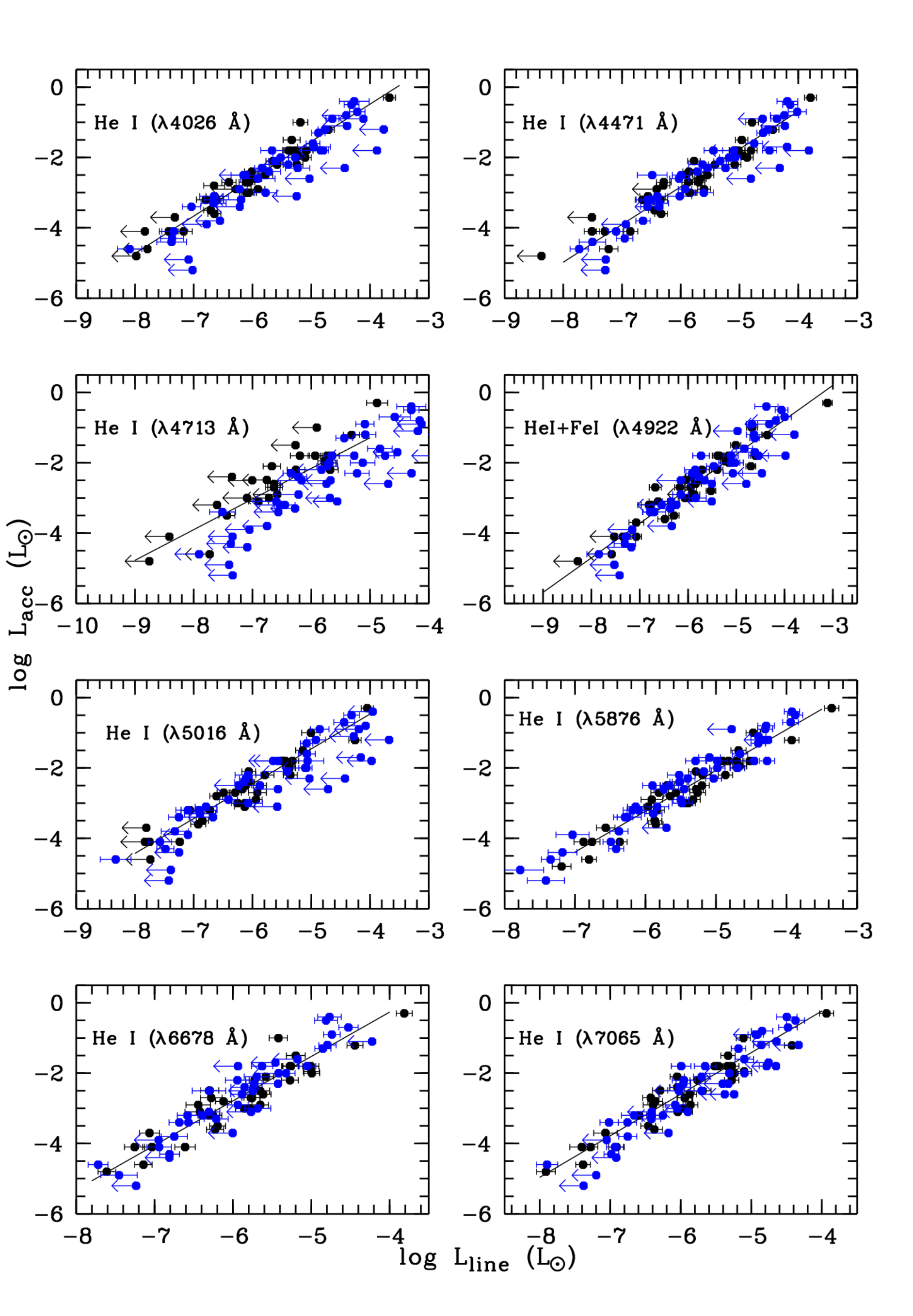}}
\caption{Relationships between accretion luminosity and line luminosity for 
      the several diagnostics as labelled in each panel. Plotting symbols
      are as in Figure~\ref{correl1}.
    \label{correl4}}
\end{figure}


\begin{figure} 
\resizebox{0.9\hsize}{!}{\includegraphics[]{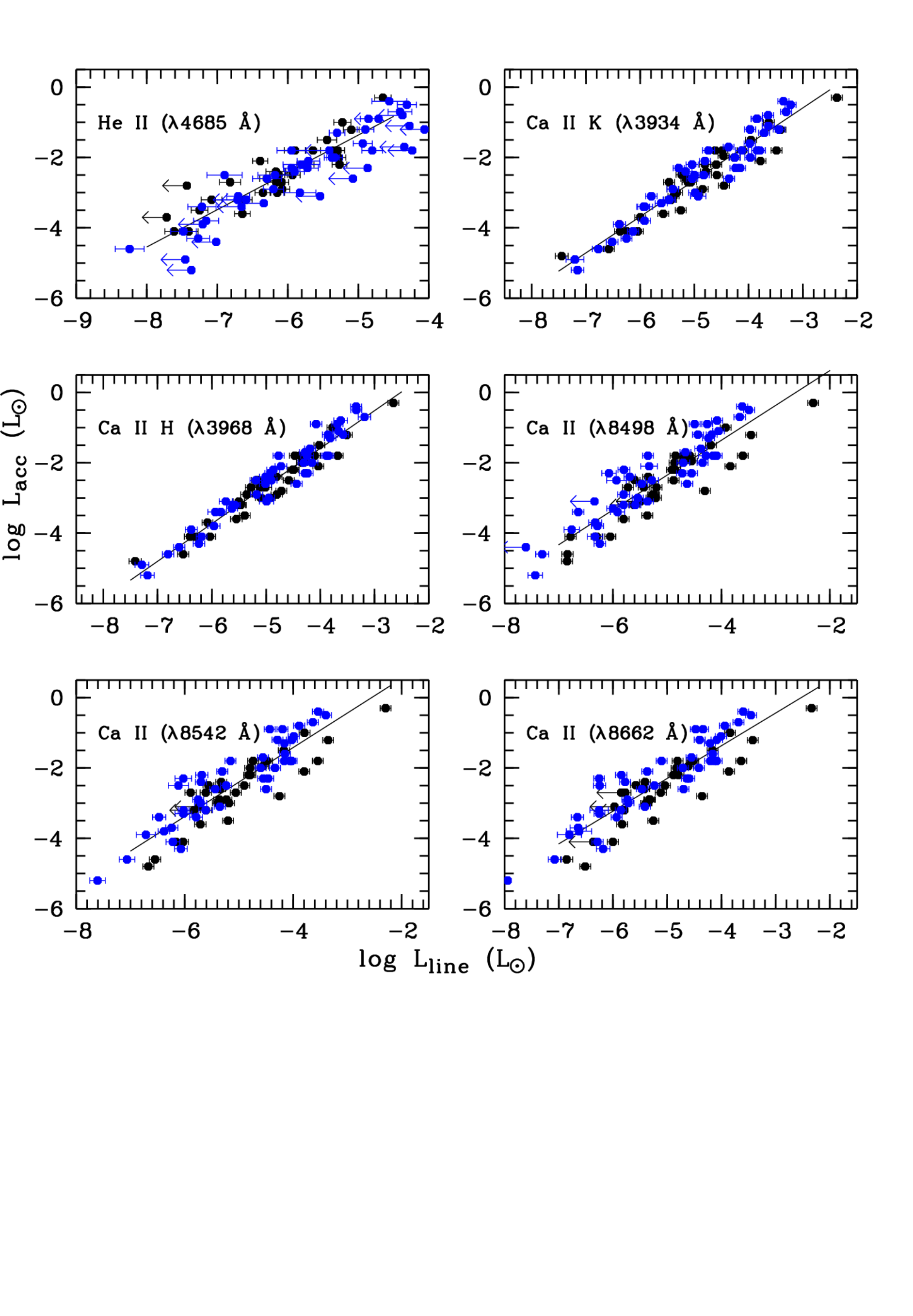}}
\caption{Relationships between accretion luminosity and line luminosity for 
     the several diagnostics as labelled in each panel. Plotting symbols
      are as in Figure~\ref{correl1}.
    \label{correl5}}
\end{figure}


\begin{figure} 
\resizebox{0.9\hsize}{!}{\includegraphics[]{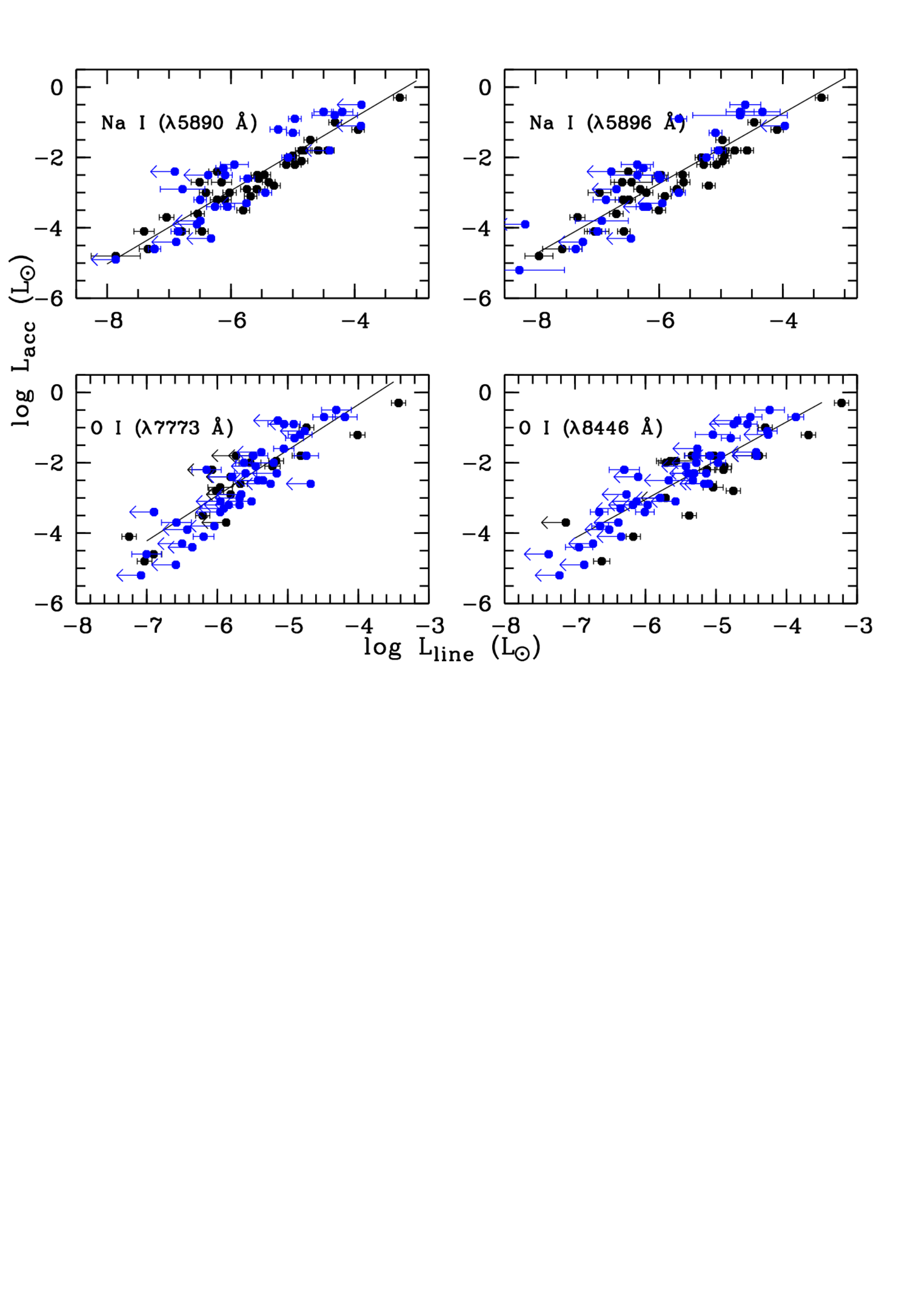}}
\caption{Relationships between accretion luminosity and line luminosity for 
     the several diagnostics as labelled in each panel. Plotting symbols
      are as in Figure~\ref{correl1}.
    \label{correl6}}
\end{figure}


\begin{figure} 
\resizebox{0.9\hsize}{!}{\includegraphics[]{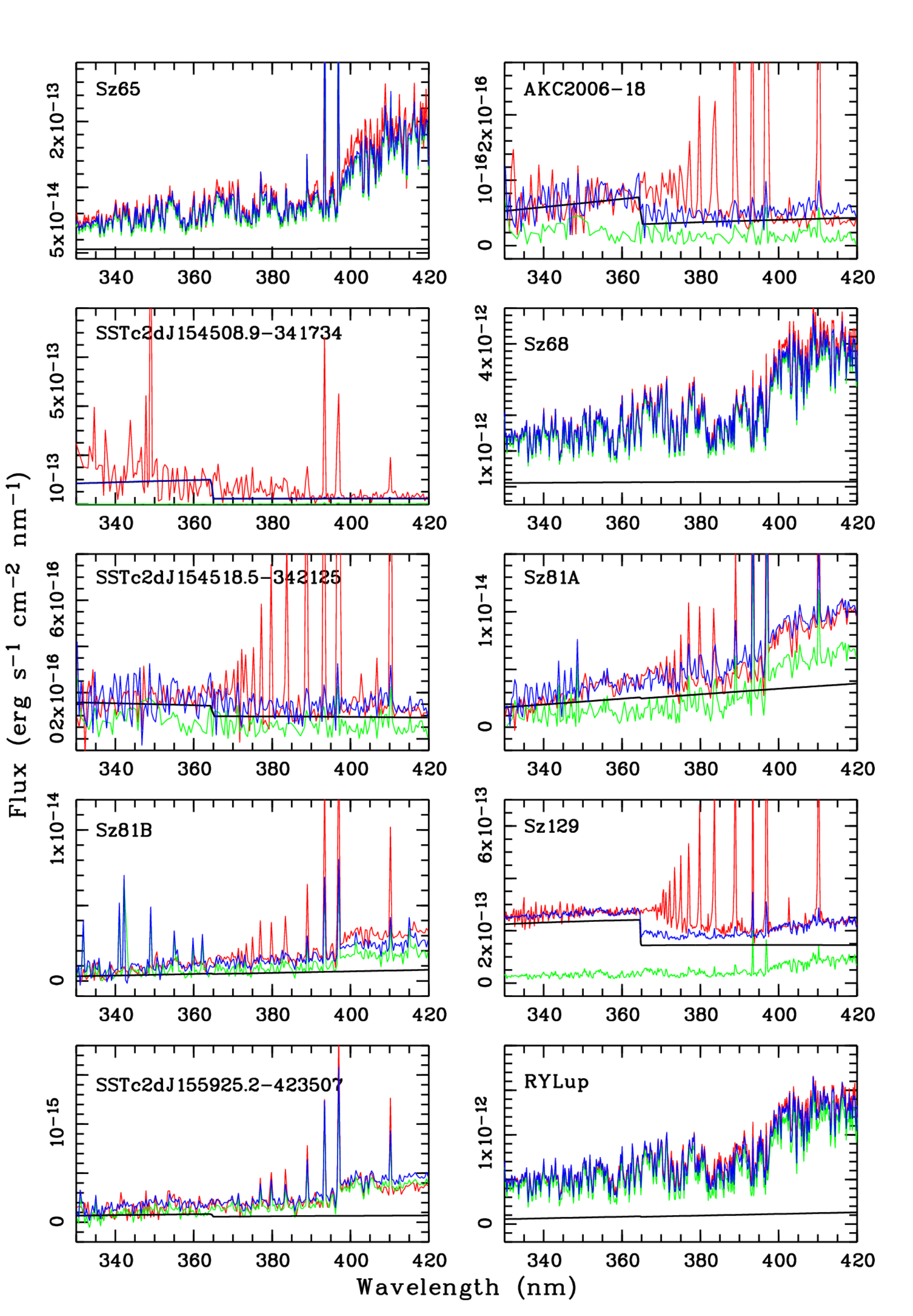}}
\caption{Extinction-corrected spectra (red) fitted with a combination of a 
 photospheric template (green) and the synthetic continuum spectrum from
 a hydrogen slab (black). The total fit is represented with the blue line.
    \label{slab1}}
\end{figure}

\begin{figure} 
\resizebox{0.9\hsize}{!}{\includegraphics[]{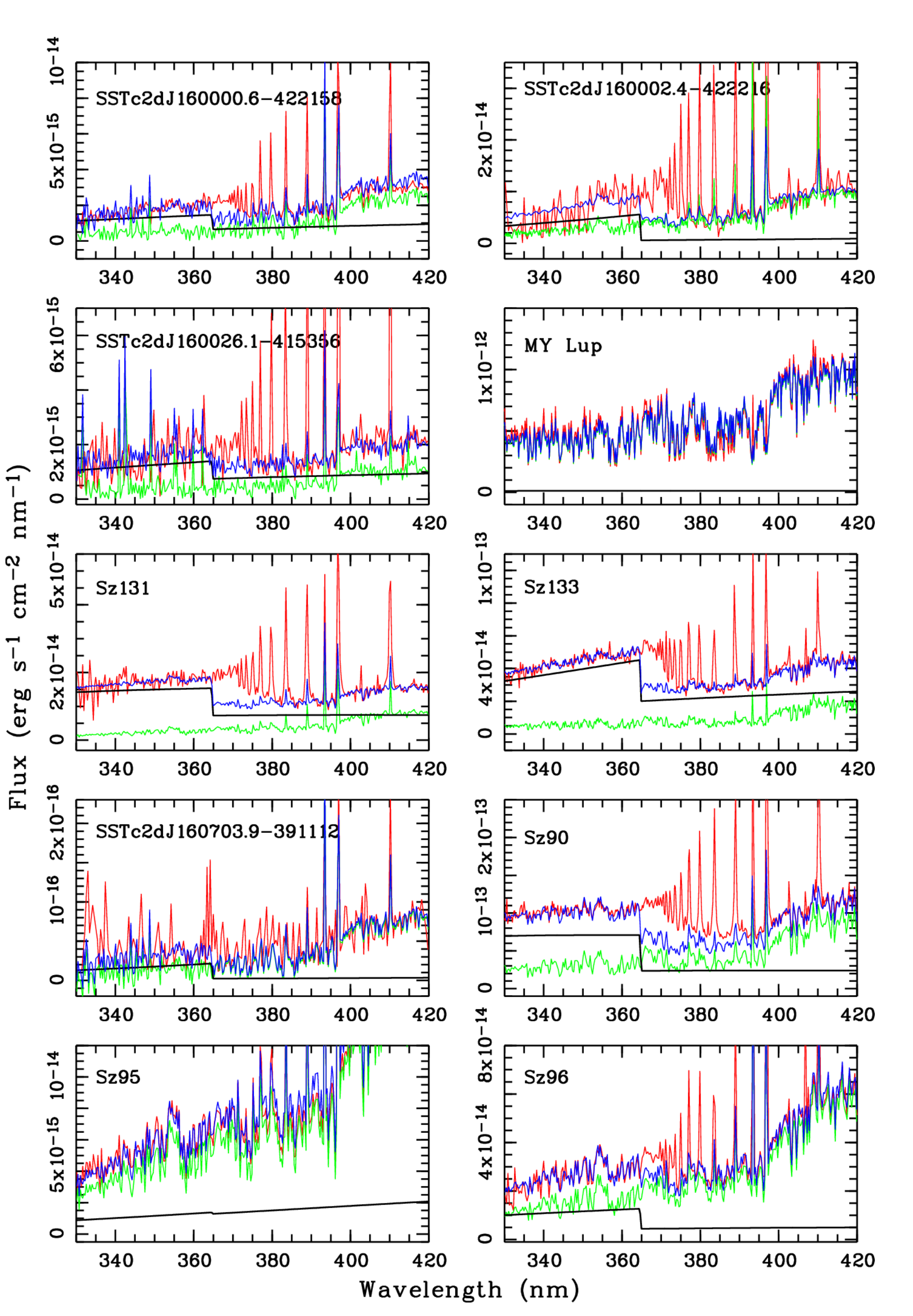}}
\caption{Extinction-corrected spectra (red) fitted with a combination of a 
 photospheric template (green) and the synthetic continuum spectrum from
 a hydrogen slab (black). The total fit is represented with the blue line.
    \label{slab2}}
\end{figure}


\begin{figure} 
\resizebox{0.9\hsize}{!}{\includegraphics[]{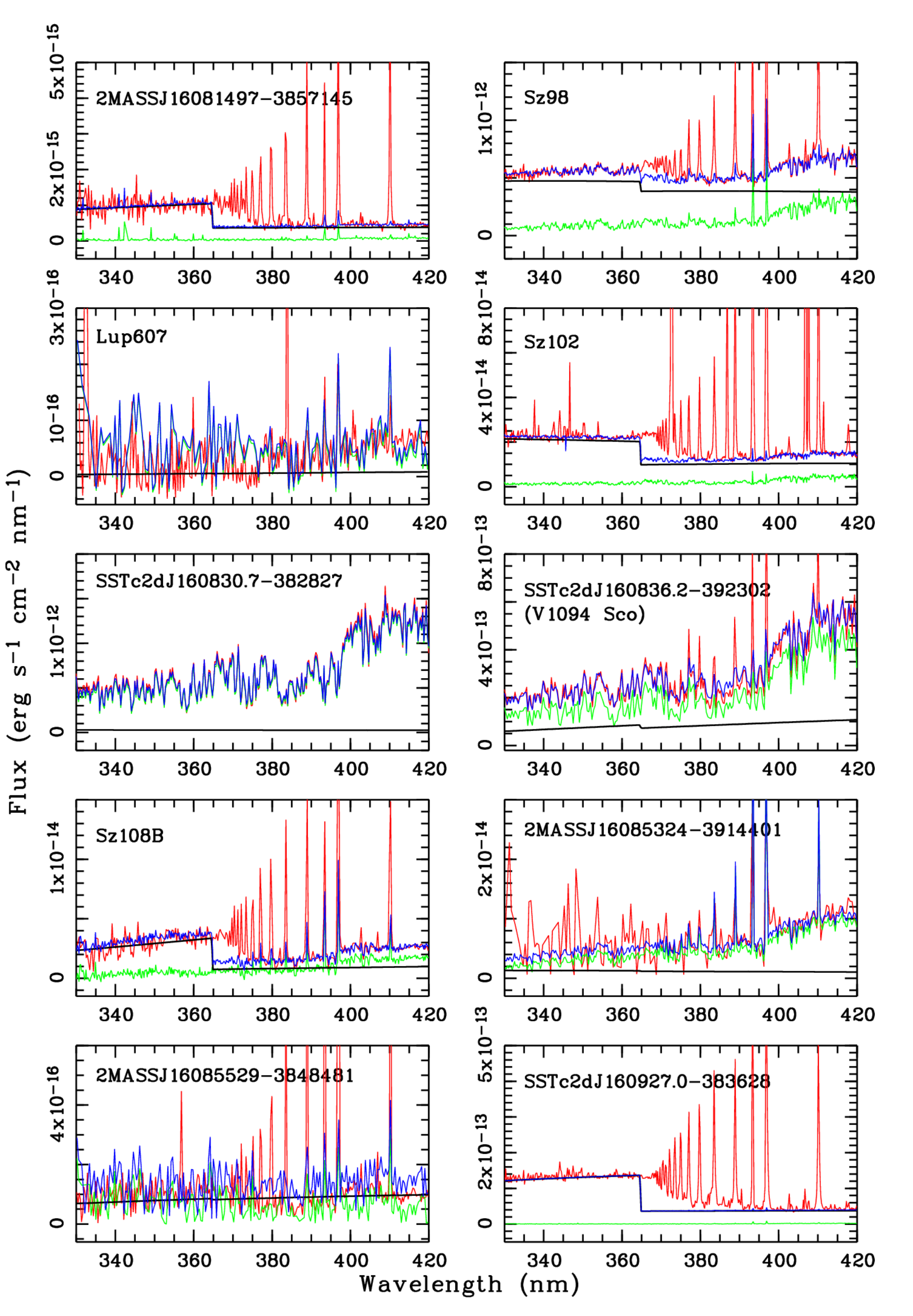}}
\caption{Extinction-corrected spectra (red) fitted with a combination of a 
 photospheric template (green) and the synthetic continuum spectrum from
 a hydrogen slab (black). The total fit is represented with the blue line.
    \label{slab3}}
\end{figure}


\begin{figure} 
\resizebox{0.9\hsize}{!}{\includegraphics[]{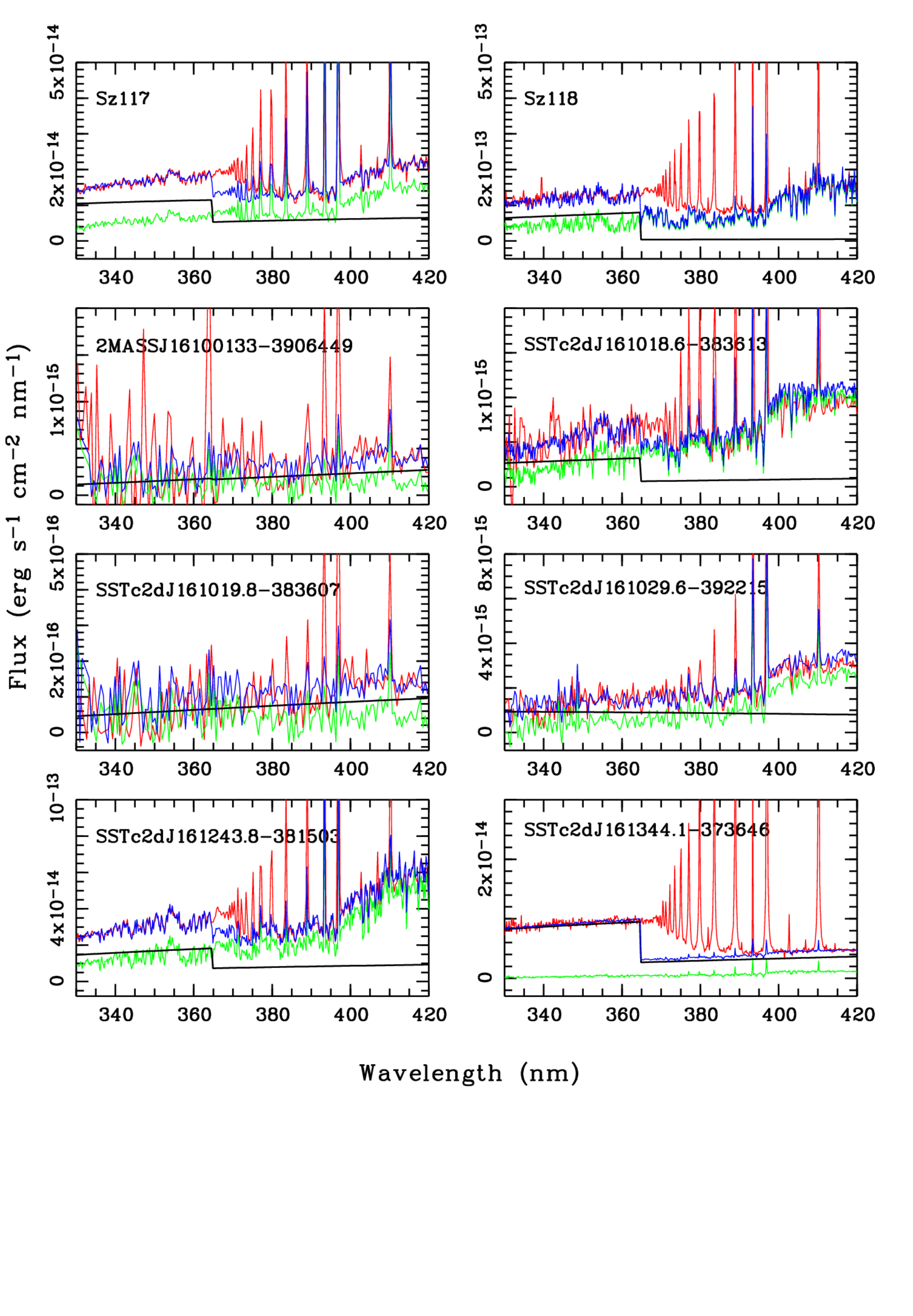}}
\caption{Extinction-corrected spectra (red) fitted with a combination of a 
 photospheric template (green) and the synthetic continuum spectrum from
 a hydrogen slab (black). The total fit is represented with the blue line.
    \label{slab4}}
\end{figure}


\begin{figure} 
\resizebox{0.9\hsize}{!}{\includegraphics[]{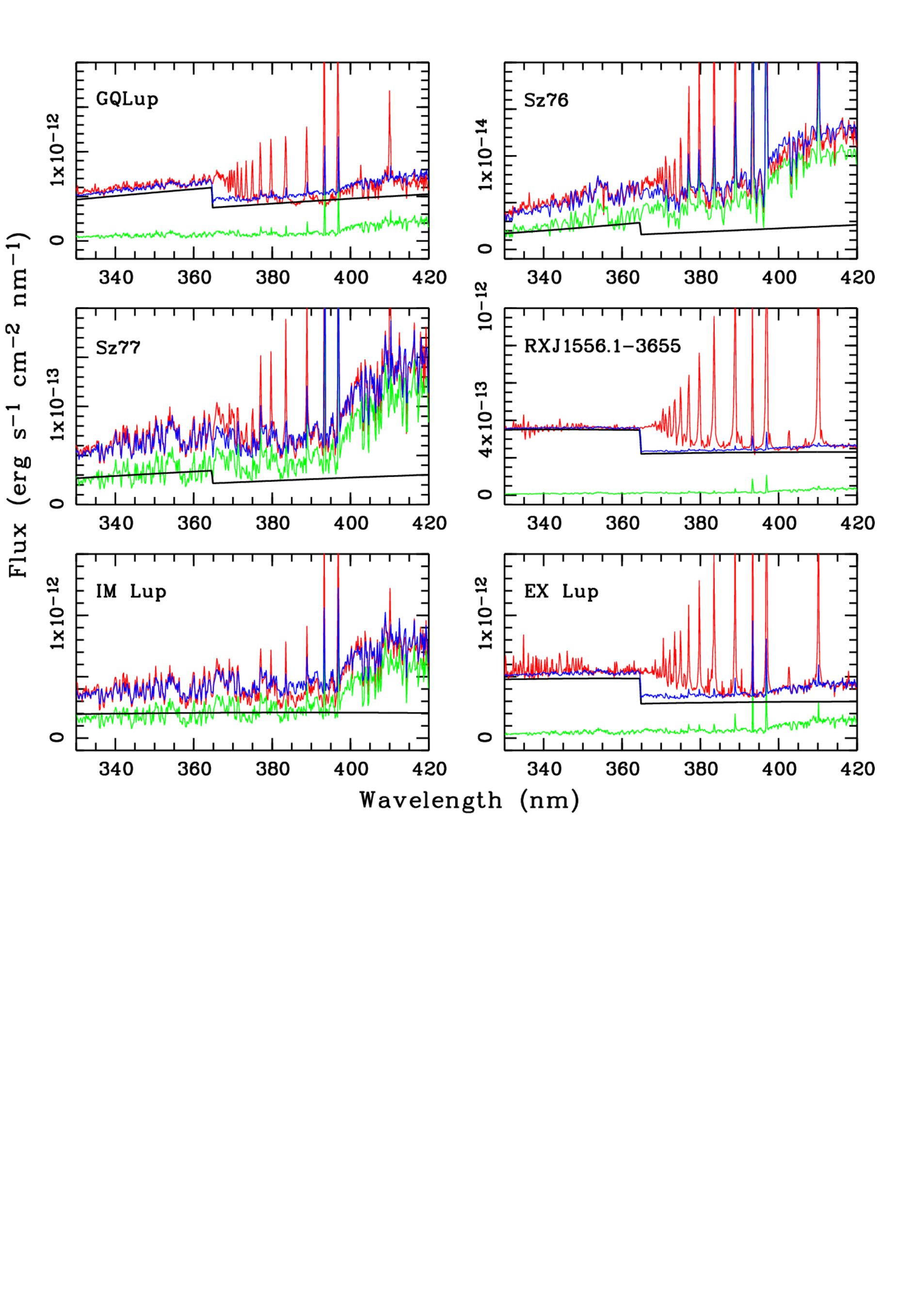}}
\caption{Extinction-corrected spectra of the objects drawn form the ESO archive (red), 
 fitted with a combination of a photospheric template (green) and the synthetic 
 continuum spectrum from a hydrogen slab (black). The total fit is represented with 
 the blue line.
    \label{slab5}}
\end{figure}

\setlength{\tabcolsep}{3pt}

\begin{landscape}
\scriptsize
\begin{longtable}{l|c|r|c|c|c|c|c|c|c|c}
\caption[ ]{\label{tab:fluxes_EWs_Hae} Extinction-corrected fluxes and equivalent widths of Balmer lines for the new sample: H$\alpha$ to H$\epsilon$.}\\
\hline\hline
Object & $f_{\rm H\alpha}$ &  $EW_{\rm H\alpha}$&$f_{\rm H\beta}$& $EW_{\rm H\beta}$&$f_{\rm H\gamma}$& $EW_{\rm H\gamma}$&$f_{\rm H\delta}$& $EW_{\rm H\delta}$&$f_{\rm H\epsilon}$& $EW_{\rm H\epsilon}$\\
    & (erg\,s$^{-1}$\,cm$^{-2}$)&  (nm)      &(erg\,s$^{-1}$\,cm$^{-2}$)& (nm)            &(erg\,s$^{-1}$\,cm$^{-2}$)& (nm)          &(erg\,s$^{-1}$\,cm$^{-2}$)& (nm)          &(erg\,s$^{-1}$\,cm$^{-2}$)& (nm)          \\
\hline
&&&&&&&&&&\\
 Sz65                     &  2.55($\pm$0.22)e$-$13  &  $-$0.348$\pm$0.041  & 2.87($\pm$1.47)e$-$14   &  $-$0.066$\pm$0.022   &            ...            & ...           & 1.45($\pm$0.85)e$-$14   &  $-$0.088$\pm$0.035   & 1.38($\pm$0.27)e$-$14  &  $-$0.207$\pm$0.083 \\ 
 AKC2006-18               &  2.05($\pm$0.10)e$-$15  &  $-$4.280$\pm$0.816  & 6.65($\pm$0.60)e$-$16   &  $-$7.174$\pm$2.670   & 3.87($\pm$0.31)e$-$16   &  $-$4.143$\pm$1.450   & 3.65($\pm$0.57)e$-$16   &  $-$7.620$\pm$3.600   & 2.22($\pm$0.59)e$-$16  &  $-$2.517$\pm$1.259        \\ 
 SSTc2dJ154508.9-341734   &  6.11($\pm$0.12)e$-$13  & $-$20.445$\pm$3.080  & 9.52($\pm$1.22)e$-$14   &  $-$6.407$\pm$1.800   & 4.62($\pm$1.24)e$-$14   &  $-$2.477$\pm$0.867   & 3.08($\pm$1.05)e$-$14   &  $-$4.200$\pm$1.680   & $<$1.78e$-$14          &       ...        \\ 
 Sz68                     &  5.06($\pm$0.20)e$-$12  &  $-$0.670$\pm$0.039  &          ...              &             ...      &             ...         &  ...           &         ...            &           ...         &        ...             &        ...          \\
\hline                                                    
\end{longtable}
 \footnotesize{Notes:
 \begin{itemize}
 \item[] $\dagger$ : not detected in the UVB spectrograph arm.
 \end{itemize}
 }
\end{landscape}

\setlength{\tabcolsep}{5pt}

\begin{landscape}
\scriptsize
\begin{longtable}{l|c|c|c|c|c|c|c|c}
\caption[ ]{\label{tab:fluxes_EWs_H812} Extinction-corrected fluxes and equivalent widths of Balmer lines: H8 to H11.}\\
\hline\hline
Object & $f_{\rm H8}$ &  $EW_{\rm H8}$&$f_{\rm H9}$& $EW_{\rm H9}$&$f_{\rm H10}$& $EW_{\rm H10}$&$f_{\rm H11}$& $EW_{\rm H11}$  \\
    & (erg\,s$^{-1}$\,cm$^{-2}$)&  (nm)      &(erg\,s$^{-1}$\,cm$^{-2}$)& (nm)            &(erg\,s$^{-1}$\,cm$^{-2}$)& (nm)          &(erg\,s$^{-1}$\,cm$^{-2}$)& (nm)     \\
\hline
&&&&&&&\\
Sz65                    & 1.27($\pm$0.38)e$-$14   & $-$0.181$\pm$0.093 &  1.12($\pm$0.41)e$-$14  &  $-$0.214$\pm$0.107  & 7.03($\pm$4.37)e$-$15  &  $-$0.112$\pm$0.056  &  ...             &       ...            \\ 
AKC2006-18              & 2.62($\pm$0.33)e$-$16   & $-$4.695$\pm$2.350 &  1.32($\pm$0.52)e$-$16  &  $-$2.230$\pm$1.115  & 7.00($\pm$1.93)e$-$17  &  $-$0.954$\pm$0.477  & 6.11($\pm$2.01)e$-$17  &  $-$1.000$\pm$0.500  \\ 
SSTc2dJ154508.9-341734  & $<$2.54e$-$14           &        ...         &  $<$3.32e$-$14          &       ...            &        ...             &       ...            & ...            &        ...           \\ 
Sz68                    &       ...               &        ...        &  1.49($\pm$1.07 )e$-$13 &  $-$0.136$\pm$0.068  & 1.25($\pm$0.69)e$-$13  &  $-$0.088$\pm$0.044  &  ...             &       ...            \\
\hline                                                    
\end{longtable}
 \footnotesize{Notes:
 \begin{itemize}
 \item[] $\dagger$ : not detected in the UVB spectrograph arm.
 \end{itemize}
 }
\end{landscape}

\setlength{\tabcolsep}{5pt}

\begin{landscape}
\scriptsize
\begin{longtable}{l|c|c|c|c|c|c|c|c}
\caption[ ]{\label{tab:fluxes_EWs_H1315} Extinction-corrected fluxes and equivalent widths of Balmer lines: H12 to H15.}\\
\hline\hline
Object & $f_{\rm H12}$ &  $EW_{\rm H12}$ & $f_{\rm H13}$ &  $EW_{\rm H13}$ &$f_{\rm H14}$& $EW_{\rm H14}$&$f_{\rm H15}$& $EW_{\rm H15}$                                \\
    & (erg\,s$^{-1}$\,cm$^{-2}$)&  (nm)    & (erg\,s$^{-1}$\,cm$^{-2}$)&  (nm)        &(erg\,s$^{-1}$\,cm$^{-2}$)& (nm)          &(erg\,s$^{-1}$\,cm$^{-2}$)& (nm)       \\
\hline
 & & & & & & & \\

Sz65                     &            ...         &         ...          & $<$2.83e$-$15          &         ...           &  $<$3.71e$-$15          &          ...        & $<$3.68e$-$15           &       ...            \\ 
AKC2006-18               &  5.91($\pm$2.06)e$-$17 &  $-$1.101$\pm$0.440   & 2.73($\pm$1.63)e$-$17  &  $-$0.570$\pm$0.228   &  1.64($\pm$1.24)e$-$17  &  $-$0.177$\pm$0.106  & 2.00($\pm$1.12)e$-$17  &  $-$0.418$\pm$0.251  \\ 
SSTc2dJ154508.9-341734   &          ...           &     ...              &       ...               &     ...               &     ...                &     ...               &     ...                &    ...               \\ 
Sz68                     &          ...           &     ...              &       ...               &     ...               &     ...                &     ...               &     ...                &    ...               \\
\hline                                                    
\end{longtable}
 \footnotesize{Notes:
 \begin{itemize}
 \item[] $\dagger$ : not detected in the UVB spectrograph arm.
 \end{itemize}
 }
\end{landscape}

\setlength{\tabcolsep}{5pt}

\begin{landscape}
\scriptsize
\begin{longtable}{l|c|c|c|c|c|c|c|c}
\caption[ ]{\label{tab:fluxes_EWs_PabPa8} Extinction-corrected fluxes and equivalent widths of Paschen lines: Pa$\beta$ to Pa8.}\\
\hline\hline
Object & $f_{\rm Pa\beta}$ &  $EW_{\rm Pa\beta}$&$f_{\rm Pa\gamma}$& $EW_{\rm Pa\gamma}$&$f_{\rm Pa\delta}$& $EW_{\rm Pa\delta}$&$f_{\rm Pa8}$& $EW_{\rm Pa8}$    \\
    & (erg\,s$^{-1}$\,cm$^{-2}$)&  (nm)      &(erg\,s$^{-1}$\,cm$^{-2}$)& (nm)            &(erg\,s$^{-1}$\,cm$^{-2}$)& (nm)          &(erg\,s$^{-1}$\,cm$^{-2}$) & (nm)\\
\hline
 & & & & & & & & \\
Sz65                    & $<$1.38e$-$14           &         ...         & $<$7.58e$-$15             &     ...             &  $<$1.45e$-$14           &     ...              &   $<$1.31e$-$14           &      ...            \\       
AKC2006-18              & $<$2.50e$-$16           &         ...         & $<$9.60e$-$17             &     ...             &  $<$1.56e$-$16           &     ...              &   $<$2.76e$-$16           &      ...            \\       
SSTc2dJ154508.9-341734  & 4.89($\pm$0.24)e$-$14   & $-$0.508$\pm$0.031  & 3.29($\pm$0.19)e$-$14   &  $-$0.391$\pm$0.028  &  2.50($\pm$0.41)e$-$14   &  $-$0.364$\pm$0.071 &   2.20($\pm$0.17)e$-$14   & $-$0.294$\pm$0.028  \\ 
Sz68                    & 4.72($\pm$0.28)e$-$13   & $-$0.152$\pm$0.010  & 2.82($\pm$0.24)e$-$13   &  $-$0.071$\pm$0.006  &  1.50($\pm$0.64)e$-$13   &  $-$0.033$\pm$0.009 &   $<$1.54e$-$13              &         ...          \\
\hline                                                    
\end{longtable}
\end{landscape}

\setlength{\tabcolsep}{5pt}

\begin{landscape}
\scriptsize
\begin{longtable}{l|c|c|c|c|c|c}
\caption[ ]{\label{tab:fluxes_EWs_Pa9Br8} Extinction-corrected fluxes and equivalent widths of Pa9, Pa10, and Br$\gamma$.}\\
\hline\hline
Object & $f_{\rm Pa9}$ &  $EW_{\rm Pa9}$   &  $f_{\rm Pa10}$ &  $EW_{\rm Pa10}$ & $f_{\rm Br\gamma}$& $EW_{\rm Br\gamma}$  \\
    & (erg\,s$^{-1}$\,cm$^{-2}$)&  (nm)  & (erg\,s$^{-1}$\,cm$^{-2}$)&  (nm)         &(erg\,s$^{-1}$\,cm$^{-2}$)& (nm)       \\
\hline
 & & & & & & \\

Sz65                      &  $<$2.81e$-$14          &    ...               &   $<$2.62e$-$14           &     ...            &  $<$8.23e$-$15           &  ...                  \\ 
AKC2006-18                &  $<$1.58e$-$16          &                      &   $<$1.60e$-$16           &     ...            &  $<$1.58e$-$16           &  ...                  \\ 
SSTc2dJ154508.9-341734    &  1.46($\pm$0.08)e$-$14  &  $-$0.193$\pm$0.013  &   1.11($\pm$0.13)e$-$14   &  $-$0.171$\pm$0.026 &  1.64($\pm$0.17)e$-$14   & $-$0.198$\pm$0.022    \\ 
Sz68                      &  $<$6.06e$-$14          &    ...               &   $<$8.83e$-$14           &     ...            &  $<$6.78e$-$14           &  ...                  \\
\hline                                                    
\end{longtable}
\end{landscape}

\setlength{\tabcolsep}{3pt}

\begin{landscape}
\scriptsize
\begin{longtable}{l|c|c|c|c|c|c|c|c|c|c}
\caption[ ]{\label{tab:fluxes_EWs_Helium1} Extinction-corrected fluxes and equivalent widths of helium lines.}\\
\hline\hline
Object &$f_{\rm \ion{He}{i}~\lambda402.6}$&$EW_{\rm \ion{He}{i}~\lambda402.6}$&$f_{\rm \ion{He}{i}~\lambda447.1}$&$EW_{\rm \ion{He}{i}~\lambda447.1}$&$f_{\rm \ion{He}{i}~\lambda471.3}$& $EW_{\rm \ion{He}{i}~\lambda471.3}$&$f_{\rm \ion{He}{i}~\lambda501.6}$&$EW_{\rm \ion{He}{i}~\lambda501.6}$&$f_{\rm \ion{He}{i}~\lambda587.6}$& $EW_{\rm \ion{He}{i}~\lambda587.6}$\\
   & (erg\,s$^{-1}$\,cm$^{-2}$)&  (nm)       &(erg\,s$^{-1}$\,cm$^{-2}$)& (nm)            &(erg\,s$^{-1}$\,cm$^{-2}$)& (nm)          &(erg\,s$^{-1}$\,cm$^{-2}$)& (nm)          &(erg\,s$^{-1}$\,cm$^{-2}$)& (nm)          \\
\hline
&&&&&&&&&&\\
 Sz65                     & $<$1.35e$-$14            &  ...               &  $<$2.26e$-$14             &      ...            &  $<$2.96e$-$14        &     ...               & $<$2.80e$-$14           &        ...            &  $<$5.12e$-$15            &       ...            \\ 
 AKC2006-18               & 1.17($\pm$0.49)e$-$17    & $-$0.290$\pm$0.145 &  2.71($\pm$0.73)e$-$17    &  $-$0.484$\pm$248   &  $<$1.79e$-$17           &      ...     & 6.80($\pm$3.77)e$-$18   &  $-$0.071$\pm$0.030    &  6.52($\pm$1.26)e$-$17    &  $-$0.530$\pm$0.186   \\ 
 SSTc2dJ154508.9-341734   & $<$2.32e$-$14            &                    &  $<$1.17e$-$14             &      ...            &  $<$7.82e$-$15        &      ...              & $<$3.97e$-$15           &        ...            &  7.42($\pm$1.70)e$-$15    &  $-$0.423$\pm$0.148   \\    
 Sz68                     & $<$2.48e$-$13            &  ...               &      ...                   &      ...            &  $<$3.68e$-$13        &       ...    & $<$3.07e$-$13            &        ...            &  $<$7.97e$-$14           &         ...            \\
\hline                                                    
\end{longtable}
 \footnotesize{Notes:
 \begin{itemize}
 \item[] $\dagger$ : not detected in the UVB spectrograph arm.
 \end{itemize}
 }
\end{landscape}

\setlength{\tabcolsep}{5pt}

\begin{landscape}
\scriptsize
\begin{longtable}{l|c|c|c|c|c|c|c|c}
\caption[ ]{\label{tab:fluxes_EWs_Helium2} Extinction-corrected fluxes and equivalent widths of helium lines.}\\
\hline\hline
Object &$f_{\rm \ion{He}{i}~\lambda667.8}$&$EW_{\rm \ion{He}{i}~\lambda667.8}$&$f_{\rm \ion{He}{i}~\lambda706.5}$ & $EW_{\rm \ion{He}{i}~\lambda706.5}$ & $f_{\rm \ion{He}{i}+\ion{Fe}{i}~\lambda492.2}$ & $EW_{\rm \ion{He}{i}+\ion{Fe}{i}~\lambda492.2}$& $f_{\rm \ion{He}{ii}~\lambda468.5}$ & $EW_{\rm \ion{He}{ii}~\lambda468.5}$    \\
       & (erg\,s$^{-1}$\,cm$^{-2}$)       &  (nm)                             &(erg\,s$^{-1}$\,cm$^{-2}$)          & (nm)                                 & (erg\,s$^{-1}$\,cm$^{-2}$)                     & (nm)                                       & (erg\,s$^{-1}$\,cm$^{-2}$)           & (nm)                                    \\
\hline
 &&&&&&&& \\
Sz65                    &       ...                &     ...             &   $<$6.10e$-$15            &  ...                  &  $<$2.33e$-$14            &   ...           &  $<$1.22e$-$14           &  ...                   \\ 
AKC2006-18              &  2.76($\pm$0.49)e$-$17   & $-$0.090$\pm$0.023  &   1.84($\pm$0.48)e$-$17     & $-$0.040$\pm$0.011   &  2.06($\pm$1.06)e$-$17    &  $-$0.148$\pm$0.067    &  8.28($\pm$3.38)e$-$18   &  $-$0.104$\pm$0.047    \\ 
SSTc2dJ154508.9-341734  &  3.41($\pm$0.73)e$-$15   & $-$0.166$\pm$0.048  &   1.47($\pm$0.32)e$-$15     & $-$0.058$\pm$0.014   &  1.11($\pm$0.30)e$-$14    &  $-$0.795$\pm$0.358    &  $<$5.63e$-$15           &   ...                 \\ 
Sz68                    &       ...                &     ...             &   $<$6.80e$-$14            &  ...                  &  $<$2.31e$-$13            &   ...           &  $<$1.26e$-$13           &   ...                  \\
\hline                                                    
\end{longtable}
\end{landscape}

\setlength{\tabcolsep}{3pt}
\begin{landscape}
\scriptsize
\begin{longtable}{l|c|c|c|c|c|c|c|c|c|c}
\caption[ ]{\label{tab:fluxes_EWs_CaII} Extinction-corrected fluxes and equivalent widths of \ion{Ca}{ii} H and K and IRT lines.}\\
\hline\hline
Object & $f_{\rm \ion{Ca}{ii}~\lambda393.4}$ & $EW_{\rm \ion{Ca}{ii}~\lambda393.4}$&$f_{\rm \ion{Ca}{ii}~\lambda396.8}$& $EW_{\rm \ion{Ca}{ii}~\lambda396.8}$&$f_{\rm \ion{Ca}{ii}~\lambda849.8}$& $EW_{\rm \ion{Ca}{ii}~\lambda849.8}$&$f_{\rm \ion{Ca}{ii}~\lambda854.2}$& $EW_{\rm \ion{Ca}{ii}~\lambda854.2}$&$f_{\rm \ion{Ca}{ii}~\lambda866.2}$& $EW_{\rm \ion{Ca}{ii}~\lambda866.2}$\\
    & (erg\,s$^{-1}$\,cm$^{-2}$)&  (nm)      &(erg\,s$^{-1}$\,cm$^{-2}$)& (nm)            &(erg\,s$^{-1}$\,cm$^{-2}$)& (nm)          &(erg\,s$^{-1}$\,cm$^{-2}$)& (nm)          &(erg\,s$^{-1}$\,cm$^{-2}$)& (nm)          \\
\hline
&&&&&&&&&&\\
Sz65                      &   6.34($\pm$0.18)e$-$14  & $-$1.470$\pm$0.204  &  5.31($\pm$0.32)e$-$14   & $-$0.836$\pm$0.235  &  3.37($\pm$0.28)e$-$14  & $-$0.037$\pm$0.002  &  4.58($\pm$0.07)e$-$14  & $-$0.051$\pm$0.003  &  2.90($\pm$0.09)e$-$14  & $-$0.033$\pm$0.002   \\ 
AKC2006-18                &   2.46($\pm$0.26)e$-$16  & $-$4.970$\pm$2.230  &  2.26($\pm$0.17)e$-$16   & $-$4.371$\pm$1.967  &  7.18($\pm$1.09)e$-$17  & $-$0.030$\pm$0.013  &  1.25($\pm$0.30)e$-$16  & $-$0.052$\pm$0.017  &  1.22($\pm$0.18)e$-$16  & $-$0.050$\pm$0.013   \\ 
SSTc2dJ154508.9-341734    &   1.09($\pm$0.15)e$-$13  & $-$4.579$\pm$2.061  &  8.26($\pm$1.09)e$-$14   & $-$2.587$\pm$1.164  &  1.18($\pm$0.03)e$-$13  & $-$2.197$\pm$0.033  &  1.27($\pm$0.01)e$-$13  & $-$2.469$\pm$0.039  &  1.18($\pm$0.02)e$-$13  & $-$2.441$\pm$0.033   \\ 
Sz68                      &   5.26($\pm$0.47)e$-$13  & $-$0.650$\pm$0.137  &  3.82($\pm$0.17)e$-$13   & $-$0.334$\pm$0.007  &  9.56($\pm$0.17)e$-$14  & $-$0.016$\pm$0.002  &  1.37($\pm$0.15)e$-$13  & $-$0.023$\pm$0.002  &  1.19($\pm$0.19)e$-$13  & $-$0.020$\pm$0.002   \\
\hline                                                    
\end{longtable}
 \footnotesize{Notes:
 \begin{itemize}
 \item[] $\dagger$ : not detected in the UVB spectrograph arm.
 \end{itemize}
 }
\end{landscape}

\setlength{\tabcolsep}{5pt}

\begin{landscape}
\scriptsize
\begin{longtable}{l|c|c|c|c}
\caption[ ]{\label{tab:fluxes_EWs_NaI} Extinction-corrected fluxes and equivalent widths of the $\ion{Na}{i}$~D lines }\\
\hline\hline
Object & $f_{\rm \ion{Na}{i}~\lambda588.99}$ &  $EW_{\rm \ion{Na}{i}~\lambda588.99}$ & $f_{\rm \ion{Na}{i}~\lambda589.59}$ &  $EW_{\rm \ion{Na}{i}~\lambda589.59}$ \\
    & (erg\,s$^{-1}$\,cm$^{-2}$)&  (nm)    & (erg\,s$^{-1}$\,cm$^{-2}$)&  (nm)         \\
\hline
 & & & &  \\
Sz65                      &     ...                &     ...              &      ...               &      ...             \\ 
AKC2006-18                & 8.45($\pm$0.15)e$-$17  & $-$1.125$\pm$0.057   & 6.50($\pm$0.45)e$-$17  & $-$3.009$\pm$1.080   \\ 
SSTc2dJ154508.9-341734    & $<$5.75e$-$14          &     ...              & 1.32($\pm$0.18)e$-$14  & $-$0.709$\pm$0.164   \\ 
Sz68                      &     ...                &     ...              &      ...               &      ...             \\
\hline                                                    
\end{longtable}
\end{landscape}

\end{appendix}

\end{document}